\documentclass[iop]{emulateapj}
\usepackage{epsfig}

\usepackage{amsfonts}
\usepackage{amsmath}
\usepackage{mathrsfs}
\usepackage{amssymb}
\usepackage{rotating}

\usepackage{color}
\usepackage{ulem}
\usepackage{xspace}
\usepackage{cancel}
\definecolor{lgray}{gray}{0.85}

\newcommand{\lya} {Ly$\alpha$\xspace}
\newcommand{\civ} {\ion{C}{4}\xspace}
\newcommand{\heii}{\ion{He}{2}\xspace}
\newcommand{\nv}{\ion{N}{5}\xspace}
\newcommand{\niv}{\ion{N}{4}]\xspace}
\newcommand{\nii}{[\ion{N}{2}]\xspace}
\newcommand{\cii}{\ion{C}{2}]\xspace}
\newcommand{\oii}{[\ion{O}{2}]\xspace}

\newcommand{\unitcgslum} {erg\,s$^{-1}$\xspace}
\newcommand{\unitcgssb}  {erg\,s$^{-1}$\,cm$^{-2}$\,arcsec$^{-2}$\xspace}
\newcommand{\unitcgsflux}{erg\,s$^{-1}$\,cm$^{-2}$\xspace}
\def\bea{\begin{eqnarray}}
\def\eea{\end{eqnarray}}

\begin{document}

\title{A deep narrowband imaging search for \civ and \heii emission from L\lowercase{y}$\alpha$ 
Blobs\footnotemark[\large $\star$] }\footnotetext[\large $\star$]{Based on observations collected at the European Southern Observatory, Chile, under programs {\tt 
085.A-0989}, {\tt 087.A-0297}.}
 
\author{Fabrizio Arrigoni Battaia\altaffilmark{1,2}, 
        Yujin Yang\altaffilmark{1,3},
        Joseph F. Hennawi\altaffilmark{1}, 
        J. Xavier Prochaska\altaffilmark{4,5}, 
	Yuichi Matsuda\altaffilmark{6,7}, 
        Toru Yamada\altaffilmark{8}, 
        Tomoki Hayashino\altaffilmark{9}
        }
\altaffiltext{1}{Max-Planck-Institut f\"ur Astronomie, K\"onigstuhl 17, D-69117 Heidelberg, Germany; arrigoni@mpia.de}
\altaffiltext{2}{Member of the International Max Planck Research School for Astronomy \& Cosmic Physics at the University of Heidelberg (IMPRS-HD)}
\altaffiltext{3}{Argelander Institut f\"ur Astronomie, Universit\"at Bonn, Auf dem H\"ugel 71, 53121 Bonn, Germany}
\altaffiltext{4}{Department of Astronomy and Astrophysics, University of California, 1156 High Street, Santa Cruz, California 95064, USA}
\altaffiltext{5}{University of California Observatories, Lick Observatory, 1156 High Street, Santa Cruz, California 95064, USA}
\altaffiltext{6}{National Astronomical Observatory of Japan, 2-21-1 Osawa, Mitaka, Tokyo 181-8588, Japan}
\altaffiltext{7}{The Graduate University for Advanced Studies (SOKENDAI), 2-21-1 Osawa, Mitaka, Tokyo 181-0015, Japan}
\altaffiltext{8}{Astronomical Institute, Tohoku University, Aramaki, Aoba-ku, Sendai, Miyagi 980-8578, Japan}
\altaffiltext{9}{Research Center for Neutrino Science, Graduate School of Science, Tohoku University, Sendai 980-8578, Japan}

\slugcomment{Submitted to ApJ}
\shorttitle{\civ and \heii Emission from \lya Blobs}
\shortauthors{Arrigoni Battaia et al.}
 
\begin{abstract}

We conduct a deep narrow-band imaging survey of 13 \lya blobs (LABs)
located in the SSA22 proto-cluster at $z\sim3.1$ in the \civ and \heii
emission lines in an effort to constrain the physical process powering
the \lya emission in LABs.  Our observations probe down to
unprecedented surface brightness limits of 2.1 -- 3.4 $\times$
$10^{-18}$ \unitcgssb per 1 arcsec$^2$ aperture (5$\sigma$) for the
\heii$\lambda$1640 and \civ$\lambda$1549 lines, respectively.  We do
not detect extended \heii and \civ emission in any of the LABs,
placing strong upper limits on the \heii/Ly$\alpha$ and
\civ/Ly$\alpha$ line ratios, of 0.11 and 0.16, for the brightest two
LABs in the field. We conduct detailed photoionization modeling of the
expected line ratios and find that, although our data constitute the
deepest ever observations of these lines, they are still not deep
enough to rule out a scenario where the \lya emission is powered by
the ionizing luminosity of an obscured AGN. Our models can accommodate
\heii/Ly$\alpha$ and \civ/Ly$\alpha$ ratios as low as $\simeq0.05$ and
$\simeq0.07$ respectively, implying that one needs to reach surface
brightness as low as 1 -- 1.5 $\times$ $10^{-18}$ \unitcgssb (at 5$\sigma$) in order to
rule out a photoionization scenario.  These depths will be achievable
with the new generation of image-slicing integral field units such as
VLT/MUSE or Keck/KCWI.  We also model the expected \heii/Ly$\alpha$
and \civ/Ly$\alpha$ in a different scenario, where \lya emission is
powered by shocks generated in a large-scale superwind, but find that
our observational constraints can only be met for shock velocities
$v_{\rm s} \gtrsim$ 250 km s$^{-1}$, which appear to be in conflict
with recent observations of quiescent kinematics in LABs.

\end{abstract}

\maketitle

\section{Introduction}

In the current $\Lambda$CDM  paradigm of structure formation,
gas collapses onto the potential wells of dark matter halos, and whether 
it shock heats to the halo virial temperature and cools slowly, or
flows in preferentially along cold filamentary streams (\citealt{Dekel2009}),
its gravitational energy is eventually radiated away, as it
settles into galactic disks and forms stars. 
This star formation results in the growth of 
galactic bulges, and in the innermost regions, the gas could also accrete onto a
supermassive black hole powering an active galactic nucleus
(AGN). Many have theorized (e.g., \citealt{SilkRees1998, Fabian1999, King2003}) 
that star-formation and/or BH
accretion could be self-regulating, such that ``feedback'' processes
inject energy back into the inter-stellar medium (ISM), heating the gas, and preventing
further star-formation or accretion.

The complex interplay of gas accreted from the intergalactic medium (IGM)
and the galactic outflows which may be the signatures of
mechanical/radiative feedback are poorly understood, particularly at
high-redshift, where the feedback processes are often invoked as being
most intense. These processes conspire to determine the structure of
the circumgalactic medium (CGM), which comprises the interface between 
galaxies and the IGM. 
At high redshift, the CGM has been extensively studied by analyzing
absorption features in the spectra of background sources. A significant 
amount of effort has been devoted to the studying of the CGM of 
the so-called Lyman
break galaxies (LBGs), star-forming galaxies at $z\sim2$ (\citealt{Adelberger2005,Steidel2010,
Crighton2011, Rakic2012, Rudie2012, Crighton2013, Crighton2014}). 
These studies have illustrated that typical star-forming galaxies exhibit
a modest $\sim 20\%$ covering factor of optically thick neutral hydrogen (\citealt{Rudie2012}), 
and  enrichment levels ranging from extremely metal-poor (\citealt{Crighton2013}) to nearly
solar (\citealt{Crighton2014}). 
On the other hand, using projected QSO pairs, \citet{Hennawi2006}
launched an innovative technique to study the properties of the gas on
scales of a few 10 kpc to several Mpc of the much more massive dark
matter halos traced by quasars, initiating the Quasars Probing Quasars
survey (\citealt{Hennawi2007, Prochaska2009, Hennawi2013,
  Prochaska2013, Prochaska2013b}). These studies have revealed a
massive $(\gtrsim 10^{10}$ M$_{\odot}$) resevoir of cool ($T\simeq 10^4\,{\rm
  K}$ gas in the CGM of massive halos (see also
\citealt{Bowen2006,Farina2013}), which appears to be in conflict with
the predictions of hydrodynamical zoom-in simulations of galaxy
formation (\citealt{Fumagalli2014}).

These absorption studies are however, limited by the paucity of bright
background sources, and by the inherently one-dimensional nature of
the technique. Complementary information can be obtained by directly
observing the CGM in emission, and this emission may be easier to
detect in AGN environments.  In particular, if an AGN illuminates the
cool CGM gas around it, the reprocessed emission (fluorescence) from
this cool medium could be detectable as extended Ly$\alpha$ emission
(e.g., \citealt{Rees1988, HaimanRees2001}).  Indeed, many searches for
emission from the CGM of QSOs have been undertaken, reporting
detections on scale of $10-50$ kpc around $z\sim2-4$ QSOs (e.g,
\citealt{HuCowie1987, Heckman1991spec, Heckman1991, Christensen2006,
  North2012}).
Recently \citet{Cantalupo2014} reported the discovery of an extraordinary 
extended ($\sim500$ kpc) Ly$\alpha$ nebula around the radio-quiet QSO UM287, believed
to be fluorescent emission powered by the QSO radiation. This discovery is part of 
a large homogenous survey of emission from the CGM of quasars which will enable 
statistical studies of this phenomenon (e.g., \citealt{FABProceeding}).  

Extended Ly$\alpha$ nebulae have also been frequently observed also
around high-redshift ($z\geq2$) radio galaxies (HzRGs; e.g.,
\citealt{McCarthy1993, vanOjik1997, Nesvadba2006, VillarM2007,
  Reuland2007}).  With an avarage \lya luminosity of $L_{{\rm
    Ly}\alpha} \sim 10^{44.5}$ erg s$^{-1}$ and a diameter $\gtrsim
100$ kpc, these nebulae tend to be brighter and larger than those
around QSOs, although current surveys are very inhomogenous. But 
an important difference between these two types of nebulae is that for
quasars a strong source of ionizing photons is directly identified,
whereas for the HzRGs this AGN is obscured from our perspective
\citep[see e.g.][]{Miley2008}), in accord with unified models of AGN
(e.g., \citealt{Antonucci1993, UP1995, Elvis2000}).  Further, the
study of the properties of the gas surrounding HzRGs has to take into
account the impact of the complicated interaction between the strong 
radio jets and the ambient gas.

Intriguingly, the so-called \lya blobs (LABs), large (50--100~kpc)
luminous ($L$(\lya) $\sim$ 10$^{43-44}$ \unitcgslum) \lya nebulae at
$z\sim 2-6$, exhibit properties similar to \lya nebulae around QSOs
and HzRGs, but without obvious evidence for the presence of an AGN
\cite[e.g.,][]{Keel1999, Steidel2000, Francis2001, Matsuda2004,
  Matsuda2011, Dey2005, Saito2006, Smith2007, Ouchi2009, Prescott2009,
  Prescott2012a, Yang2009, Yang2010}. LABs are believed to be the
sites of massive galaxy formation, where strong feedback processes may
be expected to occur (\citealt{Yang2010}). However, despite intense
interest and multi-wavelength studies, the physical mechanism powering
the Ly$\alpha$ emission in the LABs is still poorly understood. The
proposed scenarios include photo-ionization by AGNs \citep{Geach2009},
shock-heated gas by galactic superwinds \citep{Taniguchi&Shioya2000},
cooling radiation from cold-mode accretion \citep{Fardal2001,
  Haiman2000, Dijkstra&Loeb2009, Goerdt2010, Faucher2010},
and resonant scattering of \lya from star-forming galaxies \citep{Steidel2011, Hayes2011}.

\begin{figure}
\epsfig{file=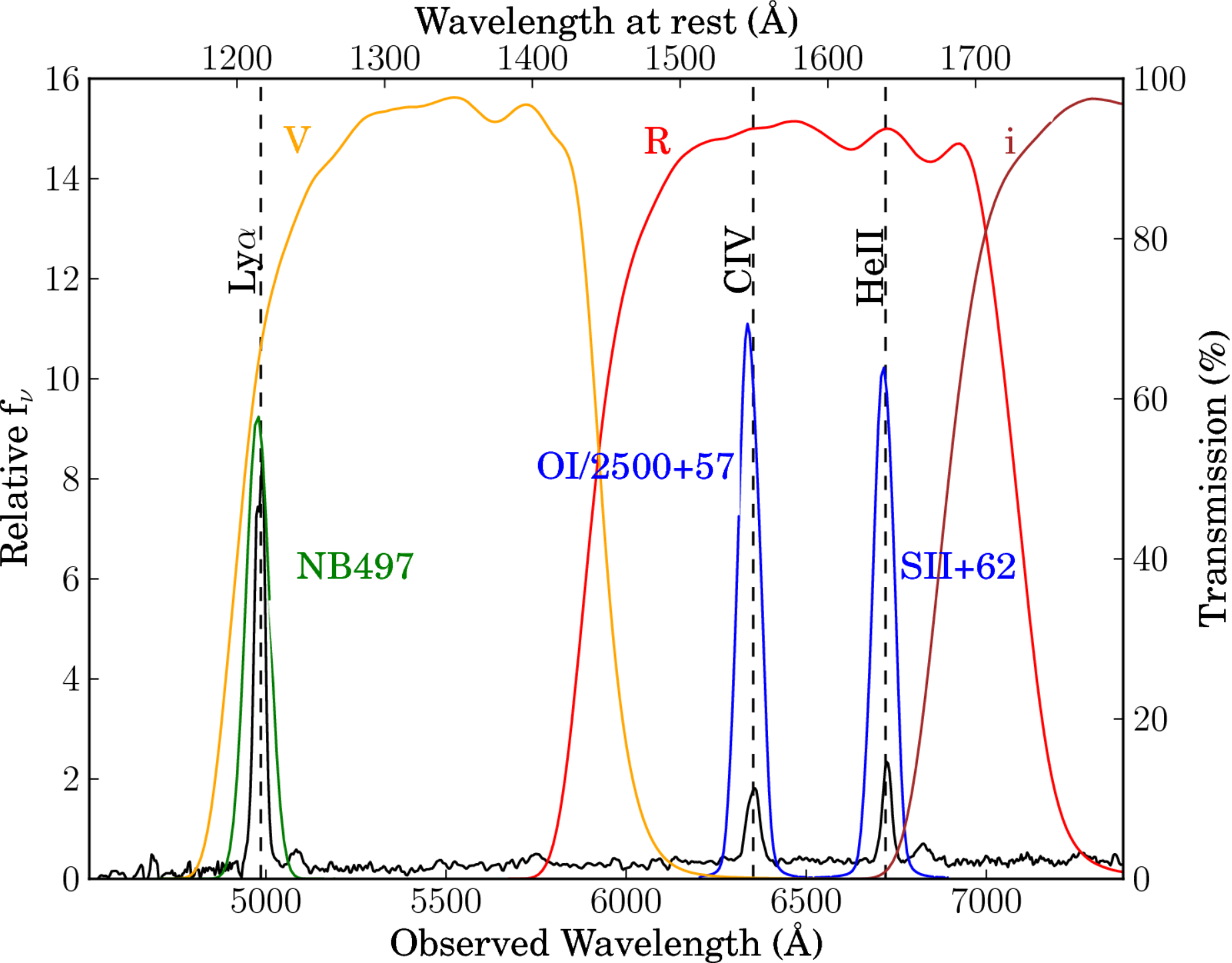, width=0.97\columnwidth}\\
\\
\epsfig{file=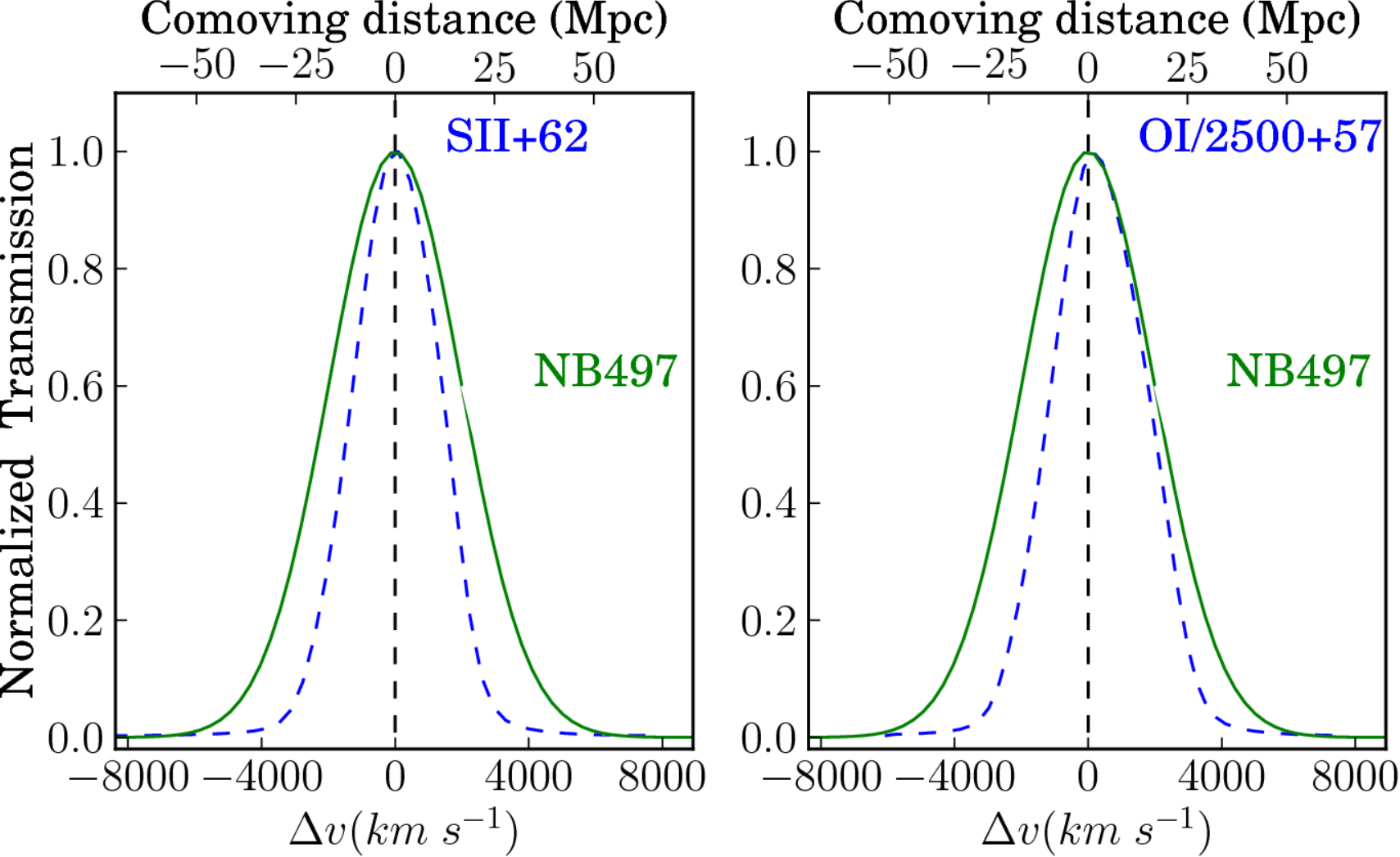, width=0.9\columnwidth} 
\caption{{\bf Top panel:} Filter response profiles for the narrowband filters NB497 (green), SII+62 and OI/2500+57 (blue) and the broad-band filters $V$ (orange), $R$ (red) and $i$ (brown) 
overplotted on a composite radio galaxy spectrum (\citealt{McCarthy1993}). {\bf Bottom panels:} Comparison between the NB497 (green) and the  SII+62 and OI/2500+57 (dashed blue) filters 
shifted to match the narrowband filter used for \lya \citep{Matsuda2004}. The filter curves are here normalized to their peak value and plotted with respect to the velocity and comoving distance probed. 
Note the nearly perfect match between the \lya narrowband filter and the two FORS2 
narrowband filters used for \civ$\lambda$1549 and \heii$\lambda$1640 in this work.  
}
\label{Fig1}
\end{figure}

Our ignorance of the physical process powering the emission 
in LABs likely results from 
the current lack of other emission-line diagnostics besides the strong
\lya line \cite[e.g.,][]{Matsuda2006}.  In this paper, we attempt to
remedy this problem, by searching for emission in two additional
rest-frame UV lines, namely \civ $\lambda$1549 and \heii
$\lambda$1640.  We present deep narrowband imaging observations tuned
to the \civ $\lambda$1549 \footnote{Throughout the paper,
  \civ\;$\lambda$1549 represents a doublet emission line, \mbox{\civ
    $\lambda\lambda$\;1548,1550}.}  and \heii $\lambda$1640 emission
lines of 13 LABs at $z\sim3.1$ in the well-known SSA22 proto-cluster
field (\citealt{Steidel2000, Hayashino2004, Matsuda2004}).  Our
observations exploit a fortuitous match between two narrowband filters
on VLT/FORS2 and the wavelengths of the redshifted CIV and HeII
emission lines of a dramatic overdensity of LABs (and Ly$\alpha$
emitters (LAEs)) in the SSA22 field (\citealt{Matsuda2004}; Figure
\ref{Fig1}), and achieve unprecedented depth.  This overdensity
results in a large multiplexing factor allowing us to carry out a
sensitive census of \civ/\lya and \heii/\lya line ratios for a
statistical sample of LABs in a single pointing.

In the following, we review four mechanisms which have been proposed to power the \lya blobs, 
which could also possibly act together, and discuss how they might generate \civ and \heii line 
emission.

\begin{enumerate}

\item {\bf Photoionization by a central AGN}: as stressed above, 
  it is well established
  that the ionizing radiation from a central AGN can power giant
  Ly$\alpha$ nebulae, with sizes up to $\sim$200 kpc, around high-$z$
  radio galaxies (e.g., \citealt{VillarM2003}; \citealt{Reuland2003};
  \citealt{Venemans2007}) and quasars (e.g., \citealt{Heckman1991};
  \citealt{Christensen2006}; \citealt{Smith2009};
  \citealt{Cantalupo2014}).   
  If the halo gas is already polluted with heavier elements (e.g., C, O) by outflows from the central source, one 
  expects to detect both \civ and \heii emission from the 
  extended \lya-emitting gas. If not, only extended \heii emission is expected.
  Indeed, extended \civ and \heii emission have been clearly detected
  in HzRGs (\citealt{VillarM2003b, Humphrey2006, VillarM2007}) and 
  tentatively detected around QSOs (\citealt{Heckman1991,Heckman1991spec, Humphrey2013})
  on scales of 10-100 kpc.  The photoionization
  scenario gains credence from a number of studies suggesting that
  LABs host an AGN which is obscured from our perspective (\citealt{Geach2009, Overzier2013, Yang2014a}, 
  but see \citealt{Nilsson2006, Smith2007}).

\item {\bf Shocks powered by galactic-scale outflows}: Several studies
  have argued that shell-like or filamentary morphologies, large
  Ly$\alpha$ line widths ($\sim$1000 km\,s$^{-1}$), and enormous \lya
  sizes ($\sim$100\,kpc) imply that extreme galactic-scale outflows,
  and specifically the ionizing photons produced by strong shocks,
  power the LABs \citep{Taniguchi&Shioya2000, Taniguchi2001,
    Ohyama2003, Wilman2005, Mori2006}.  If violent star-formation
  feedback powers a large-scale superwind, the halo should be
  highly enriched, and with a significant amount of gas at
  $T\sim 10^5\,{\rm K}$. One would therefore also expect to detect
  extended \heii and \civ emission, but with potentially different
  line ratios than the simple photoionization case. Note that
  collisional excitations of singly ionized helium peaks at $T\sim
  10^5\,{\rm K}$, making the \heii line one of the dominant observable
  coolants at this temperature (\citealt{Yang2006}). Note however, 
  that  the relatively
  quiescent ISM kinematics of star-forming galaxies embedded within
  LABs appear to be at odds with this scenario (\citealt{McLinden2013,
    Yang2011, Yang2014b}).

\item {\bf Gravitational cooling radiation}: A large body of
  theoretical work has suggested that \lya emission nebulae could
  result from \lya cooling radiation powered by gravitational collapse
  \citep{Haiman2000,Furlanetto05,Dijkstra06,Faucher2010,Rosdahl12}.
  In the absence of significant metal-enrichment, collisionally
  excited Ly$\alpha$ is the primary coolant of $T \sim 10^4\,{\rm K}$
  gas; hence cool gas steadily accreting onto halos hosting \lya blobs
  may radiate away their gravitational potential energy in the \lya
  line. However, the predictions of the \lya emission from these
  studies are uncertain by orders of magnitude
  \citep[e.g.][]{Furlanetto05,Faucher2010,Rosdahl12} because the
  emissivity of collisionally excited \lya is exponentially
  sensitive to gas temperature. Accurate prediction of the temperature
  requires solving a coupled radiative transfer and hydrodynamics
  problem which is not currently computational feasible \citep[but
    see][]{Rosdahl12}. While \citet{Yang2006} suggest that the \heii
  cooling emission could be as high as 10$\%$ of Ly$\alpha$ near the
  embedded galaxies (i.e. point-source emission) where the density of
  IGM/CGM is highest, the extended ($\gtrsim20$ kpc) \heii emission
  may be challenging to detect with current facilities (${\rm
    HeII/Ly\alpha} < 0.1$). Note that if \lya emission arises from
  cooling radiation of pristine gas, no {\it extended} \civ emission
  is expected.

\item {\bf Resonant scattering of \lya from embedded sources}: In this
  scenario, \lya photons are produced in star-forming galaxies or AGNs
  embedded in the LABs, but the extended sizes of the \lya halos
  result from resonant scattering of Ly$\alpha$ photons as they
  propagate outwards \citep{Dijkstra2008, Hayes2011, Cen2013, Cantalupo2014}.
  In this picture, non-resonant \heii emission ({\it if} produced in the galaxies or
  AGN) should be compact, in contrast with the extended \lya halos. 
  In other words, if extended \heii is detected on the same 
  scale as the extended \lya emission, this implies that resonant scattering does not play a significant role
  in determining the extent of the \lya nebulae. 
  Conversely, as the \civ line is a resonant line, it
  is conceivable that extended emission could arise due to scattering
  by the same medium scattering Ly$\alpha$, provided that the halo gas
  is optically thick to \civ, which in turn depends on the metallicity
  and ionization state of the halo gas. In this context, it is
  interesting to note that \citet{Prochaska2014sub} find a high
  covering factor of optically thick \ion{C}{2}\ and \civ absorption
  line systems out to $> 200$ kpc around $z\sim2$ QSOs, implying that
  the CGM of massive halos is significantly enriched.

\end{enumerate}

In summary, a detection of extended emission in the \civ line will
provide us information on the intensity and hardness of an ionizing
source or the speed of shocks in a superwind
\cite[e.g.,][]{Ferland1984, Nagao2006, Allen2008}, the metallicity of
gas in the CGM of LABs, and the sizes of metal-enriched halos.  A
detection of extended (non-resonant) \heii emission similarly
constrains the ionizing spectrum or the speed of shocks, and can be
used to test whether Ly$\alpha$ photons are resonantly scattered, as
well as constrain the amount of material in a warm $T\sim 10^5\,{\rm
  K}$ phase.  To date, there are five detections of extended \civ and
\heii emission from LABs reported in the literature (\citealt{Dey2005}
and \citealt{Prescott2009,Prescott2013}).  
The extended \civ and \heii emission from these \lya nebulae has
fluxes up to $F_{\rm CIV}\sim4\times10^{-17}$ \unitcgsflux and $F_{\rm
  HeII}\sim6\times10^{-17}$ \unitcgsflux, implying
\civ/\lya$\lesssim0.13$ and \heii/\lya$\lesssim0.13$. Publication
bias, i.e. the fact that searches for these lines that resulted in
non-detections are likely to have gone unpublished, makes it
challenging to assess rate of detections in LABs, which is one of the
goals of the present work.

This paper is organized as follows.  In \S\ref{sec:data}, we describe
our VLT/FORS2 narrowband imaging observations, the data reduction
procedures, and the surface brightness limits of our images.  In
\S\ref{sec:results}, we present our measurements for \civ and \heii
lines.  \S\ref{sec:Pre-data} describes previous measurements for \civ
and \heii in the literature.  In \S\ref{sec:Discussion}, we discuss
photoionization models and shock models for LABs, and compare them
with our observations and other sources in the literature.
\S\ref{sec:Conclusion} summarizes our conclusions.  Throughout this
paper, we adopt the cosmological parameters $H_0 = 70$ km s$^{-1}$
Mpc$^{-1}$, $\Omega_M=0.3$ and $\Omega_{\Lambda}=0.7$. In this
cosmology, 1\arcsec\ corresponds to 7.6 physical kpc at $z=3.1$. All
magnitudes are in the AB system (\citealt{Oke1974}).

\section{Observations and data reduction}
\label{sec:data}

\subsection{VLT/FORS2 observations and data reduction}

\begin{figure}
\centering{\epsfig{file=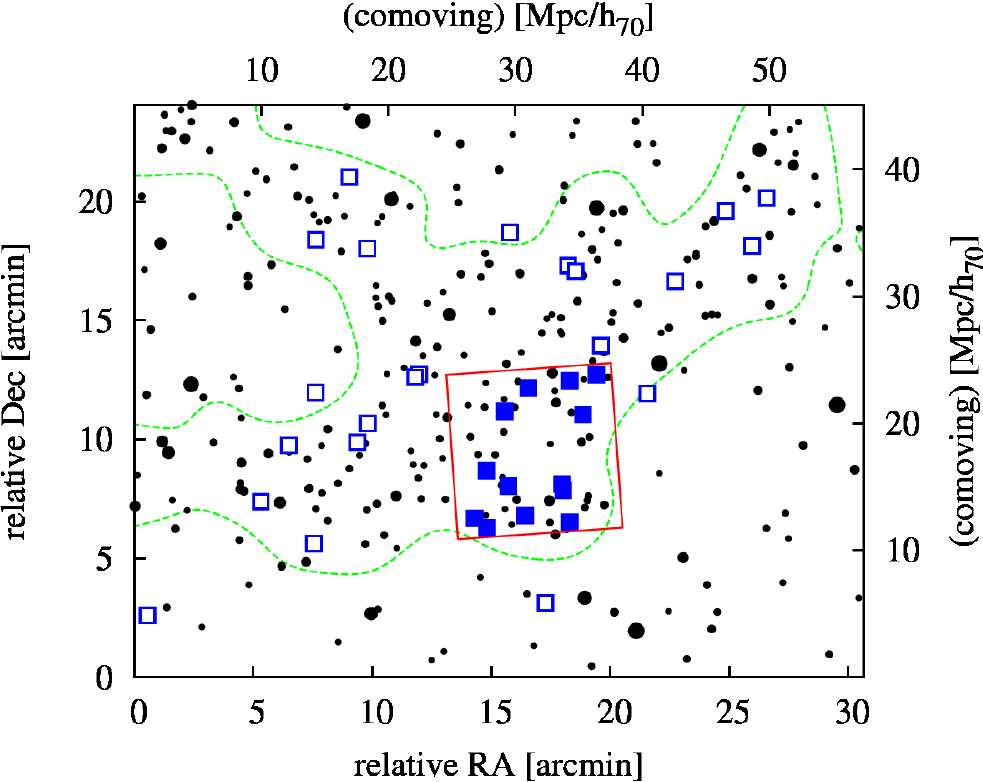, width=0.93\columnwidth}} 
\caption{Spatial distribution of the Ly$\alpha$ emitters (black filled circles) and \lya blobs (blue squares) in the SSA22 proto-cluster \citep{Hayashino2004, Matsuda2004}. 
The red box is the FOV of our FORS2 imaging (7\arcmin $\times$ 7\arcmin) which includes 13 LABs (blue filled squares). The green dashed line indicates high-density region traced by the \lya emitters.}
\label{Fig2}
\end{figure}

We obtained deep \civ and \heii narrowband images of 13 LABs in the
SSA22 proto-cluster field, including the two largest LABs that were
originally discovered by \citet{Steidel2000}.  Data were taken in
service-mode using the FORS2 instrument on the VLT 8.2m telescope Antu
(UT1) on 2010 August, September, October and 2011 September over 25
nights.  We used two narrowband filters, OI/2500+57 and SII+62
matching the redshifted \civ$\lambda$1549 and \heii$\lambda$1640 at $z
= 3.1$, respectively. The OI/2500+57 filter has a central wavelength
of $\lambda_c\approx 6354$\,\AA\ and has a FWHM of
$\Delta\lambda_{\rm FWHM} \approx 59$\,\AA, while the SII+62 filter
has $\lambda_c\approx 6714$\,\AA\ and $\Delta\lambda_{\rm FWHM}
\approx 69$\,\AA\ (Fig.~\ref{Fig1}). The FORS2 has a pixel scale of
0\farcs25 pixel$^{-1}$ and a field of view (FOV) of 7\arcmin$\times$7\arcmin\ that
allow us to observe a total of 13 LABs in a single pointing. The
pointing was chosen to maximize the number of \lya blobs while
including the two brightest 
LABs,  LAB1 and LAB2 (\citealt{Steidel2000}). We show the spatial
distribution of $\sim$300 LAEs and 35 LABs in the SSA22 region and
mark the LABs within FORS2 narrowband images in Figure \ref{Fig2}.

\begin{figure*}
\centering{\epsfig{file=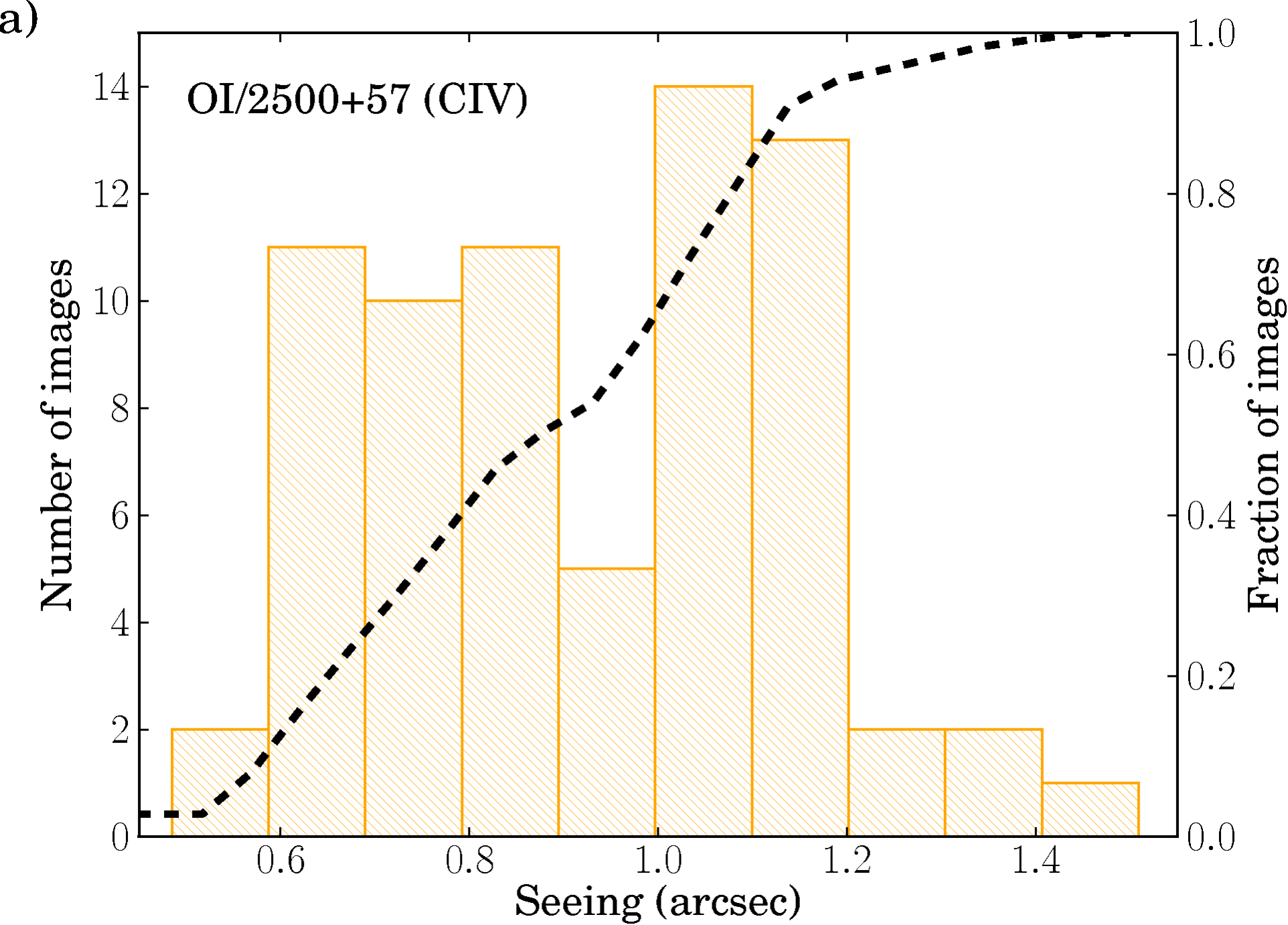, width=0.9\columnwidth}}
\hspace{0.5cm}
\centering{\epsfig{file=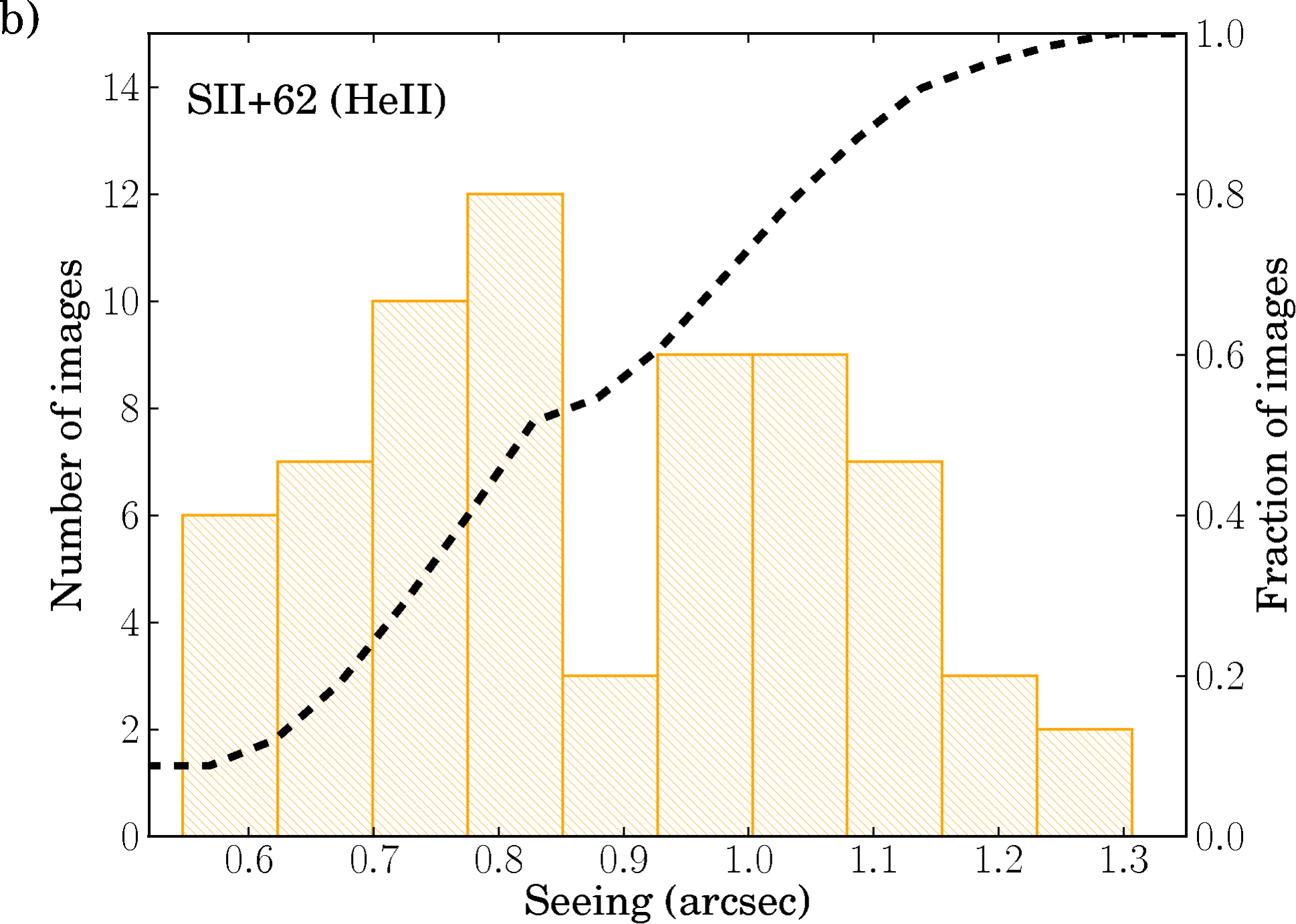, width=0.9\columnwidth}}
\caption{{\bf (a)} Distribution of seeings for the OI/2500+57 (\civ $\lambda$1549) images. {\bf (b)} Same for the SII+62 (\heii $\lambda$1640) images. The black dashed lines indicate the cumulative distribution. The median seeing is $\sim$0\farcs8 for both \civ and \heii images.}
\label{Fig3}
\end{figure*}

The total exposure time was 19.9 and 19.0 hours for \civ and \heii
lines, respectively. These exposures consist of 71 and 68 individual
exposures of $\sim$17 minutes, taken with a dither pattern to fill in
a gap between the two chips, and to facilitate the removal of cosmic
rays. Because our targets are extended over
5\arcsec--17\arcsec\ diameter and our primary goal is to detect the
extended features rather than compact embedded galaxies, we carried
out our observations under any seeing conditions (program ID: {\tt 
085.A-0989}, {\tt 087.A-0297}).
Figure \ref{Fig3} shows the distribution of FWHMs measured from stars in individual exposures. Although the observations were carried out under poor or variable seeing condition, 
the seeing ranges from 0\farcs5 to 1\farcs4 depending on the nights and the median seeing is $\sim$0\farcs8 in both filters.  
In Table \ref{Table1}, we summarize our VLT/FORS2 narrowband observations.

The data were reduced with standard routines using {\tt
  IRAF}\footnote{IRAF is the Image Analysis and Reduction Facility
  made available to the astronomical community by the National Optical
  Astronomy Observatories, which are operated by AURA, Inc., under
  contract with the U.S. National Science Foundation. STSDAS is
  distributed by the Space Telescope Science Institute, which is
  operated by the Association of Universities for Research in
  Astronomy (AURA), Inc., under NASA contract NAS 5--26555.}.  The
images were bias-subtracted and flat-fielded using twilight flats. To
improve the flat-fielding essential for detecting faint extended
emission across the fields, we further correct for the illumination
patterns using night-sky flats.  The night-sky flats were produced by
combining the unregistered science frames with an average sigma-clipping
algorithm after masking out all the objects. Satellite trails, CCD
edges, bad pixels, and saturated pixels are masked. Each individual
frame is cleaned from cosmic rays using the L.A.Cosmic algorithm
(\citealt{vanDokkum2001}). The astrometry was calibrated with the
SDSS-DR7 $r$-band catalogue using {\tt SExtractor} and {\tt SCAMP}
(\citealt{Bertin2006}). The RMS uncertainties in our astrometric
calibration are $\sim$0\farcs2 for both \civ and \heii images.

The final stacks for each filter (\civ and \heii) were obtained using
{\tt SWarp} (\citealt{Bertin2002}): the individual frames were
sky-subtracted using a background mesh size of 256 pixels ($\approx
64$\arcsec), then projected onto a common WCS using a {\it
  Lanczos3}\ interpolation kernel, and average-combined with weights
proportional to flat and night-sky flat images. Note that we choose
the mesh size to be large enough to ensure that we do not mistakenly
subtract any extended emission as sky background. For flux
calibration, we use four spectrophotometric standard stars (Feige110,
EG274, LDS749B, and G158-100) that were repeatedly observed during our
observations. Typical uncertainties in the derived zero-points are
$\approx$0.03 mag.

\subsection{Subaru Suprime-Cam Data}
\label{sec:SUBARU}

To subtract continuum from our narrowband images and compare the \civ and \heii line fluxes with those of \lya, we rely on previous Subaru observations. The SSA22 field has been extensively 
observed in $B$, $V$, $R$, $i'$, and NB497 bands (\citealt{Hayashino2004}, \citealt{Matsuda2004}) with the Subaru Suprime-Cam (\citealt{Miyazaki2002}). These images have a pixel scale of 0\farcs20 
and a FOV of 34\arcmin $\times$ 27\arcmin. The NB497 narrowband filter, tuned to \lya line at $z$ $\sim$ 3.1, has a central wavelength of 4977\,\AA\ and a FWHM of 77\,\AA. The total exposure time 
for the \lya narrowband image was 7.2 hours with a $5\sigma$ sensitivity of $5.5\times10^{-18}$ \unitcgssb per 1 arcsec$^2$ aperture, which is roughly 1.5 -- 2.5 times shallower than those of FORS2 
\heii and \civ images.  In Table 1, we summarize the Subaru broadband and narrowband images that were used in this work. 


\begin{deluxetable*}{lccccc ccc}
\tablewidth{0pt}
\tabletypesize{\small}
\tabletypesize{\scriptsize}
\tablecaption{VLT FORS2 Observations and Subaru Data}
\tablehead{
\colhead{Telescope                 }&
\colhead{Instrument                }&
\colhead{Filter (target line)      }&
\colhead{$\lambda_{\rm Central} ^a$         }&
\colhead{$\Delta\lambda_{\rm FWHM} ^b$}&
\colhead{Seeing$^c$                    }&
\colhead{Exp. Time                 }&
\colhead{Depth$^d$                 }&
\colhead{Pixel Scale               }\\[0.5ex]
\colhead{                      }&
\colhead{                      }&
\colhead{                      }&
\colhead{(\AA)                 }&
\colhead{(\AA)                 }&
\colhead{(arcsec)              }&
\colhead{(hours)               }&
\colhead{(mag)                 }&
\colhead{(arcsec)              }
}
\startdata
  VLT        & FORS2      & OI/2500+57 (CIV)      & 6354              &                59      &      0.8   & 19.9      &               25.9  & 0.25         \\
  VLT        & FORS2      & SII+62 (HeII)         & 6714              &                69      &      0.8   & 19.0      &               26.5  & 0.25         \\
\hline\\[-1.5ex]
  Subaru$^e$ & S-Cam      & NB497 (Ly$\alpha$)    & 4977              &                77      &      1.0   & 7.2       &               26.2  & 0.20         \\
  Subaru$^e$ & S-Cam      & $R$                   & 6460              &                1177    &      1.0   & 2.9       &               26.7  & 0.20         
\enddata
\tablenotetext{a}{Central wavelength of the filter.} 
\tablenotetext{b}{FWHM of the filter.} 
\tablenotetext{c}{Median seeing of our FORS2 observations and average seeing of the Subaru data (\citealt{Matsuda2004}).} 
\tablenotetext{d}{5$\sigma$ detection limit for 2\arcsec-diameter aperture.} 
\tablenotetext{e}{Images from \citet{Hayashino2004,Matsuda2004}.} 
\label{Table1}
\end{deluxetable*}



Using these deep Subaru data, \citet{Matsuda2004} found 
35 LABs, defined to be \lya emitters with the observed EW(\lya) $>$ 80 \AA\ and
an isophotal area larger than 16 arcsec$^2$, which
corresponds to a spatial extent of 30 kpc at $z = 3$.
The isophotal area was measured above the $2\sigma$ surface brightness limit
($2.2\times10^{-18}$ erg s$^{-1}$ cm$^{-2}$ arcsec$^{-2}$).  In Table
\ref{Table2}, we list the properties (e.g., \lya luminosity and
isophotal area) of the 13 LABs that were observed with VLT/FORS2. We
refer readers to \citet{Matsuda2004} for more details of this \lya
blob sample.


\begin{deluxetable*}{lcccccccc}
\tablewidth{0pt}
\tabletypesize{\small}
\tabletypesize{\scriptsize}
\tablecaption{Properties of the 13 LABs in our sample.}
\tablehead{
\colhead{Object                    }&
\colhead{$F$(Ly$\alpha$)           }&
\colhead{$L$(Ly$\alpha$)           }&
\colhead{Area                      }&
\colhead{SB (Ly$\alpha$)           }&
\colhead{SB (CIV)                  }&
\colhead{SB (HeII)                 }&
\colhead{CIV/Ly$\alpha$            }&
\colhead{HeII/Ly$\alpha$           }\\[0.5ex]
\colhead{}&
\colhead{(1)}&
\colhead{(2)}&
\colhead{(3)}&
\colhead{(4)}&
\colhead{(5)}&
\colhead{(6)}&
\colhead{(7)}&
\colhead{(8)}
}
\startdata
LAB1      &  9.4                &    7.8        & 200      &   4.7                   &  $<$0.74 	       & $<$0.50		     &   $<$0.16	       &       $<$0.11  	\\
LAB2      &  8.2                &    6.8        & 145      &   5.6                   &  $<$0.89 	       & $<$0.63		     &   $<$0.16	       &       $<$0.11  	\\
LAB7      &  1.3                &    1.1        & 36       &   3.6                   &  $<$1.19 	       & $<$0.99		     &   $<$0.33	       &       $<$0.27  	\\
LAB8      &  1.5                &    1.3        & 36       &   4.2                   &  $<$1.24 	       & $<$0.93		     &   $<$0.29	       &       $<$0.22  	\\
LAB11     &  0.8                &    0.6        & 28       &   2.8                   &  $<$1.23 	       & $<$1.08		     &   $<$0.44	       &       $<$0.38  	\\
LAB12     &  0.7                &    0.6        & 27       &   2.7                   &  $<$1.29 	       & $<$1.06		     &   $<$0.48	       &       $<$0.39  	\\
LAB14     &  1.1                &    0.9        & 25       &   4.5                   &  $<$1.38 	       & $<$1.10		     &   $<$0.31	       &       $<$0.24  	\\
LAB16     &  1.0                &    0.9        & 25       &   4.1                   &  $<$1.39 	       & $<$1.07		     &   $<$0.34	       &       $<$0.26  	\\
LAB20     &  0.6                &    0.5        & 22       &   2.8                   &  $<$1.35 	       & $<$1.16		     &   $<$0.48	       &       $<$0.41  	\\
LAB25     &  0.6                &    0.5        & 22       &   2.7                   &  $<$1.36 	       & $<$1.12		     &   $<$0.50	       &       $<$0.41  	\\
LAB30     &  0.9                &    0.8        & 17       &   5.8                   &  $<$1.45 	       & $<$1.36		     &   $<$0.25	       &       $<$0.23  	\\
LAB31     &  1.2                &    1.0        & 19       &   6.6                   &  $<$1.44 	       & $<$1.18		     &   $<$0.22	       &       $<$0.18  	\\
LAB35     &  1.0                &    0.8        & 17       &   5.9                   &  $<$1.52 	       & $<$1.29		     &   $<$0.26	       &       $<$0.22
\enddata
\tablecomments{
(1) \lya line flux within the isophote in $10^{-16}$ \unitcgsflux, 
(2) \lya luminosity in 10$^{43}$ \unitcgslum,
(3) isophotal area in arcsec$^2$ above $2.2\times10^{-18}$ \unitcgssb,
(4) average surface brightness within the isophote,
(5) 5$\sigma$ upper limits on \civ surface brightness,
(6) 5$\sigma$ upper limits on \heii surface brightness,
(7--8) 5$\sigma$ upper limits \civ/\lya and \heii/\lya line ratios.
All surface brighnesses are given in unit of 10$^{-18}$ \unitcgssb.
}
\label{Table2}
\end{deluxetable*}




\subsection{Continuum subtraction}
\label{sec:Csub}

To identify the emission in the \civ$\lambda$1549 and \heii$\lambda$1640 lines we subtract the continuum emission underlying the OI/2500+57 and SII+62 filter. We estimate the continuum using the deep 
Subaru $R$ band image. 
Because the Subaru and FORS2 images have different pixel scales, we resample the $R$-band image to the FORS2 pixel scale and register them to our WCS in order to compare all 
the images pixel by pixel. We do not match the point spread functions (PSFs) given that FORS2 images were obtained with a wide range of seeing and we are mostly interested 
in the extended emission. 
We produce the  continuum subtracted image for each filter (\civ and \heii) using the following relations (\citealt{Yang2009}):

\begin{equation}
f_{\lambda,\, {\rm cont}}^{B\!B} = \frac{F_{B\!B} - F_{N\!B}}{\Delta\lambda_{B\!B} - \Delta\lambda_{N\!B} }
\end{equation} 

\begin{equation}
F_{\rm line} = F_{N\!B} - f_{\lambda,\, {\rm cont}}^{B\!B}\Delta\lambda_{N\!B},
\end{equation} 
where $F_{B\!B}$ is the flux in the $R$ band, $F_{N\!B}$ is the flux in one of the narrowband filters. $\Delta\lambda_{B\!B}$ and $\Delta\lambda_{N\!B}$ represent the FWHM of the $R$ and narrowband filters, 
respectively. $f_{\lambda,\, {\rm cont}}^{B\!B}$ is the flux density of the continuum within the $R$ band, and $F_{\rm line}$ is the line flux (\civ or \heii). 

\subsection{Surface Brightness Limits}
\label{sec:SB}

We compute a global surface brightness limit for detecting \heii and
\civ lines using a global root-mean-square (rms) of the images.  To
calculate the global rms per pixel, we first mask out the sources, in
particular the scattered light and halos of bright foreground stars,
and compute the standard deviation of sky regions using a
sigma-clipping algorithm. We convert these rms values into the surface
brightness (SB) limits {\it per} 1 sq.\,arcsec aperture. We find that
the 1$\sigma$ detection limit per 1\,arcsec$^2$ aperture ($ {\rm SB}_1$) is
$4.2\times 10^{-19}$ and $6.8 \times 10^{-19}$ erg s$^{-1}$ cm$^{-2}$
arcsec$^{-2}$ for \heii and \civ, respectively. These represent the 
deepest \heii$\lambda$1640 and \civ$\lambda$1549 
narrow-band images ever taken.

\begin{figure*}
\centering{\epsfig{file=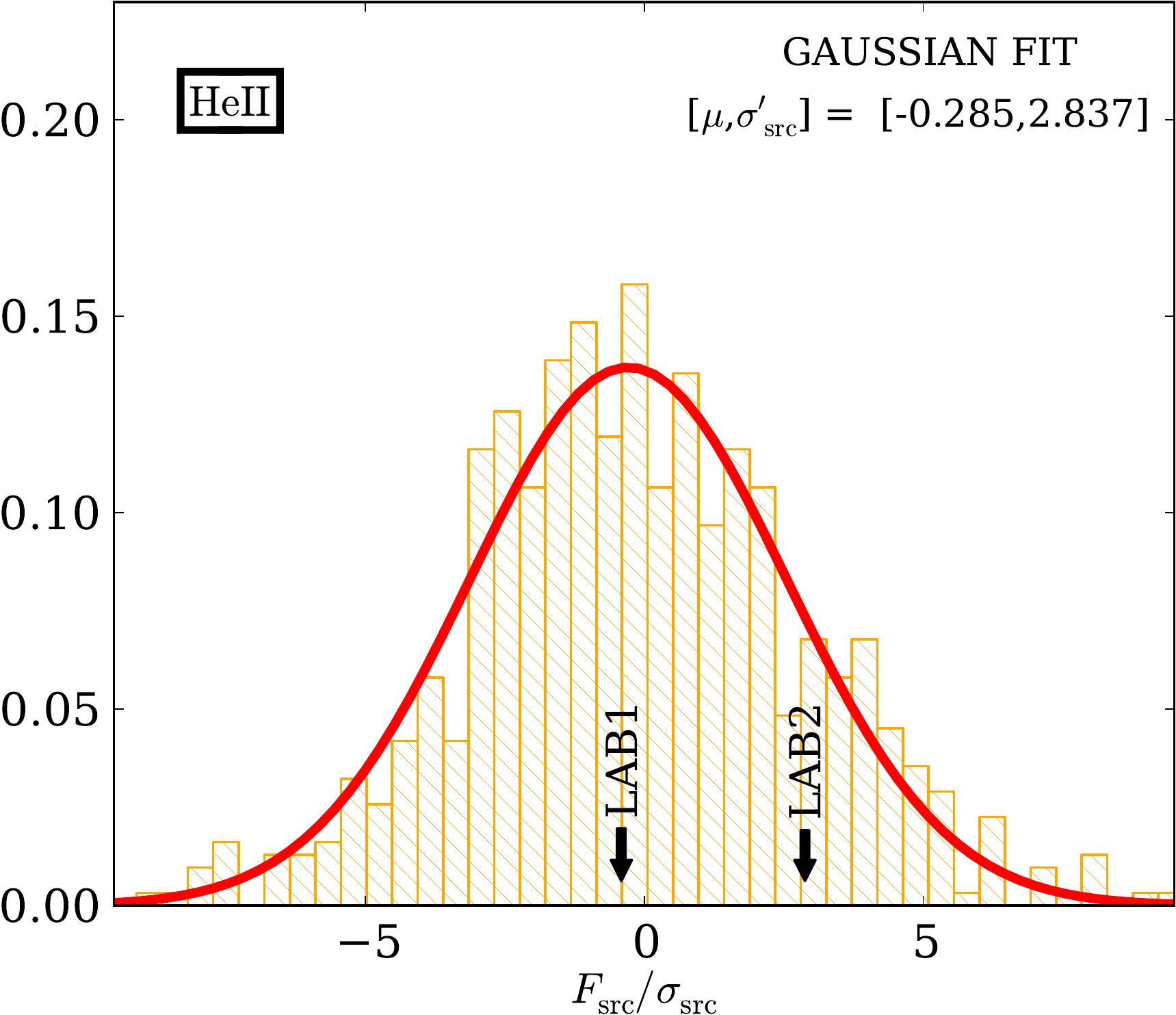, width=0.88\columnwidth}}
\hspace{0.5cm}
\centering{\epsfig{file=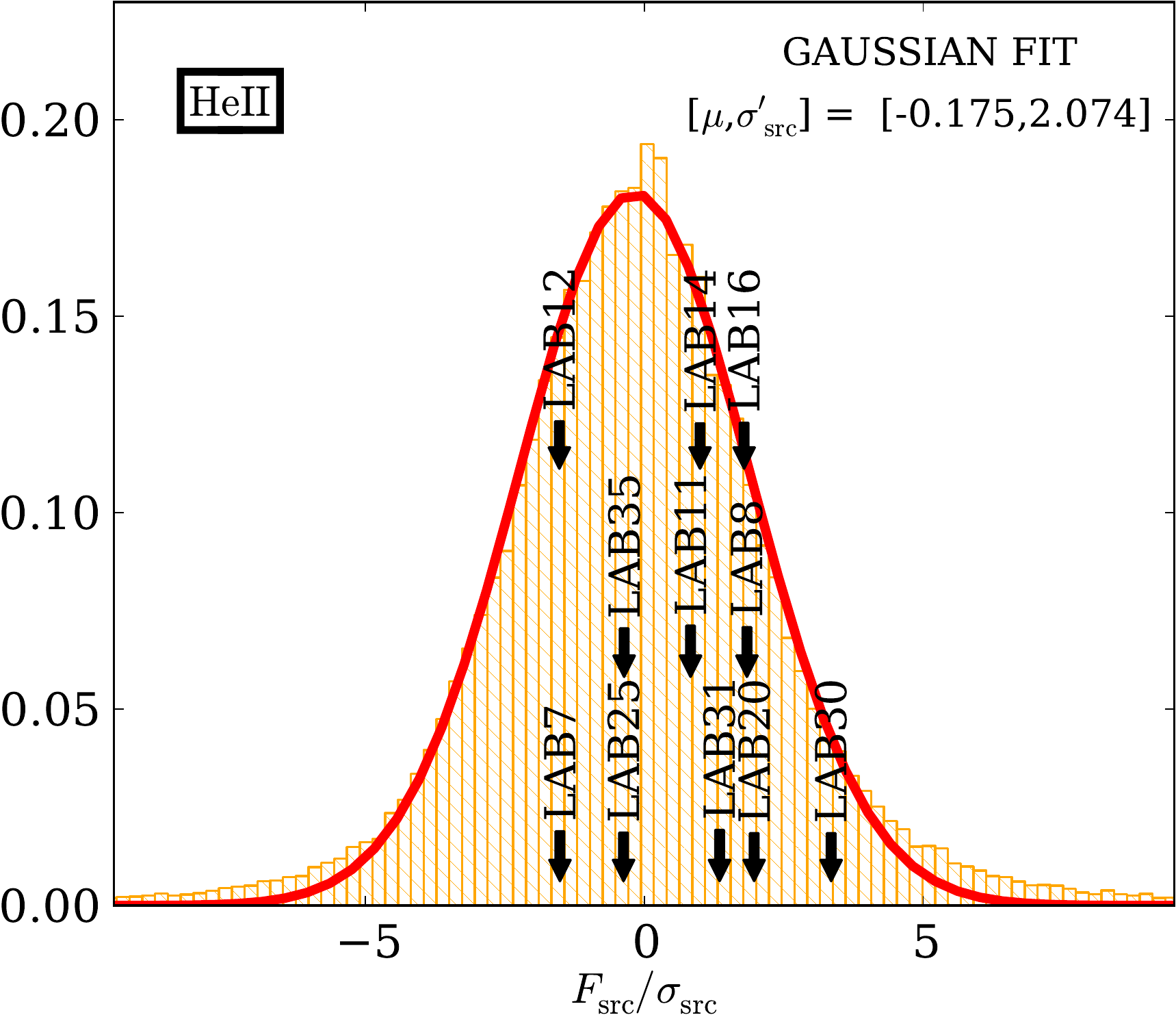, width=0.88\columnwidth}}
\caption{Analysis of the systematics in the \heii line image. {\bf Left:} Distribution of the normalized flux, $F_{\rm src}$/$\sigma_{\rm src}$ for random circular apertures with the 
same extent as LAB1 and LAB2. Here, $F_{\rm src}$ is a total flux 
within an aperture and $\sigma_{\rm src}$ is the expected 1$\sigma$ flux limit in an ideal case with uniform noise properties, i.e., $\sigma_{\rm src}$ = $ {\rm SB}_1\sqrt{A_{\rm src}}$. 
The Gaussian fit to the histogram is highlighted in red. The observed values for LAB1 and LAB2 are shown by the black arrows. {\bf Right:} Same for all the other LABs with $A_{\rm src}$ $<$ 40 arcsec$^2$ in 
our sample. The black arrows indicate the value of each LAB. Note that in the absence of systematics, i.e., in ideal conditions when the sky and continuum subtractions are perfect, these histograms should be 
a Gaussian with unit variance, but they are $\approx 3$ or $\approx 2$ times broader, i.e., $\sigma'_{\rm src}$ $\approx$ 2--3 $\sigma_{\rm src}$.} 
\label{Fig4}
\end{figure*}

\begin{figure*}
\centering{\epsfig{file=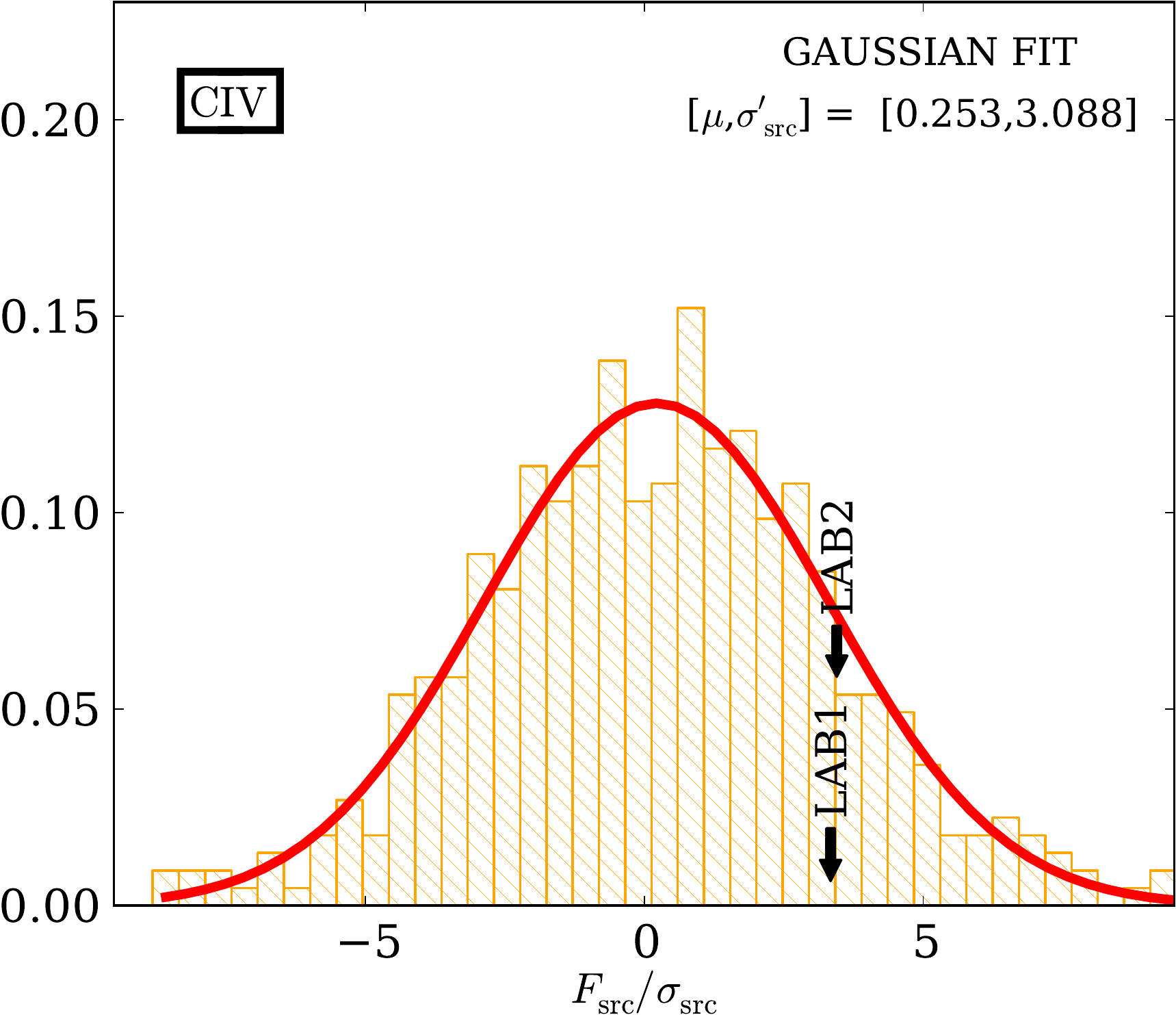, width=0.88\columnwidth}}
\hspace{0.5cm}
\centering{\epsfig{file=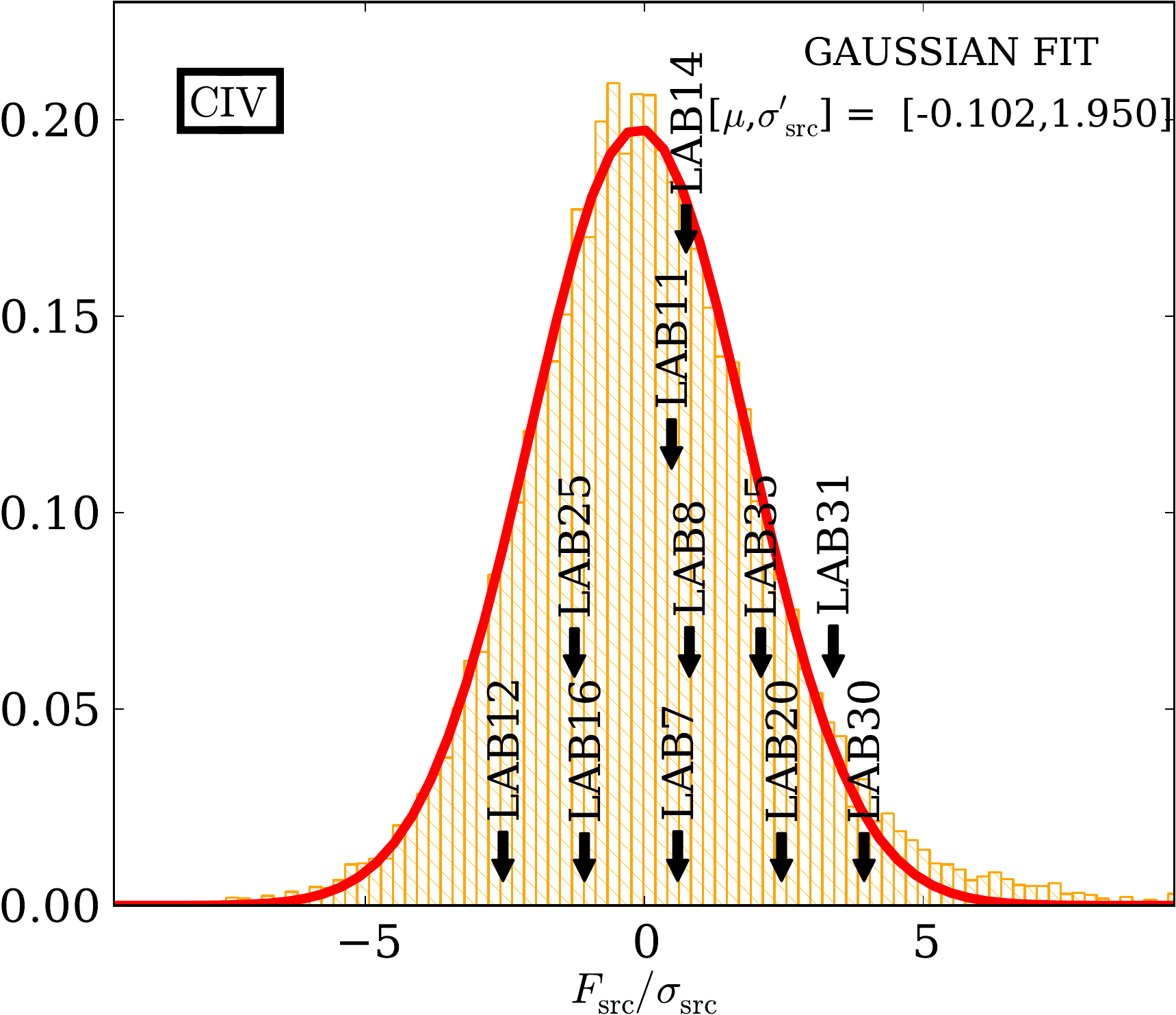, width=0.88\columnwidth}}
\caption{Analysis of the systematics in the \civ line image. {\bf Left:} Distribution of the normalized flux, $F_{\rm src}$/$\sigma_{\rm src}$ for random circular apertures with the same extent as LAB1 and LAB2. 
Here, $F_{\rm src}$ is a total flux 
within an aperture and $\sigma_{\rm src}$ is the expected 1$\sigma$ flux limit in an ideal case with uniform noise properties, i.e., $\sigma_{\rm src}$ = ${\rm SB}_1\sqrt{A_{\rm src}}$. 
The Gaussian fit to the histogram is highlighted in red. The observed values for LAB1 and LAB2 are shown by the black arrows. {\bf Right:} Same for all the other LABs with $A_{\rm src}$ $<$ 40 arcsec$^2$ in 
our sample. The black arrows indicate the value of each LAB. Note that in the absence of systematics, i.e., in ideal conditions when the sky and continuum subtractions are perfect, these histograms should be 
a Gaussian with unit variance, but they are $\approx 3$ or $\approx 2$ times broader, i.e., $\sigma'_{\rm src}$ $\approx$ 2--3 $\sigma_{\rm src}$.} 
\label{Fig5}
\end{figure*}

The sensitivity required to detect an extended source depends on its
size because one can reach lower surface brightness levels by
spatially averaging. In an ideal case of perfect sky and continuum 
subtraction, the 1$\sigma$ SB limit for an
extended source is given by ${\rm SB}_{1}$/$\sqrt{A_{\rm src}}$, where
$A_{\rm src}$ is the isophotal area in arcsec$^2$ and ${\rm SB}_{1}$ is
the surface brightness limit per 1 arcsec$^2$ aperture.  However, in practice
the actual detection limits are limited by systematics 
resulting from imperfect sky and continuum subtraction.  Therefore,
we empirically determine the detection limits for extended sources
with different sizes as follows.

In the continuum-subtracted line images, we mask all the artifacts
(e.g., CCD edges and scattered light from bright stars) and also the
locations of the LABs. For each LAB that we consider, we randomly
place circular apertures with the same area of the LAB and extract the
fluxes ($F_{\rm src}$) within these apertures.  If the images have
uniform noise properties in the absence of systematics, the fluxes
($F_{\rm src}$) from many random apertures should follow a Gaussian
distribution with a width of $\sigma_{\rm src}$ $\equiv$ ${\rm SB}_{1}
\sqrt{A_{\rm src}}$.  We find that the actual Gaussian width
($\sigma'_{\rm src}$) of the distribution is much broader than
$\sigma_{\rm src}$ (Fig.~\ref{Fig4} and \ref{Fig5}).  We adopt $F_{\rm
  limit}$ $\equiv$ $\sigma'_{\rm src}$ as a 1$\sigma$ upper limit on
the total line flux of each LAB. The corresponding upper limit for the
surface brightness is given by ${\rm SB}_{\rm limit}$ $\equiv$ $F_{\rm
  limit}$/$A_{\rm src}$.

Figure \ref{Fig4} and \ref{Fig5} show the distribution of $F_{\rm
  src}$/$\sigma_{\rm src}$ for \heii and \civ images, respectively.
Note that we normalize the extracted fluxes to the $\sigma_{\rm src}$
in order to show the distributions for LABs with similar sizes in one
plot.  As the size of the LABs in our sample spans a large range, we
show the distributions for two sub-samples: one for LAB1 and LAB2 with
$A_{\rm src}$ $>$ 100 arcsec$^2$ and the other for the remaining LABs
with $A_{\rm src}$ $<$ 40 arcsec$^2$.  As previously stated, in the
ideal case of no systematics, $\sigma_{\rm src}$ characterizes the
noise in $F_{\rm src}$, and thus the distribution of the quantity
$F_{\rm src}$/$\sigma_{\rm src}$ should be a Gaussian with unit
variance.
For both sub-samples, we find that $F_{\rm src}$/$\sigma_{\rm src}$
histograms show a variance greater than unity, suggesting that
imperfect sky and continuum subtraction dominates our error budget.
The normalized histograms have a standard deviation of $\approx$\,3 on
the scale of the bigger LABs (LAB1 and LAB2), and $\approx$\,2 on the
scale of the smaller LABs.  Thus, as our 1$\sigma$ limit on the total
line flux of the largest LABs in our sample (LAB1 and LAB2), we adopt
$F_{\rm limit} \equiv \sigma'_{\rm src}=3\sigma_{\rm src}$, where
$\sigma_{\rm src}$ $\equiv$ ${\rm SB}_{1} \sqrt{A_{\rm src}}$ is computed
using the area of the blob. 
For all of the other blobs in our sample,
we follow the same approach but use a value $F_{\rm limit}$ $\equiv$
$\sigma'_{\rm src}=2\sigma_{\rm src}$ 
We conservatively define our detection threshold to be $5 \sigma'_{\rm src}$, which 
formally means 15$\sigma_{\rm src}$ for LAB1 and LAB2, and 10$\sigma_{\rm src}$
for all the other blobs.
In each histogram, we show the values extracted inside the
isophotal contours of each LAB (black arrows).  These values are well
within the distribution of $F_{\rm src}$/$\sigma_{\rm src}$ determined from random
apertures (see Table \ref{Table2a}).


\begin{deluxetable}{lrr}
\tablewidth{0pt}
\tablewidth{0.4\textwidth}
\tablewidth{0.9\columnwidth}
\tabletypesize{\scriptsize}
\tablecaption{Extracted fluxes and significance for the 13 LABs in our sample.}
\tablehead{
\colhead{Object     }&
\colhead{$F$(\heii) }&
\colhead{$F$(\civ)  }\\[0.5ex]
\colhead{   }&
\colhead{(1)}&
\colhead{(2)}
}
\startdata
LAB1      &  -2.98 (-0.41)	 &   31.19 ( 3.34)	      \\
LAB2      &  17.81 ( 2.88)	 &   27.47 ( 3.45)	      \\
LAB7      &  -4.63 (-1.51)	 &    2.38 ( 0.60)	      \\
LAB8      &   5.69 ( 1.84)	 &    3.22 ( 0.81)	      \\
LAB11     &   2.56 ( 0.83)	 &    1.96 ( 0.49)	      \\
LAB12     &  -4.04 (-1.52)	 &   -8.68 (-2.53)	       \\
LAB14     &   2.59 ( 1.00)	 &    2.49 ( 0.75)	       \\
LAB16     &   4.64 ( 1.79)	 &   -3.56 (-1.07)	       \\
LAB20     &   4.78 ( 1.97)	 &    7.69 ( 2.46)	       \\
LAB25     &  -0.89 (-0.37)	 &   -3.91 (-1.25)	       \\
LAB30     &   7.06 ( 3.35)	 &   10.67 ( 3.94)	       \\
LAB31     &   3.02 ( 1.35)	 &    9.74 ( 3.39)	       \\
LAB35     &  -0.76 (-0.36)	 &    5.75 ( 2.09)
\enddata
\tablecomments{
(1) \heii line flux in $10^{-18}$ \unitcgsflux extracted within the isophotal area defined in \citet{Matsuda2004},
(2) \civ line flux in $10^{-18}$ \unitcgsflux. 
For each value is given in brackets the statistical significance with respect to the $\sigma_{\rm src}$. 
}
\label{Table2a}
\end{deluxetable}


To test if our derived detection limits are reasonable, we visually confirm the detectability as a function of size by placing artificial model sources in \heii and \civ narrowband images.
We adopt circular top-hat sources with a uniform surface brightness corresponding to 1, 2, 3, 4, 5, 8, 10, 20 ${\rm SB}_{\rm limit}$ , and an area of 200, 100, 40 and 20 arcsec$^2$, comparable 
to the size of the LABs in our sample (see Table \ref{Table2}). 
After placing the simulated sources in the narrowband images, we subtract the continuum in the same way as explained in Section \S\ref{sec:Csub}. Because the detectability strongly depends on the residual structure 
of the continuum subtraction, we place the model sources at different locations in the narrowband images after masking all the bad regions as explained above. Following \citet{Hennawi2013}, we construct a $\chi$ image by 
dividing the continuum-subtracted image by a ``sigma'' image. Here, the sigma image (or the square root of the variance image) is calculated by taking into account our stacking procedure, e.g., bad pixels, 
satellite trails and sky subtraction. In other words, this variance image is the theoretical photon counting noise variance, taking into account all the bad-behaving pixels.
In this calculation, we do not include the variance due to $R$-band continuum, i.e., we ignore the photon counting noise from $R$-band image, thus it is likely that our sigma image might slightly 
underestimate the noise. Note however that the shallower NB images are very likely dominating the noise, thus the 
$R$-band contribution to the variance is a small correction.

To test the detectability of extended emission, we compute a smoothed $\chi$ image following the technique in \citet{Hennawi2013}. First, we smooth an image:
\begin{equation}
I_{\rm smth}={\rm CONVOL[NB-CONTINUUM]},
\end{equation}
where the CONVOL operation denotes convolution of the stacked images with a Gaussian kernel with FWHM=2.35\arcsec.  
Then, we calculate the sigma image ($\sigma_{\rm smth}$) for the smoothed image ($I_{\rm smth}$) by propagating the variance image of the unsmoothed data: 
\begin{equation}
\sigma_{\rm smth}=\sqrt{{\rm CONVOL}^2{\rm [}\sigma^2_{\rm unsmth}{\rm]}},
\end{equation}
where the CONVOL$^2$ operation denotes the convolution of variance image with the square of the Gaussian kernel.
Thus, the smoothed $\chi$ image is defined by
\begin{equation}
\chi_{\rm smth}=\frac{I_{\rm smth}}{\sigma_{\rm smth}}.
\end{equation}
This $\chi_{\rm smth}$ is more effective in visualizing the presence of extended emission.

\begin{figure*}
\centering{\epsfig{file=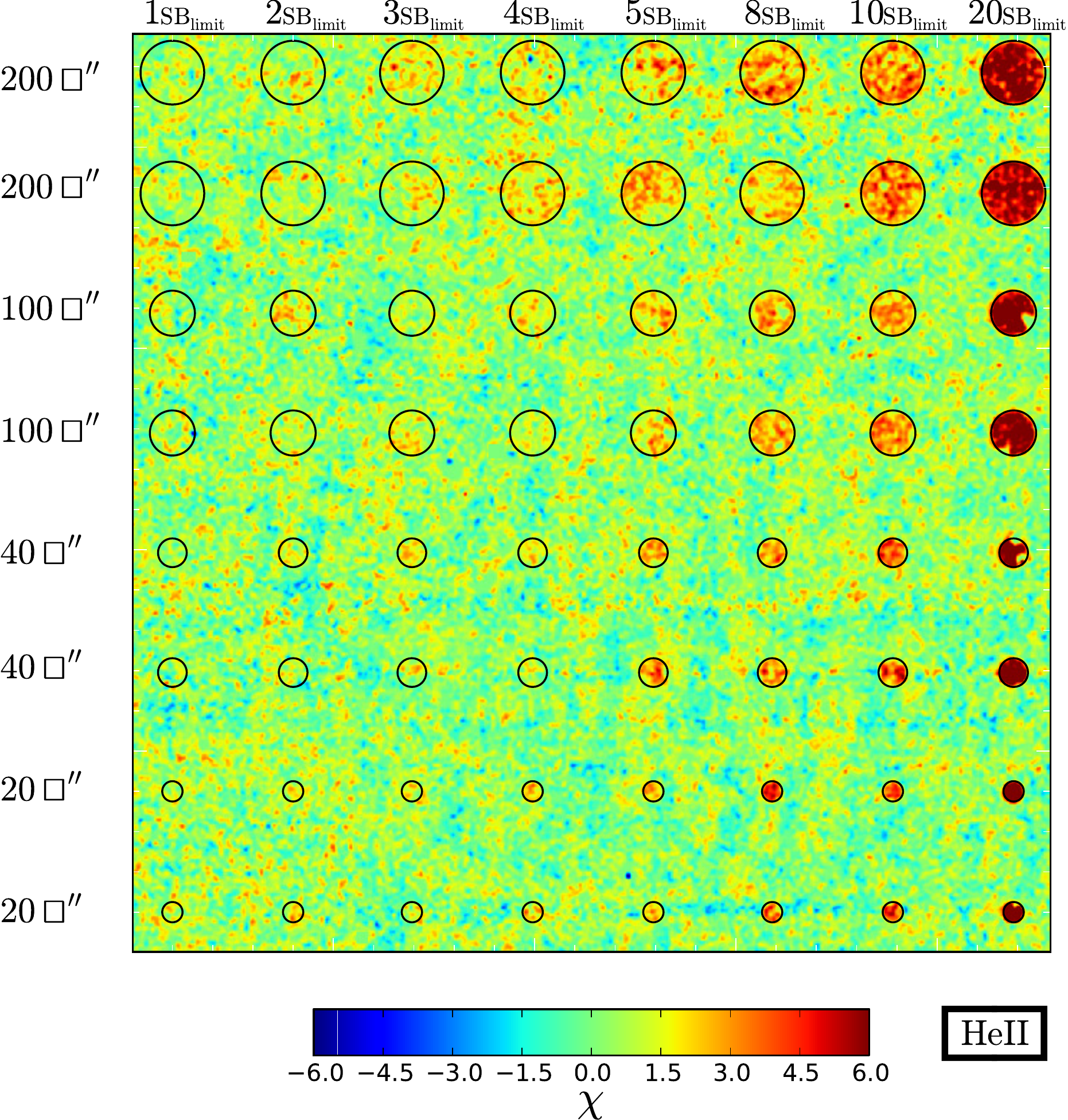, viewport=0 0 590 600, clip, width=\textwidth}}
\caption{Illustration of detection significance of the simulated sources as a function of sizes in \heii line. The panel shows the $\chi_{smth}$ image for the 
simulated sources with circular top-hat profile with uniform surface brightness. From top to bottom, the simulated sources are placed as follow: two rows for 
each area (200, 100, 40, 20 arcsec$^2$) with a surface brightness level of 1, 2, 3, 4, 5, 8, 10, 20 ${\rm SB}_{\rm limit}$. 
The black circles indicate the position of the simulated sources. Note that we should be able to 
detect sources down to a sensitivity limit of $5{\rm SB}_{\rm limit}$, which corresponds to SB(HeII) = $5.02\times10^{-19}$ erg s$^{-1}$ cm$^{-2}$ arcsec$^{-2}$ for an 
area of 200 arcsec$^2$ (i.e. LAB1). The same stretch and color schemes are adopted in Figures \ref{Fig7} and \ref{Fig10}. 
}   
\label{Fig6}
\end{figure*}

\begin{figure*}
\centering{\epsfig{file=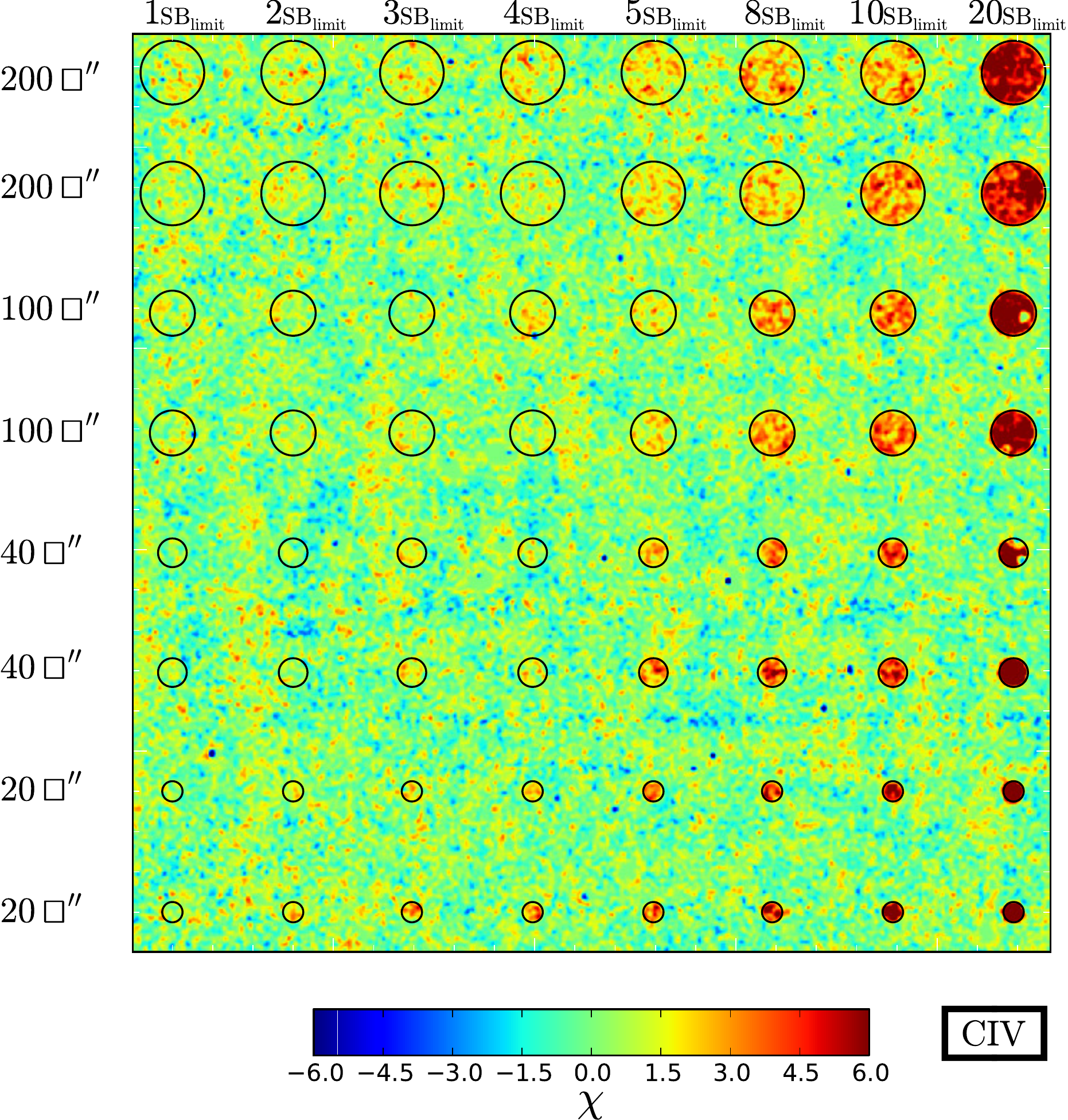, viewport=0 0 590 600, clip, width=\textwidth}}
\caption{Illustration of detection significance of the simulated sources as a function of sizes in \civ line. The panel shows the $\chi_{smth}$ image for the 
simulated sources with circular top-hat profile with uniform surface brightness. From top to bottom, the simulated sources are placed as follow: two rows for 
each area (200, 100, 40, 20 arcsec$^2$) with a surface brightness level of 1, 2, 3, 4, 5, 8, 10, 20 ${\rm SB}_{\rm limit}$. 
The black circles indicate the position of the simulated sources. Note that we should be able to 
detect sources down to a sensitivity limit of $5{\rm SB}_{\rm limit}$, which corresponds to SB(CIV) = $7.36\times10^{-19}$ erg s$^{-1}$ cm$^{-2}$ arcsec$^{-2}$ for an 
area of 200 arcsec$^2$ (i.e. LAB1). The same stretch and color schemes are adopted in Figures \ref{Fig6} and \ref{Fig10}. 
}   
\label{Fig7}
\end{figure*}

Figure \ref{Fig6} and \ref{Fig7} show the $\chi_{\rm smth}$ for the simulated sources for \heii and \civ images, respectively. 
For each detection significance and source size, the simulated sources are shown for two different positions within the \heii or the \civ images. To guide the eye, 
these positions are highlighted by a black circle.
These simulated $\chi_{\rm smth}$ images confirm that we should be able to detect extended emission down to a level of $5{\rm SB}_{\rm limit}$, 
justifying our choice for this detection threshold. Note again that ${\rm SB}_{\rm limit}$ includes the correction we made to take into account the systematics.

In addition to the previous analysis, in order to further test our continuum subtraction, we also performed the continuum subtraction 
using two off-band images ($V$ and $i'$; \citealt{Hayashino2004}), finding that the results remain unchanged.
Note however, that due to the differences in the telescope PSFs and seeing of the observations, the use of two bands increases the noise. 
Thus, we prefer to estimate the continuum using only the $R$-band image.

\section{Observational Results}
\label{sec:results}

\begin{figure*}
\centering{\epsfig{file=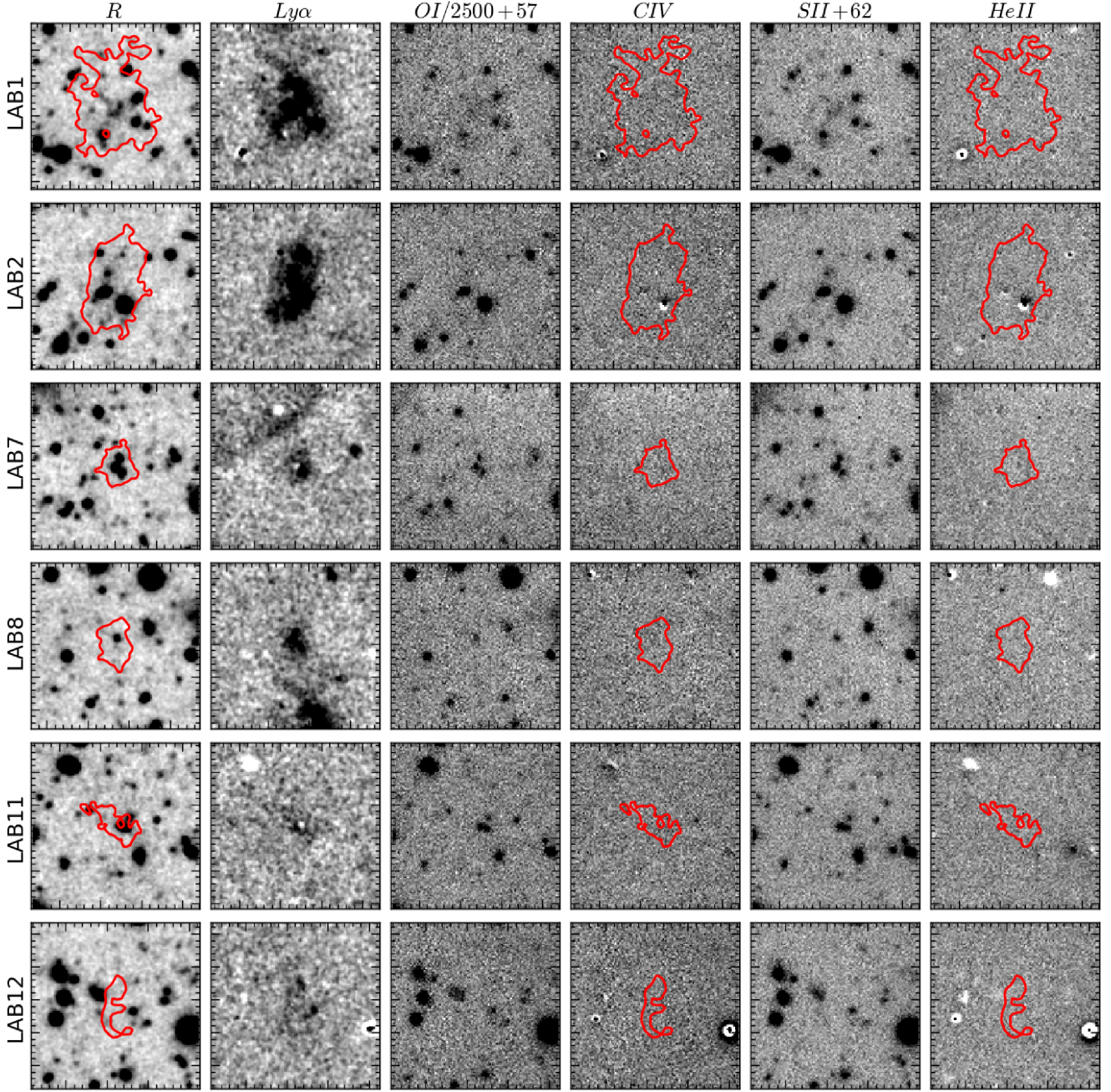, viewport=0 100 600 685, clip, width=\textwidth, angle=270}}
\caption{Postage-stamp images of 30\arcsec $\times$ 30\arcsec (corresponding to about 230 kpc $\times$ 230 kpc at $z=3.1$) centered on LAB1, LAB2, LAB7, LAB8, LAB11, and LAB12. 
From left to right: $R$-band, Ly$\alpha$, OI/2500+57 (NB CIV), CIV$\lambda$1549, SII+62 (NB HeII), and HeII$\lambda$1640. On the $R$-band, CIV$\lambda$1549, and HeII$\lambda$1640, 
is over-plotted the 2$\sigma$ isophotal aperture of the Ly$\alpha$ emission (red line) as adopted by \citet{Matsuda2004}. Note the lack of extended emission in the CIV$\lambda$1549 and HeII$\lambda$1640 
in comparison with the outstanding Ly$\alpha$ line. North is up, East is left.}
\label{FigATLAS1}
\end{figure*}

\begin{figure*}
\centering{\epsfig{file=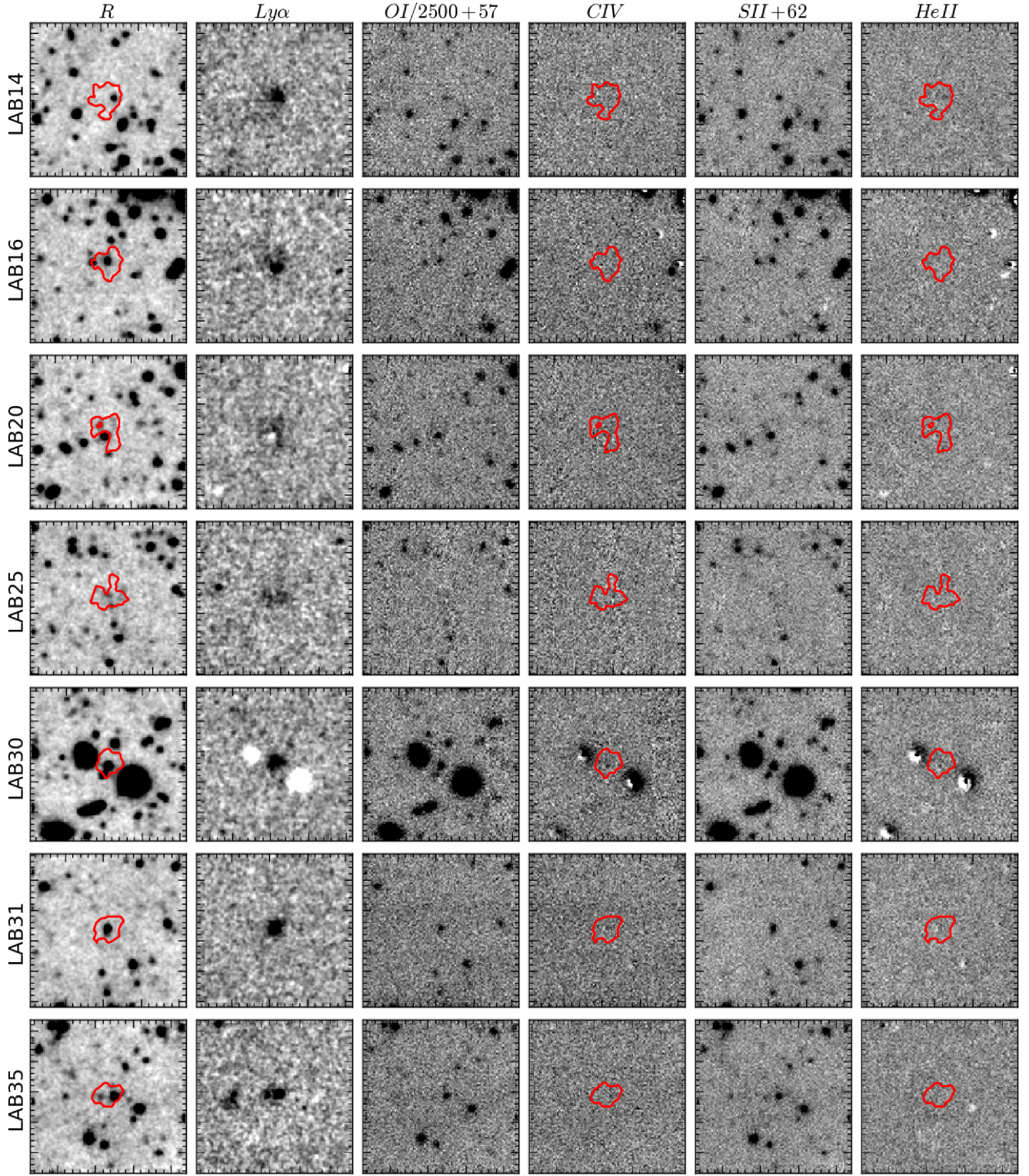, viewport=10 60 600 760, clip, width=\textwidth}} 
\caption{Postage-stamp images of 30\arcsec $\times$ 30\arcsec (corresponding to about 230 kpc $\times$ 230 kpc at $z=3.1$) centered on LAB14, LAB16, LAB20, LAB25, LAB30, LAB31, and LAB35. 
From left to right: $R$-band, Ly$\alpha$, OI/2500+57 (NB CIV), CIV$\lambda$1549, SII+62 (NB HeII), and HeII$\lambda$1640. On the $R$-band, CIV$\lambda$1549, and HeII$\lambda$1640, 
is over-plotted the 2$\sigma$ isophotal aperture of the Ly$\alpha$ emission (red line) as adopted by \citet{Matsuda2004}. Note the lack of extended emission in the CIV$\lambda$1549 and HeII$\lambda$1640 
in comparison with the outstanding Ly$\alpha$ line. North is up, East is left.}
\label{FigATLAS}
\end{figure*}

In Figure \ref{FigATLAS1} and \ref{FigATLAS}, we show the postage-stamp images for the 13
LABs in our sample. Each row displays the $R$-band, the
continuum-subtracted \lya line image, the narrowband image of the
\civ$\lambda$1549 line, the continuum-subtracted \civ line image, the
\heii$\lambda$1640 narrowband image, and the continuum-subtracted
\heii line image, respectively. The red contours indicate the
isophotal aperture of LABs defined as the area above 2$\sigma$
detection limit for the Ly$\alpha$ emission as originally adopted by
\citet{Matsuda2004}, i.e. $2.2\times 10^{-18}$ \unitcgssb. 
The continuum-subtracted \civ and \heii line images are
nearly flat and lack significant large-scale residuals, indicating
good continuum and background subtraction.  
Note that there could be still some
residuals within the isophotal apertures (e.g., LAB2) because of minor
mis-alignment between $R$-band and our narrowband images. However,
these residuals do not affect our flux and surface brightness
measurements. We do not detect any extended \civ or \heii emission on
the scale of the Ly$\alpha$ line in any of the LABs.

\begin{figure*}
\centering{\epsfig{file=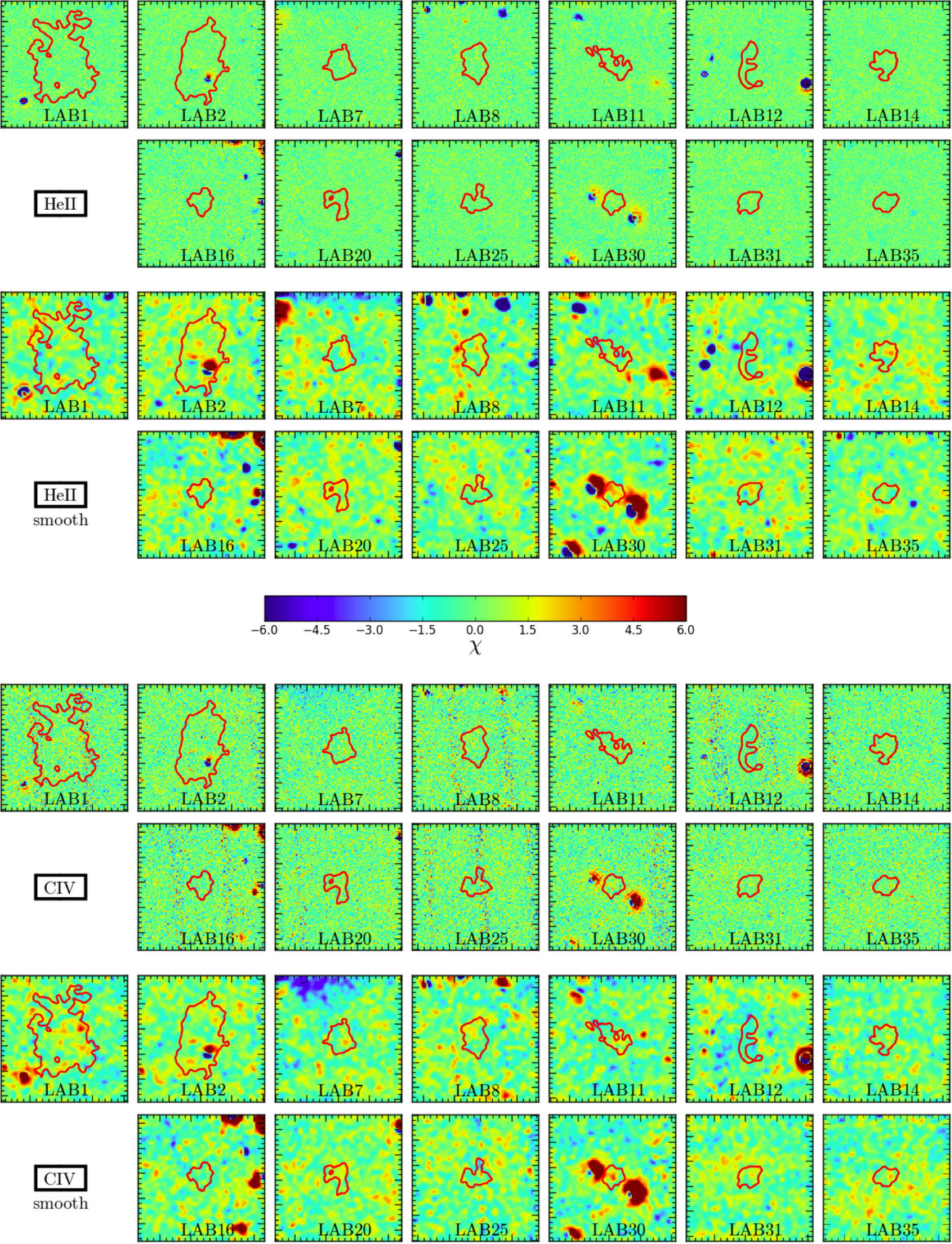, viewport=0 0 530 690, clip, width=0.95\textwidth}} 
\caption{Postage-stamp $\chi$ and $\chi_{\rm smth}$ images of the 13 LABs in our sample (\S\ref{sec:SB}). Each postage-stamp has a size of $30\arcsec\times30\arcsec$ (corresponding to about 230 kpc $\times$ 230 kpc at $z=3.1$). 
To guide the eye, on each image is overplotted the 2$\sigma$ isophotal aperture of the Ly$\alpha$ emission (red line) as adopted by \citet{Matsuda2004}. A comparison with Figures \ref{Fig6}-\ref{Fig7} 
suggest that we did not detect any extended emission from any of the sources in our sample. Note that we used the same stretch and colormap as in Figure \ref{Fig6} and \ref{Fig7}. Residuals from bright foreground objects due to
minor mis-alignment between our data and SUBARU data are clearly visible. North is up, East is left.}
\label{Fig10}
\end{figure*}

In order to better visualize these non-detections, we compute the
$\chi$ and $\chi_{\rm smth}$ described in \S\ref{sec:SB} for each LAB
(using the pure photon counting noise estimates).  Figure \ref{Fig10}
shows the $\chi$ images and the $\chi_{\rm smth}$ images of
$30\arcsec\times30\arcsec$ (corresponding to 230 kpc $\times$ 230 kpc
at $z=3.1$) centered on each LAB. 
A comparison of
  the $\chi_{\rm smth}$ images of the individual \lya blobs with the
  simulated images in Figures \ref{Fig6} and \ref{Fig7} shows that we
  do not detect any extended emission in the HeII and CIV lines for
  the 13 LABs down to our sensitivity limits of $5{\rm SB}_{\rm limit}$
  defined in Section \S\ref{sec:SB}.  Note that we show images
  in Figures \ref{Fig6}, \ref{Fig7} and \ref{Fig10} with the same
  stretch and color scheme for a fair comparison. 

We thus place conservative upper limits, i.e. $5{\rm SB}_{\rm
  limit}$, on both CIV$\lambda$1549 and HeII$\lambda$1640 surface
brightness for each of the LABs.  For LAB1 (area 200 arcsec$^2$),
these limits correspond to SB(\heii) = $5.02\times10^{-19}$ erg
s$^{-1}$ cm$^{-2}$ arcsec$^{-2}$ and SB(\civ) = $7.36\times10^{-19}$
erg s$^{-1}$ cm$^{-2}$ arcsec$^{-2}$. In Table \ref{Table2}, we summarize
all of our upper limits, the properties of Ly$\alpha$ lines, and the
resulting upper limits on the \civ$\lambda$1549/\lya and
\heii$\lambda$1640/\lya flux ratios. Note that the most stringent
limits on these ratios are obtained for the brightest LAB1 and LAB2
given their larger \lya isophotal area and
luminosities. Coincidentally, these two LABs show the same values,
$F$(\heii)/$F$(\lya) $<$ 0.11 and $F$(\civ)/$F$(\lya) $<$ 0.16,
because the difference in the area (LAB1 is larger than LAB2) is
compensated by the difference in \lya SB (LAB2 has a SB higher than
LAB1). In what follows, we compare our limits to previous
constraints on HeII and CIV in other nebulae, and then discuss the
implications of our non-detections.

\section{Previous Observations of {\rm \heii} and {\rm \civ}}
\label{sec:Pre-data}

We compile \heii and \civ line observations of extended Ly$\alpha$
nebulae from the literature,  
finding data for five \lya blobs 
(\citealt{Dey2005}; \citealt{Prescott2009, Prescott2013}, summarized in Table~\ref{Table3} in the Appendix), 
\lya nebulae associated with 53 high redshift radio galaxies
(\citealt{Humphrey2006}; \citealt{VillarM2007}), and five radio-loud
QSOs (\citealt{Heckman1991spec, Heckman1991};
\citealt{Humphrey2013}).
However, a straightforward comparison is restrained by the following issues.
First of all, these data are obtained with various
different techniques (e.g., narrowband imaging, longslit spectroscopy,
integral-field unit spectroscopy), and employ varied analysis methods
(e.g., different extraction apertures), which result in different
definitions of SB limits. Thus, a major uncertainty in comparing our data with the previous
measurements are differences in the aperture for which these line
fluxes or ratios are reported. In particular, our upper limits are computed over the
entire \lya nebulae defined by the $2\sigma$ \lya isophotal apertures of \citet{Matsuda2004} 
(e.g. see Figures \ref{FigATLAS1} and \ref{FigATLAS}), above a 
\lya surface brightness limit of $2.2\times 10^{-18}$ \unitcgssb, and 
because of the use of narrow-band imaging, we can probe the whole extent
of the source. On the other hand, in the case of LABs (\citealt{Prescott2013,Dey2005}) 
and HzRGs (\citealt{VillarM2007}), the lines are extracted from smaller 
aperture forcedly defined by the slit, sampling a particular position within the nebula. 
For example, in the case of HzRGs (\citealt{DeBreuck2000}), the lines are
typically measured from a one-dimensional spectra extracted by choosing 
the aperture which includes the most extended emission line, and typically the 
slit is oriented along the radio axis. 

To further complicate the comparison, for HzRGs and QSOs where a bright
central source is clearly detected, it is difficult to separate the emission 
from the central source and from the nebula itself. For example, for 
the radio-loud QSOs, \citet{Heckman1991spec, Heckman1991} 
carefully removed the contribution from the central QSOs in both the
imaging and the spectroscopic analysis, thus these line ratios should
only reflect the line emission in the extended
nebulae\footnote{\citet{Heckman1991spec, Heckman1991} removed the
  continuum from the narrowband images and estimated the contribution
  of the QSO to the Ly$\alpha$ nebula by subtracting a scaled PSF. In
  the spectroscopic analysis, they iteratively subtracted a scaled
  version of the nuclear spectrum from the off-nuclear ones, until all
  traces of continuum flux near Ly$\alpha$ vanished.}.
However, in the measurements for HzRGs no attempt is made 
to exclude a possible contribution from the central obscured AGN.  
While in the case of the LABs, the neglect of the contribution of the sources within the \lya emission 
is not relevant because the star-forming galaxies embedded in the nebulae should scarcely emit 
in \civ and \heii lines (e.g. \citealt{Shapley2003}), and constitute 
only a small fraction of the area in the aperture.

\begin{figure}
\vspace{1cm}
\centering{\epsfig{file=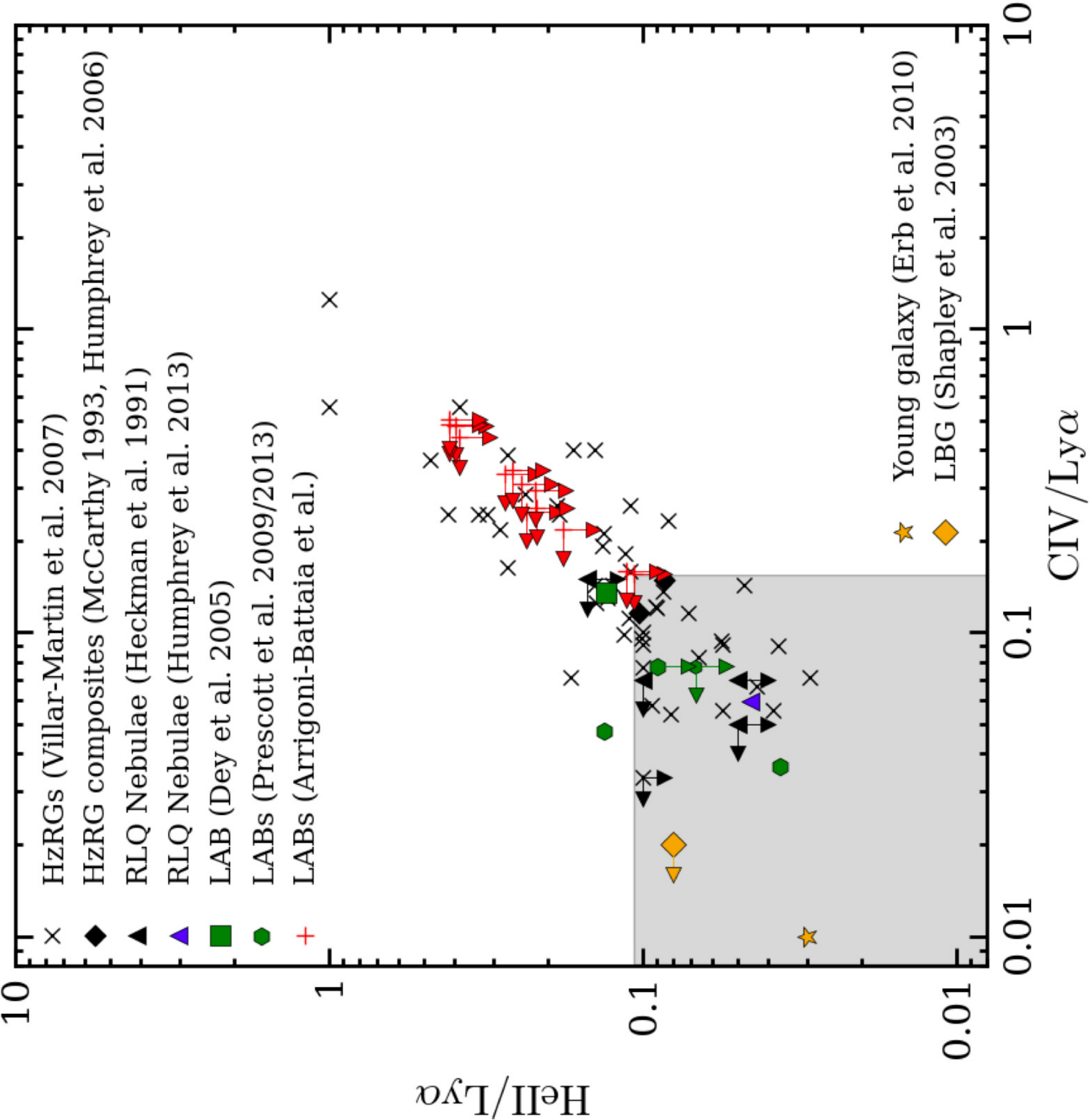, width=0.95\columnwidth, viewport=0 0 520 560, angle=270}}
\caption{HeII/Ly$\alpha$ versus CIV/Ly$\alpha$ log-log plot. Our upper limits on the HeII/Ly$\alpha$ and CIV/Ly$\alpha$ ratios are compared with the values quoted in the literature for 
HzRG, QSOs, and LABs (see text for references). Due to their larger extent, LAB1 and LAB2 define the strongest limits on these ratios: the gray shaded area highlights the regime constrained by these limits. 
Note however that these data are quite difficult to compare because of their heterogeneity. 
}
\label{Fig11}
\end{figure}

Despite these caveats, in Figure \ref{Fig11} we plot all the available data in 
the literature for completeness to show the ranges 
spanned by these different types of sources in a \heii/\lya versus \civ/\lya diagram. But 
we caution again the reader that a direct comparison of objects from different
studies in this plot could be problematic. The upper limits for the 13 \lya blobs in our sample
are shown in red. 

Figure \ref{Fig11} illustrates that our upper limits are consistent with the previous 
measurements and more interestingly, that there are sources in the
literature with line ratios even lower than our strongest upper limits (LAB1 and LAB2, gray
shaded region).  Indeed, although our narrow band images
constitute the deepest absolute SB limits ever achieved in the \ion{C}{4}
and \ion{He}{2} emission lines, 
some previous searches probed to smaller values of the line ratios because
they observed brighter Ly$\alpha$ nebulae (e.g. in the case of HzRGs) or because they probed only the 
central part of the nebula where the \lya emission is expected to be brighter.
For example, \citet{Prescott2013} probed down to lower line ratios (e.g. the lowest green point in the plot, i.e. the LAB PRG2) 
because they focus on the brightest part of the blob in \lya. Indeed, while 
the approximate isophotal area for this LAB is 103 arcsec$^2$, they covered only a smaller aperture
(1.5\arcsec$\times$7.84\arcsec) with their long-slit spectra.
Thus, notwithstanding our efforts, Figure \ref{Fig11} is clearly indicating that
in order to explore the full range of line ratios, one requires either deeper 
observations, or brighter samples of Ly$\alpha$ emission nebulae (see e.g. \citealt{Cantalupo2014}). 

In addition to the sources with giant \lya emission nebulae, Figure
\ref{Fig11} also shows line ratios for star-forming galaxies at $z = 2-3$,
for which the CIV and HeII line ratio is not powered by an AGN. 
In particular, we show the line ratios determined from the composite spectrum 
of  Lyman break galaxies (LBGs) from \citet{Shapley2003} \footnote{We 
use the values quoted for their subsample of LBGs that have strong \lya emission, i.e. EW(\lya) $=52.63\pm2.74$ (\citealt{Shapley2003}).} 
and for a peculiar galaxy (Q2343-BX418) studied in detail by
\citet{Erb2010} which exhibits particularly strong \heii emission.
We show the corresponding line-ratios for LBGs because it 
has been proposed that some LABs could be powered by
star-formation \\(\citealt{Ouchi2009}), albeit with extreme
star-formation rates $\simeq 1000\, {\rm M}_{\odot}/{\rm yr}$.
Indeed, the stacked \lya narrowband images of LBGs also exhibit
diffuse \lya emission extending as far as $\sim$50\,kpc
(\citealt{Steidel2011}), although the \lya luminosity and surface
brightness of these halos is $\gtrsim 10 \times$ fainter than the LABs
and the \lya nebulae associated with HzRGs and QSOs. However, if the
LABs represent some rare mode of spatially extended star-formation,
then the \civ and \heii line ratios of star-forming galaxies could
thus be relevant.

The origin of the \heii and \civ emission observed in the spectra of
star-forming galaxies is not completely
understood. \citet{Shapley2003} noted relatively broad (FWHM
$\sim$1500 km s$^{-1}$) \heii emission in the composite spectrum of
LBGs, and speculated that it arises from the hot, dense stellar winds
of Wolf-Rayet (W-R) stars, which descend from O stars with masses of
$M$ $>$ 20--30 M$_{\odot}$. The \civ line in LBGs exhibits a
characteristic P Cygni-type profile, which presumably arises from a
combination of stellar wind and photospheric absorption, plus a strong
interstellar absorption component due to outflows
(\citealt{Shapley2003}). There could also be a narrow nebular emission
component powered by a hard ionizing source. In Figure \ref{Fig11} we
adopt the strict upper limit of \civ/\lya $<0.02$ of the non-AGN
subsample in \citealt{Shapley2003}, whereas for the \heii/\lya ratio 
we use the global value for  the first quartile with the \lya line in emission 
because no \heii/\lya value was quoted for the non-AGN subsample.
\citet{Erb2010} studied a young ($<100$Myr), low
metallicity ($Z\sim1/6Z_{\odot}$) galaxy at $z=2.3$ which exhibits
exceptionally strong \heii emission, which they however argued is not
powered by an AGN.  \citet{Erb2010} interpreted the \heii emission as
a combination of a broad component due to W-R stars and a narrow
nebular component, powered by a hard ionizing spectrum.
Although the \heii emission is strong in comparison with other typical $z \sim 2-3$ LBGs, 
indicative of a harder ionizing spectrum, 
the \heii/\lya ratio of this galaxy is in fact
lower than that of the average LBG owing to its 
extremely strong  \lya line.

\medskip
\medskip

\section{Discussion}
\label{sec:Discussion}

In what follows we discuss our upper limits in light of a photoionization or a shock scenario.
Here, we briefly outline the physics underlying the models and the parameters used, 
but we refer the reader to \citet{Hennawi2013} and our subsequent paper (Arrigoni-Battaia et al. in prep.) 
for further details and a complete analysis.

\subsection{Comparison with Photoionization Models}
\label{sec:model}

\medskip

It is well established that the ionizing radiation from a central AGN can power giant Ly$\alpha$ nebulae, 
with sizes up to $\sim$ 200 kpc, around high-z radio galaxies (HzRG) 
(e.g., \citealt{VillarM2003}; \citealt{Reuland2003}; \citealt{Venemans2007}) 
and quasars (e.g., \citealt{Heckman1991}; \citealt{Christensen2006}; \citealt{Smith2009}),
together with extended \heii and \civ emission (\citealt{VillarM2003b}).  
Although HzRGs are more rare ($n \sim 10^{-8}$ Mpc$^{-3}$; \citealt{Miley2008}), 
the similarity between the volume density of LABs ($n \sim 10^{-5}$ Mpc$^{-3}$; \citealt{Yang2010}) 
and luminous QSOs ($n \sim 10^{-5}$ Mpc$^{-3}$; \citealt{HRH2007}), suggests that the LABs
could represent the same photoionization process around obscured QSOs.
Unified models of AGN invoke an obscuring medium which could extinguish 
a bright source of ionizing photons along our line of sight (e.g., \citealt{UP1995}). 
Indeed, evidence for obscured AGNs have been reported for several LABs (e.g., \citealt{Basu2004, Dey2005, Geach2007, Barrio2008, Geach2009, Overzier2013, Yang2014a}), 
lending credibility to a photoionization scenario; however, this is not always the case (\citealt{Nilsson2006, Smith2007, Ouchi2009}).

Despite these circumstantial evidences in favor of the photoionization
scenario, detailed modeling for \heii and \civ lines due to AGN
photoionization in the context of large Ly$\alpha$ nebulae has not
been carried out in the literature, with the exceptions of some
studies focusing on the modeling of emission lines in the case of
extended emission line regions (EELR) of HzRGs (e.g.,
\citealt{Humphrey2008}).  Although many authors have modeled the
narrow-line regions (NLR) of AGNs (e.g., \citealt{Groves2004},
\citealt{Nagao2006}, \citealt{Stern2014}), the physical conditions on
these small scales $\sim 100$ pc (i.e. gas density, ionization
parameter) are expected to be very different than the $\sim100$ kpc
scale emission of interest to us here. As such, we model the photoionization of gas on scales of $100$ kpc from a central
AGN to predict the resulting level of the \heii and \civ lines,
relative to the Ly$\alpha$ emission.

To select the parameters of the models in order to recover the \lya SB
of LABs, we follow the simple picture described by
\citet{Hennawi2013}, and assume a LAB to be powered by an obscured QSO 
with a certain luminosity at the Lyman limit ($L_{\nu_{\rm LL}}$). 
In this picture, the QSO halo is populated with spherical clouds of cool
gas ($T \sim 10^4$\,K) at a single uniform hydrogen volume density
$n_{\rm H}$ and with an average column density $N_{\rm H}$, and uniformly 
distributed throughout a halo of radius $R$, such as they have a cloud 
covering factor $f_C$ (see \citealt{Hennawi2013} for details).
We consider two limiting regimes for recombination: the optically thin ($N_{\rm HI} < 10^{17.2}$ cm$^{-2}$) 
and thick ($N_{\rm HI} > 10^{17.2}$ cm$^{-2}$) to the Lyman continuum photons, 
where $N_{\rm HI}$ is the neutral column density of a single spherical cloud. 
In this scenario, once the size of the halo is fixed, in the optically thick case  
the Ly$\alpha$ surface brightness
scales with the luminosity at the Lyman limit of the central source,
$SB_{{\rm Ly}\alpha}^{\rm thick} \propto f_C L_{\nu_{\rm LL}}$, while in
the optically thin regime ($N_{\rm HI} < 10^{17.2}$ cm$^{-2}$) the SB
does not depend on $L_{\nu_{\rm LL}}$, $SB_{{\rm Ly}\alpha}^{\rm thin}
\propto f_C n_{\rm H} N_{\rm H}$, provided the AGN is bright enough to
keep the gas in the halo ionized.

To cover the full range of possibilities, we thus construct a grid 
of $\sim$5000 Cloudy models with parameters in the following range 
(see the appendix for additional information on how the parameters were chosen):
\begin{itemize}
\item[---] $n_{\rm H} = 0.01$ to $100$ cm$^{-3}$ (steps of 0.2 dex);
\item[---] $\log N_{\rm H}= 18$ to $22$ (steps of 0.2 dex);
\item[---] $\log L_{\nu_{\rm LL}} = 29.3$ to $32.2$ (steps of 0.4
  dex).
\end{itemize}

Finally, we decide to fix the covering factor
to unity $f_C=1.0$. The assumption of a high or unit covering factor
is driven by the observed diffuse morphology of the \lya nebulae,
which do not show evidence for clumpiness arising from the presence of
a population of small unresolved clouds. We directly test this
assumption as follows. We randomly populate an area of 200 arcsec$^2$
(area of LAB1) with point sources such that $f_C=0.1-1.0$, and we
convolve the images with a Gaussian kernel with a FWHM equal to our
median seeing value, in order to mimic the effect of seeing in the
observations. We find that the smooth morphology observed for LABs cannot be
reproduced by images with $f_C<0.5$, as they appear too clumpy.

We preform photoionization calculations using the Cloudy
photoionization code (v10.01), last described by \citet{Ferland2013}. 
As the LABs are extended over $\sim 100$ kpc, whereas the radius of the 
emitting clouds is expected to be much smaller, we
assume a standard plane-parallel geometry for the emitting clouds illuminated
by the distant central source. Note that we evaluate the ionizing flux at a single location for input into Cloudy, 
specifically at $R/\sqrt{3}$ (where $R=100$ kpc). Capturing the variation of the physical 
properties of the nebula with radius is beyond the purpose of this work. 
Indeed, given that for the objects in the literature are not reported 
radial trends for the \civ/\lya and \heii/\lya ratios, and 
given that we have non detections, modeling the emission as coming from a single
radius is an acceptable first order approximation. 
We consider only models with solar metallicity, and we assume that the ionizing continuum has a power law
form $L_{\nu} = L_{\nu_{\rm LL}}(\nu\slash \nu_{\rm LL})^{\alpha}$, where 
$\nu_{\rm LL}$ is the frequency of the Lyman limit, and we take the slope of the ionizing
continuum set to be $\alpha_{EUV}=-1.57$ following
\citet{Telfer2002}. Note that our assumption of this power law ionizing
continuum amounts to assuming that the central AGN powering the LAB
has a spectrum similar to a Type-1 QSO; of course this UV ionizing source is not
directly observed because it is presumed to be obscured from our vantage
point. Note that unlike the case where a QSO is clearly powering a nebula, for LABs we
do not have a constraint on the ionizing luminosity of the central
source $L_{\nu_{\rm LL}}$. As we have assumed that a Type-1 QSO spectrum powers the
nebulae, we can convert $L_{\nu_{\rm LL}}$ into an $i$-band apparent
magnitude following the procedure described in
\citet{Hennawi2006}\footnote{This procedure simply ties the
  \citet{Telfer2002} power-law spectrum to the composite quasar
  spectrum of \citet{VandenBerk2001}, so that $i$-band magnitude can
  be computed.}. The $L_{\nu_{\rm LL}}$ that we consider correspond
to $i$-band apparent magnitudes of $i=16-23$, in steps of unity.

In addition to the \lya due to the recombination, the resonant scattering of \lya becomes important when the gas is optically 
thick at Ly$\alpha$ line, roughly for $N_{\rm HI} \gtrsim 10^{14}$ cm$^{-2}$. 
The Ly$\alpha$ emission from scattering will follow the relation (\citealt{Hennawi2013})
\begin{equation}
\label{SBscatt}
SB_{{\rm Ly}\alpha}^{\rm scatt} = \frac{h\nu_{{\rm Ly}\alpha}}{4\pi(1+z)^4}f_{\rm C}\Phi_{{\rm Ly}\alpha}(R\slash \sqrt{3}),  
\end{equation}
where $\Phi_{{\rm Ly}\alpha}$ is the flux of continuum photons emitted
close enough to the Ly$\alpha$ resonance to be scattered by gas in
motion around the quasar (we assume that the rest-frame equivalent width of Ly$\alpha$ absorption is close to 1\,\AA, \citealt{Hennawi2013}).
To take into account this effect, we simply add this scattering contribution to the photoionization \lya SB of the models.
Note that the scattering emission is not relevant in the optically thick regime because the flux of ionizing photons is larger than the 
flux of Lya photons, i.e. $\Phi_{\rm LL}/\Phi_{{\rm Ly}\alpha}\sim150$ (\citealt{Hennawi2013}). 
On the other hand, in the case of the optically thin regime, the
SB$_{{\rm Ly}\alpha}$ for scattering is comparable with the emission from
the recombination if the central source is bright enough (in this work for $i < 18$).  
Finally, from our model grid, we select 
only models with SB$_{{\rm Ly}\alpha}$ =
(1--9)$\times$$10^{-18}$ erg s$^{-1}$ cm$^{-2}$ arcsec$^{-2}$, comparable to LABs.

\begin{figure*}
\vspace*{0.5cm}
\epsfig{file=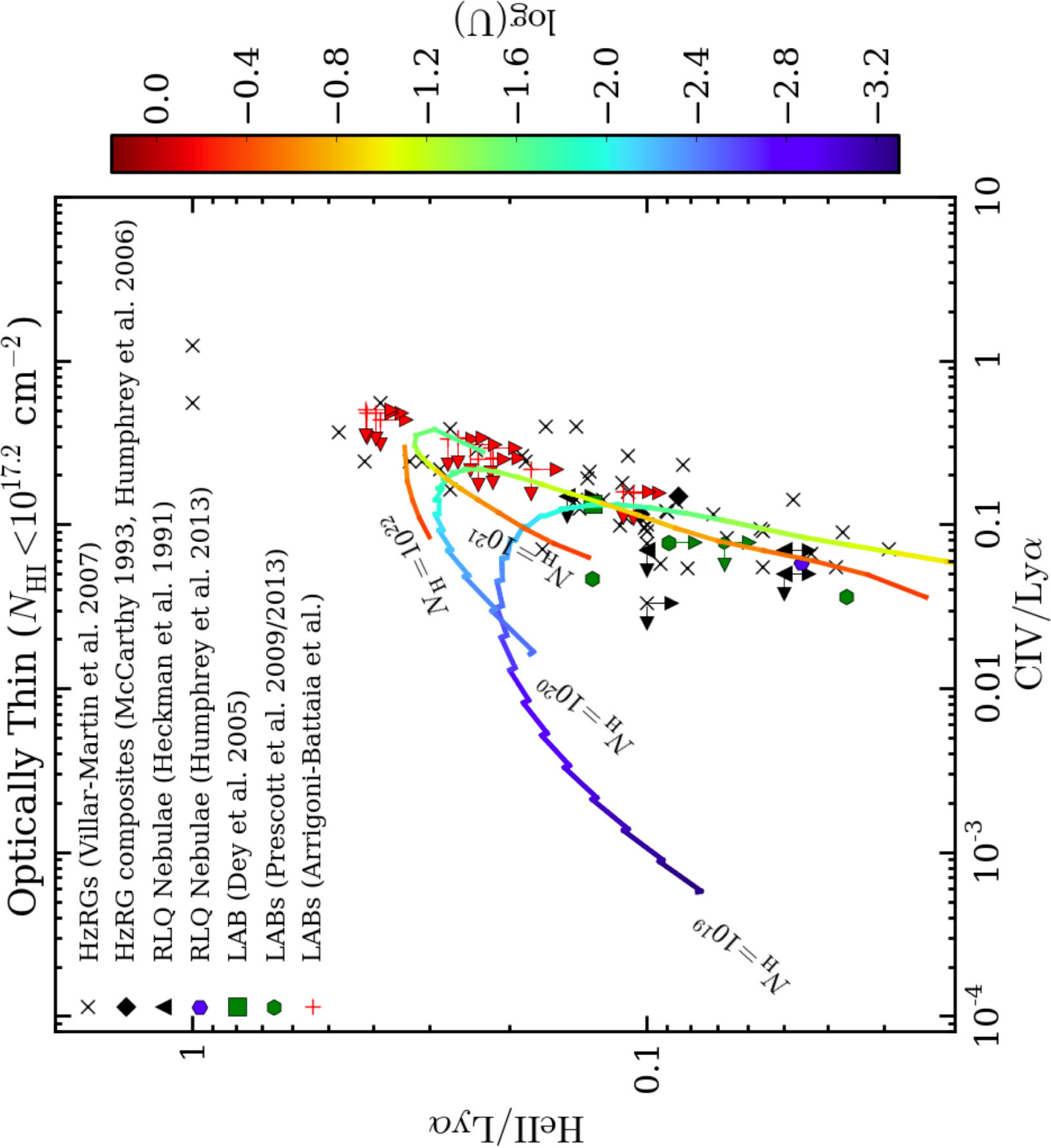, width=0.92\columnwidth, angle=270}  
\hspace{0.5cm}
\epsfig{file=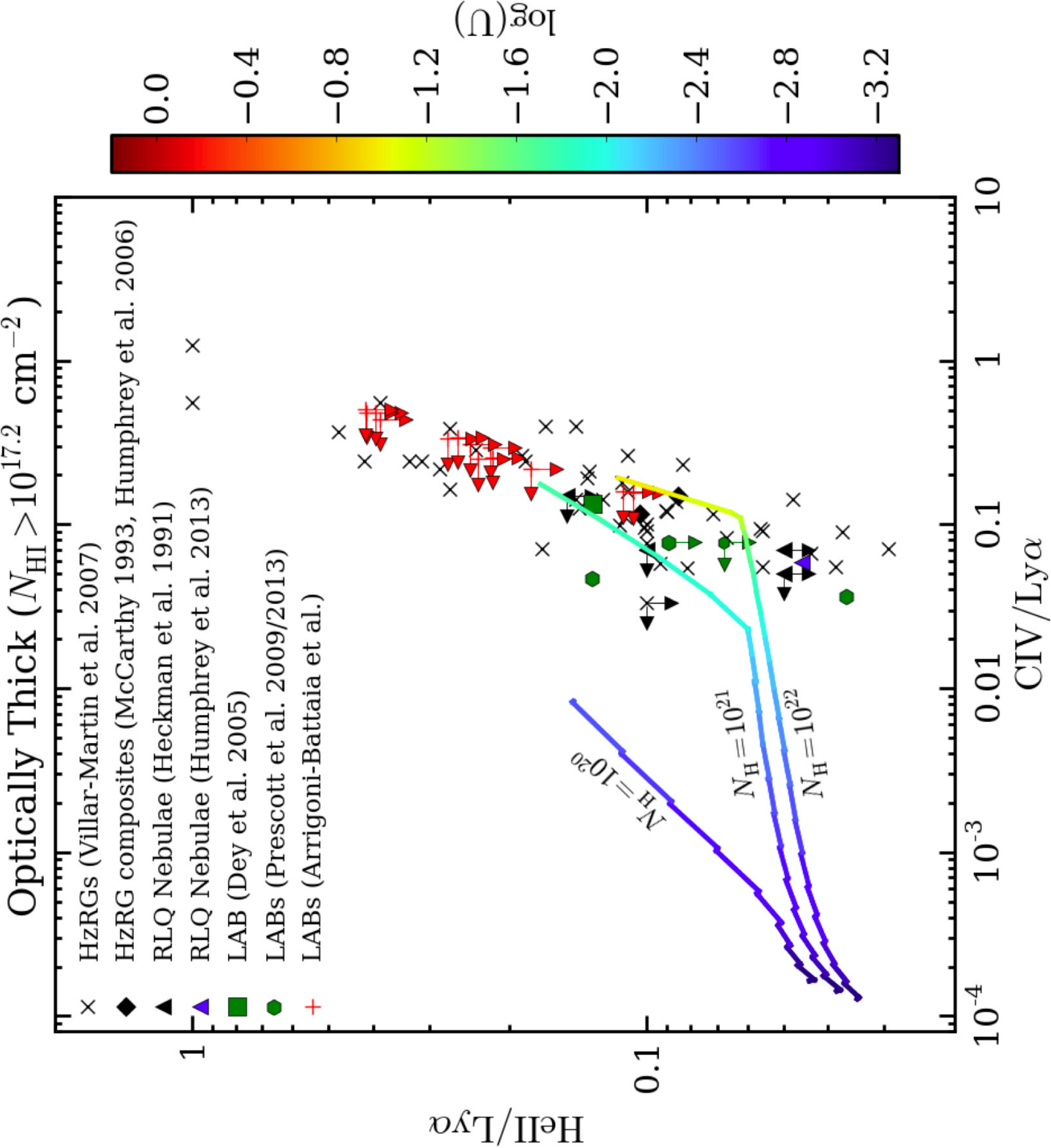, width=0.92\columnwidth, angle=270} 
\caption{HeII/Ly$\alpha$ versus CIV/Ly$\alpha$ log-log plot. Same data points as in Figure \ref{Fig11}. Our upper limits on the HeII/Ly$\alpha$ and CIV/Ly$\alpha$ ratios are compared with 
the Cloudy photoionization models. In the left panel we plot the optically thin models, while in the right panel is shown the optically thick regime. For clarity, we plot only the models with 
$N_{\rm H}=10^{19},10^{20},10^{21},10^{22}$ cm$^{-2}$. The grids are color coded following the ionization parameter (see colorbar on the right) and the value of hydrogen column density is indicated.
Note that there are no optically thick models with $N_{\rm H}=10^{19}$ cm$^{-2}$. 
Note that the x-axis is on different scale than Figure \ref{Fig11}.
}
\label{Fig12}
\end{figure*}

In Figure \ref{Fig12} we compare our photoionization model predictions in
the \heii/Ly$\alpha$ versus \civ/Ly$\alpha$ diagram to our LAB limits
and the data points from the literature. The left panel and right
panels show the optically thin and optically thick regimes,
respectively.  Note that this division into optically thin and thick
models, corresponds to a division in the ionizing luminosity of the
central source (which in the case of LABs and HzRGs is obscured from
our vantage point and is thus unknown).  Specifically, in the
optically thin regime we find that for the range of ${\rm SB}_{\rm
  Ly\alpha}$ considered, the central source must have $L_{\nu_{\rm
    LL}}\gtrsim 10^{30.5}$ erg s$^{-1}$ Hz$^{-1}$ or $i\lesssim
20$\ \footnote{This constraint follows from the definition of an
  optically thin cloud, i.e. $N_{\rm HI} \ll 10^{17.2}$ cm$^{-2}$.}.  On
the other hand, because in the optically thick limit ${\rm SB}_{\rm
  Ly\alpha}\propto L_{\nu_{\rm LL}}$, the ionizing luminosity is fixed
to be in a relatively narrow range $L_{\nu_{\rm LL}}\simeq
10^{29.7}-10^{29.3}$ erg s$^{-1}$ Hz$^{-1}$ ($i\simeq 22-23$).

For clarity, in Figure \ref{Fig12} we show only the models with
$N_{\rm H}=10^{19},10^{20},10^{21},10^{22}$ cm$^{-2}$.  The model grids are
color-coded according to the ionization parameter $U$, which is
defined to be the ratio of the number density of ionizing photons to
hydrogen atoms ($U \equiv \Phi_{LL}/c n_{\rm H} \propto L_{\nu_{\rm LL}}/n_{\rm H}$), and 
provides a useful characterization of the
ionization state of the nebulae. 
Because photoionization models are self-similar in this parameter
\citep{FerlandARAAarticle2003}, our models will exhibit a degeneracy
between $n_{\rm H}$ and $L_{\nu_{\rm LL}}$.  Nevertheless, we decided
to construct our model grid in terms of $N_{\rm H}$ and $L_{\nu_{\rm LL}}$,
in order to explore the possible ranges of both parameters.

Figure~\ref{Fig12} illustrates that, overall, our photoionization
models can cover the full range of HeII/Ly$\alpha$ and CIV/Ly$\alpha$
line ratios that are observed in the data.  The optically thin regime
(see left panel) seems to better reproduce the range of line ratios
set by our most stringent upper limits (LAB1 and LAB2), as well the
locus of measurements in the \civ/\lya -- \heii/\lya diagram for
HzRGs, QSOs, and LABs.  In particular, models with log$U\approx-1.5$
and $10^{19} \leq N_{\rm H} < 10^{20}$ cm$^{-2}$ populate the region below
our LAB limits, whereas models with log$U\gtrsim-2.0$ and $10^{19}
\leq N_{\rm H} < 10^{21}$ cm$^{-2}$ would be broadly consistent with most of
the detections.  Note that previous studies of EELR around HzRGs
favored models with log$U\sim-1.46$ (e.g. \citealt{Humphrey2008}),
which are consistent with our results.

Note however that two HzRGs with \heii/Ly$\alpha$ $\approx$ 1 and
\civ/Ly$\alpha$ $\approx$ 1, are not covered by our models. For both
of these data, emission from the central source has not been excluded,
and thus we speculate that these very high line ratios arise because
of contamination from the narrow-line region of the obscured AGN,
where \lya photons have been destroyed by dust.  Indeed, both of these
objects, MG1019+0535 and TXS0211-122, have a \civ/\heii ratio similar
to the bulk of the HzRGs population, but they exhibit unusually weak
\lya lines (\citealt{Dey1995}, \citealt{vanOjik1994}).  Note however,
that while destruction of \lya by dust grains can have a large impact
on these line ratios for emission emerging from the much smaller scale
narrow line region, dust is not expected to significantly attenuate
the \lya\ emission in the extended nebulae around QSOs (see discussion
in Appendix A of \citealt{Hennawi2013}) given the physical conditions
characteristic of the CGM, and thus we neglect destruction of \lya
photons by dust in our modeling.

The trajectory of the optically thin models through the
HeII/Ly$\alpha$ and CIV/Ly$\alpha$ diagram can be understood as
follows.  We first focus on the curve for $N_{\rm
  H}=10^{19}\,{\rm cm^{-2}}$ and follow it from low to high $U$.
Recall that in the optically thin regime $SB_{Lya}\propto n_{\rm H}
N_{\rm H}$, but is roughly independent of the source luminosity
$L_{\nu_{\rm LL}}$ \footnote{Note that in this regime the \lya emission 
is not completely independent on the luminosity of the central source. 
Indeed, this scaling neglects small variations due to temperature effects,
which Cloudy is able to trace.}. 
Thus by fixing $N_{\rm H}=10^{19}\,{\rm cm^{-2}}$, and requiring that $SB_{{\rm
    Ly}\alpha}=$(1--9)$\times10^{-18}$ erg s$^{-1}$ cm$^{-2}$, we also
fix $n_{\rm H}$.  Thus $U$ is increases along this track because the
central source luminosity is increasing $L_{\nu_{\rm LL}}$, which hardly changes
the \lya emission, but results in significant variation in  both \heii
and \civ. 

First consider the trend of the \heii/\lya\ ratio.  \heii is a
recombination line and thus, once the density is fixed, its emission
depends basically on what fraction of Helium is doubly ionized.  For
this reason, the \heii/\lya\ ratio is increasing from log$U=-3.3$ and
reaches a peak at log$U\sim-2.0$, corresponding to an increase in the
fraction of the He$^{++}$ phase from about 20\% to 90\% of the total
Helium. Further increases $U$, result in only modest changes to the
He$^{++}$ fraction, but result in an increase in gas temperature. These
higher temperatures result in a decrease of the He$^{++}$ recombination rate. 
In addition this higher temperature impacts the \lya line in the same way, but  
continuum pumping due to the increased luminosity of the central source 
further increases the \lya emission, with the net
effect that \heii emission is reduced relative to \lya 
(as discuss in Arrigoni-Battaia et al. in prep.).

Our photoionization models indicate that the \civ emission line is an
important coolant and is powered primarily by collisional excitation.
Figure~\ref{Fig12} shows that our models span a much wider range in
the \civ/\lya ($\sim 3$ dex) ratio than in \heii/\lya ($\lesssim 2$
dex). The strong evolution in \civ/\lya results from a combination of
two effects. First, increasing $U$ increases the temperature of the
gas, and the \civ collisional excitation rate coefficient has a strong
temperature dependence (\citealt{Groves2004}). Second, the efficacy of
\civ as a coolant depends on the amount of Carbon in the C$^{+3}$
ionic state. As log$U$ increases from $\simeq -3.3$ to $\simeq -2$,
the C$^{+3}$ fraction increases from 1\% to 37\%.  These two effects
conspire to give rise to nearly three orders of magnitude of variation in the 
\civ emission. 

Although our analysis suggests that the optically thin models are
favored, the optically thick models (see right panel of
Figure~\ref{Fig12}) can also populate the area below the upper limits for
LAB1 and LAB2, and at least the lower part of the observed
\heii/Ly$\alpha$ -- \civ/Ly$\alpha$ diagram.  Note that given the
range of $L_{\nu_{\rm LL}}$ and $n_{\rm H}$ in our parameter grid,
models with $N_{\rm H} = 10^{19}$ cm$^{-2}$ are never optically
thick\footnote{We found optically thick models for $N_{\rm H} > 10^{19.2}$
  cm$^{-2}$.}, which explains why we only show optically thick models
with $N_{\rm H} = 10^{20}, 10^{21},10^{22}$ cm$^{-2}$.  The bulk of these
models reside on a sequence with almost constant HeII/Ly$\alpha$
(around HeII/Ly$\alpha = 0.04-0.05$) for a wide range of
CIV/Ly$\alpha$, which is driven by variation in $U$. The models departing from
this sequence are characterized by $N_{\rm HI}$ slightly greater than
$10^{17.2}$ cm$^{-2}$ and they can thus be seen as a transition
between the optically thick case and the optically thin case.

To summarize, the photoionization models produce line ratios which are
consistent with our upper limits and which span the values observed in the 
literature, although we favor the optically thin scenario. In the next 
section we consider the degree to which shock powered emission can explain 
line ratios in \lya nebulae.

\subsection{Comparison with Shock Models}
\noindent
\label{shock}

\citet{Taniguchi&Shioya2000} and \citet{Mori2006} have speculated
that intense star-formation accompanied by successive supernova explosions
could power a large scale galactic superwind, and radiation generated
by overlapping shock fronts could power the \lya emission in the LABs.
However, it is well known that it is difficult to distinguish between
photoionization and fast-shocks using line-ratio diagnostic diagrams
(e.g. \citealt{Allen1998}).  Furthermore, for AGN narrow line regions,
the \lya line is
typically avoided in these diagrams because of its resonant nature and
the fact that it may be more likely to be destroyed by dust, although 
we have argued that it is not an issue for CGM gas.  It is thus interesting
to study how shock models
populate the \heii/\lya versus \civ/\lya diagram in comparison with
photoionization models and our observational limits.

To build intuition about the line ratios expected in a shock scenario
we rely on the modeling of fast shocks by \citet{Allen2008}.  We thus
imagine the \lya emission as the sum of overlapping shock fronts with
shock velocity $v_{\rm s}$, moving into a medium with preshock density
$n_{\rm H}$.  In the case of such shocks, \citet{Allen2008} showed
that the \lya emission depends strongly on $v_{\rm s}$, i.e. $F_{{\rm
    Ly}\alpha}\propto n_{\rm H} v_s^{3}$ (their Table 6).
In order to test a realistic set of parameters in the case of LABs, 
we limit the grid of models presented by \citet{Allen2008} to:
\begin{itemize}
\item $n_{\rm H}=0.01, 0.1, 1.0, 10, 100$ cm$^{-3}$,
\item shock velocities, $v_{\rm s}$, from 100 km s$^{-1}$ to 1000 km s$^{-1}$ in steps of 25 km s$^{-1}$.
\end{itemize}
We consider only models with solar metallicity \footnote{Note that the
  solar values used by \citet{Allen2008} are slightly different from
  what is used in Cloudy (and thus in our previous section).}. The
magnetic parameter $B/n^{1/2}$, where B is the magnetic field in $\mu$G, determines the relative 
strength of the thermal and magnetic pressure. We adopt a magnetic parameter $B/n^{1/2}=3.23$ $\mu$G cm$^{3/2}$,
which represents a value expected for ISM gas assuming equipartition of
magnetic and thermal energy. 
However, note that, given the very strong dependence of the ionizing flux on the shock
velocity $F_{\rm UV}\propto v_{\rm s}^3$, the line ratios do not vary
so markedly with either the metallicity or the magnetic field 
(see \citealt{Allen2008} for further details).

In Figure \ref{shock} we show two sets of shock models. On the left,
we plot the models for which the emission is coming solely from the
shocked region, where the gas is ionized and excited to high
temperatures by the shock. Temperatures ahead of the shock-front are
of the order of 10$^4$\,K , whereas temperatures as high as 10$^6$ K
can be reached in the post-shock gas (\citealt{Allen2008}).  On the
right, we plot a combination of the emission coming from the shocked
gas and from the precursor, i.e. the pre-shock region which is
photoionized by the radiation emitted upstream from the shocked
region.  The trends of the models can be explained as follows.  The
models for the shock component (left panel of Figure \ref{shock}) show
a rapid decrease in the \civ/\lya ratio for increasing $v_{\rm s}$. This is
due to a rapid increase in the \lya line due to the strong scaling of
the ionizing flux with $v_{\rm s}$, and to a decrease in the CIV line due to
the lack of carbon in the C$^{3+}$ phase for high velocities
(i.e. carbon is in higher ionization species, see Figure 9 of
\citealt{Allen2008}).  The \heii/\lya ratio depends more strongly on
the gas density because $n_{\rm H}$ sets the volume of the shocked
region and thus the recombination luminosity of Helium, i.e. at fixed
$v_{\rm s}$, a higher density corresponds to a smaller shocked volume and
less Helium emission (see Figure 6 of \citealt{Allen2008}).

The combination of shock and precursor models mainly alter the ratios
for models with high $v_{\rm s}$ (see right panel of Figure \ref{shock}).
This is because the precursor component is adding
the contribution of a photoionized gas at temperature of the order of
10$^4$ K, and the ionizing flux scales strongly with shock velocity $F_{\rm UV}\propto v_{\rm s}^3$. 
For velocities $v_{\rm s}\gtrsim 400$ km s$^{-1}$, the resulting hard
radiation field results in a large fraction of double ionized Helium He$^{++}$
over a significant volume of the precursor, significantly 
increasing the \heii emission and the \heii/\lya ratio. This photoionized
precursor similarly increases the abundance of the C$^{3+}$ phase 
giving rise to a higher \civ/\lya ratio. Thus,
adding the precursor contribution to the shock models causes the
models to fold over each other at high velocities.

Figure \ref{shock} illustrates that the shock models are capable of
populating the line ratio diagram below our tightest upper limits
(i.e. LAB1 and LAB2).  However, note that our limits on the \civ line
imply velocities above $\sim$250 km s$^{-1}$, in potential
disagreement with independent constraints on the outflow velocities in
LABs in the literature. 
Indeed, using the velocity offset between the \lya and the non-resonant [O
III] or H$\alpha$ line, the offset of stacked interstellar metal absorption
lines, and the [O III] line profile, \citet{Yang2011,Yang2014b} find that the
kinematics of gas along the line of sight to galaxies in LABs are
consistent with a simple picture in which the gas is stationary or slowly
outflowing at a few hundred km s$^{-1}$ from the embedded galaxies 
in contrast with the $\sim$1000 km s$^{-1}$ velocities necessary to power LABs via
superwind outflows (\citealt{Taniguchi&Shioya2000}).
In addition, \citet{Prescott2009} showed that the \heii line detected for 
a LAB at $z=1.67$, is narrow, i.e. $v_{\rm FWHM}\lesssim500$ km s$^{-1}$.
If shocks are the mechanism powering the nebula, this observation 
is inconsistent with strong shock velocities, i.e. $v_{\rm s} \lesssim500$ km s$^{-1}$.   
Thus, these observations seem to rule out an extreme wind scenario in these 
LABs.

\begin{figure*}
\vspace*{0.5cm}
\epsfig{file=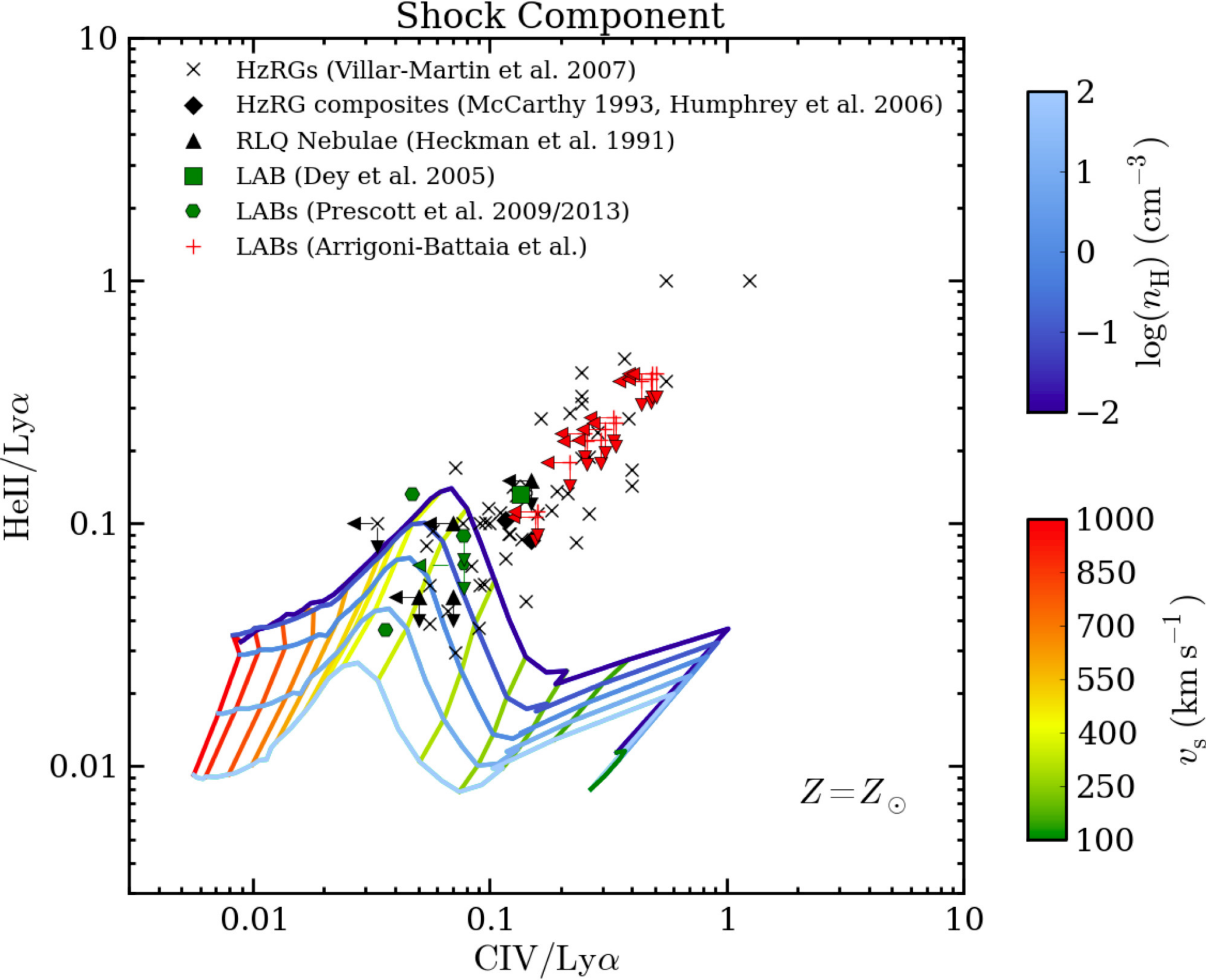, width=0.86\columnwidth, angle=270} 
\hspace{0.1cm}
\epsfig{file=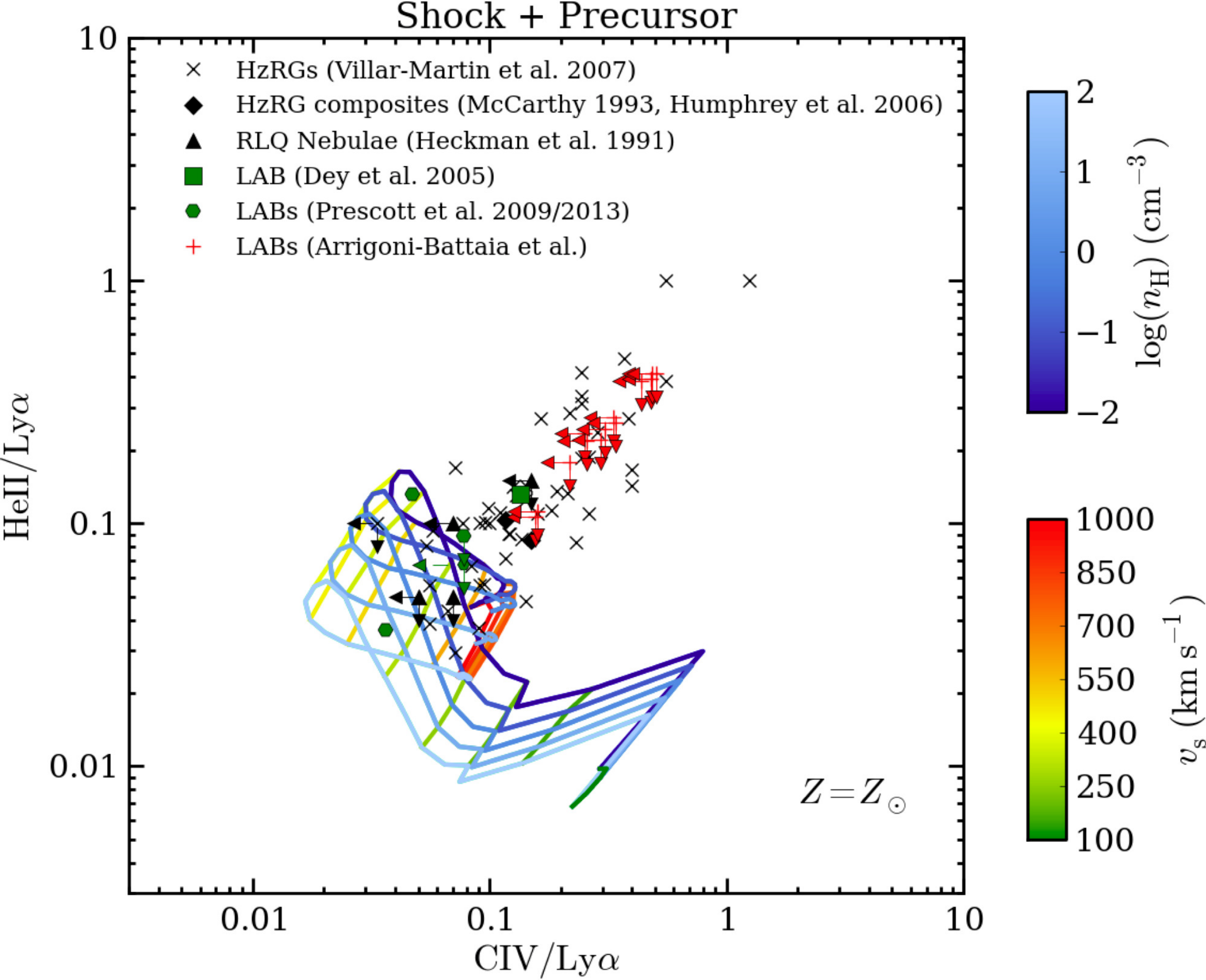, width=0.86\columnwidth, angle=270}
\caption{HeII/Ly$\alpha$ versus CIV/Ly$\alpha$ log-log plot. Same data points as in Figure \ref{Fig11}. Our upper limits on the HeII/Ly$\alpha$ and CIV/Ly$\alpha$ ratios are compared with 
the models by \citet{Allen2008}. In the left panel we plot the shock models, while in the right panel is shown the combination of shock and precursor. The grids are color coded following 
the density of the pre-shock region, $N_{\rm H}$, and the velocity of the shock, $v_{\rm s}$.
The models are not taking into account the possible additional contribution due to Ly$\alpha$ scattering.
}
\label{shock}
\end{figure*}

It is worth to stress again here, that these models suffer from uncertainty in the \lya calculation. In particular, the additional contribution from scattering is not taken into account, 
thus making the \lya line weaker. As a consequence, these grids may be shifted to lower values on both axes. 
Note also that we fix the metallicity to the solar value. However, a decrease in the \civ emission is expected for sub-solar metallicity, 
weakening the constraints on the shock velocities.
The trends with metallicity are beyond the scope of this work and we are going to address them in a subsequent paper (Arrigoni-Battaia et al. in prep.).
Thus, even though our models can give us a rough idea of the line emission in the shock scenario, these plots should be treated with caution.

\subsection{Comparison to Previous Modeling of Extended Ly$\alpha$ Emission Nebulae}
\noindent

As stated in the previous sections, rigorous modeling of photoionization of large
\lya nebulae in the context of LABs has never been performed. However,
\citet{Prescott2009} reported a detection of extended HeII and modeled
simple, constant density gas clouds assuming illumination from an AGN,
Pop III, and Pop II stars. They are not quoting all the parameters of
their Cloudy models (e.g., $N_{\rm H}$) and thus it is not possible to make
a direct comparison.  However, they found that the data are in
agreement with photoionization from a hard ionizing source, either due
to an AGN or a very low metallicity stellar population ($Z <
10^{-2}-10^{-3} Z_{\odot}$). They conclude that, in the case of an
AGN, this source must be highly obscured along the line of sight.
They also showed that their observed ratios are inconsistent with
shock ionization in solar metallicity gas.

On smaller scales, photoionization has been modeled in the case of EELR 
of HzRGs. In particular, \citet{Humphrey2008} using the code {\it MAPPINGS Ic} (\citealt{Binette1985}),
shows that the data are best described by AGN photoionization with the ionization parameter U varying 
between objects, in a range comparable with our grid.
However, they found that a single-slab photoionization model is unable to 
explain adequately the high-ionization (e.g. \nv) and low-ionization 
(e.g. \cii, \nii, \oii) lines simultaneously, 
with higher U favored by the higher ionization lines.
They also demonstrated that shock models alone are overall worse 
than photoionization models in reproducing HzRGs data.
In the shock scenario is required an additional source of ionizing photons, i.e. the obscured AGN,
in order to match most of the line ratios studied by \citet{Humphrey2008}.      
However, note that shock with precursor models can explain some ratios, e.g. \nv/\niv ,
which are hardly explained by a single-slab photoionization model (\citealt{Humphrey2008}).

\section{Summary and Conclusions}
\label{sec:Conclusion}

We obtained the deepest ever narrowband images of \heii and \civ
emission from 13 \lya blobs in the SSA22 proto-cluster region to study
the poorly understood mechanism powering the \lya blobs.  By
exploiting the overdensity of LABs in the SSA22 field, we were able to
conduct the first statistical multi emission line analysis for a
sample of 13 LABs, and compared their emission line ratios to \lya
nebulae associated with other Ly$\alpha$ blobs, high-z radio galaxies
(HzRGs), and QSOs.  We compared these results to detailed models of \heii/\lya 
and \civ/\lya line ratios assuming that the \lya emission is powered by a) 
photoionization from an AGN (including the contribution of scattering) 
or b) in a shock scenario. The primary results of our analysis are:  

\begin{itemize}

\item We do not detect extended emission in the \heii and \civ lines in any
of the 13 LABs down to our sensitivity limits, $2.1\times$ and $3.4 \times 10^{-18}$ erg s$^{-1}$ cm$^{-2}$ arcsec$^{-2}$ (5$\sigma$ in 1 arcsec$^2$) 
for \heii and \civ, respectively.

\item Our strongest constraints on emission line ratios 
  are obtained for the brightest LABs in our field
  (LAB1 and LAB2), and are thus constrained to be lower than 0.11 and 0.16
  ($5\sigma$),  for HeII/Ly$\alpha$ and CIV/Ly$\alpha$, respectively.
  
\item Photoionization models, accompanied by a reasonable variation of
  the parameters ($N_{\rm H}$, $n_{\rm H}$, $i$) describing the gas distribution
  and the ionizing source, are able to produce line ratios smaller
  than our upper limits in the HeII/Ly$\alpha$ versus CIV/Ly$\alpha$
  diagram. Although our data constitute the deepest ever observations
  of these lines, they are still not deep enough to rule out
  photoionization by an obscured AGN as the power source in
  LABs. These same photoionization models can also accommodate the
  range of line ratios in the literature for other \lya
  nebulae. Models with a population of optically thin clouds $N_{\rm
    HI} < 10^{17.2}\,{\rm cm^{-2}}$ seem to be favored over optically
  thick models $N_{\rm HI} > 10^{17.2}\,{\rm cm^{-2}}$.  In
  particular, optically thin models with log$U\approx-1.5$ and
  $10^{19} \leq N_{\rm H} < 10^{20}$ cm$^{-2}$ populate the region below our
  LAB limits, whereas models with log$U\gtrsim-2.0$ and $10^{19} \leq
  N_{\rm H} < 10^{21}$ cm$^{-2}$ would be broadly consistent with most of
  the HeII and CIV detections in the literature.

\item Shock models can populate a HeII/Ly$\alpha$ versus
  CIV/Ly$\alpha$ diagram below our LAB limits only if high velocity
  are assumed, i.e. $v_{\rm s} \gtrsim 250$ km s$^{-1}$, but they do not
  reproduce the higher line ratios implied by detections of HeII and
  CIV in the HzRGs.  Observations of relatively weak 
  outflow kinematics in the central galaxies embedded in LABs
  appear to rule out such high shock velocities
  (\citealt{Prescott2009,Yang2011,Yang2014b}).

\end{itemize}

Deeper observations of the HeII and CIV emission lines in the SSA22
field are required in order to make more definitive statements about
the mechanism powering the LABs. For example, our photoionization
modeling suggests that line ratios as low as HeII/\lya $\simeq 0.05$ and
CIV/\lya $\simeq 0.07$ can be produced by combinations of physical
parameters ($N_{\rm H} = 10^{19} - 10^{20}$ cm$^{-2}$, $n_{\rm H} = 1 - 10$ cm$^{-3}$, $i=17$) which are still
plausible. This implies that SBs as low as $1\times$ and $1.5\times 10^{-18}$ erg s$^{-1}$ cm$^{-2}$ arcsec$^{-2}$ 
per 1 arcsec$^2$ aperture (5$\sigma$) must be
achieved to start to rule out photoionization. For bright giant \lya
nebulae around QSOs, as have been recently discovered (\citealt{Cantalupo2014}),
photoionization modeling are much more constrained, because the
ionizing luminosity of the central source is known. Sensitive
measurements of line ratios from deep observations can thus constrain
the properties of gas in the CGM, as we will discuss in a future paper
(Arrigoni-Battaia in prep.).  These questions will be addressed by a new
generation of image-slicing integral field units, such as the Multi Unit Spectroscopic Explorer 
(MUSE, \citealt{Bacon2004}) on VLT or the Keck Cosmic Web
Imager (KCWI). By probing an order of magnitude deeper than our
current observations, this new instrumentation will usher in a new era
of emission studies of the CGM. This unprecedented sensitivity 
combined with the modeling methodology described here, will constitute
an important step forward in solving the mystery of the LABs.

\acknowledgments 

We thank the members of the ENIGMA
group\footnote{http://www.mpia-hd.mpg.de/ENIGMA/} at the Max Planck
Institute for Astronomy (MPIA) for helpful discussions, in particular
Jonathan Stern. JFH acknowledges generous support from the Alexander 
von Humboldt foundation in the context of the Sofja Kovalevskaja Award. 
The Humboldt foundation is funded by the German Federal Ministry 
for Education and Research. 
YM acknowledges support from JSPS KAKENHI Grant Number 20647268.

\bibliographystyle{apj}
\bibliography{LABs_s}

\begin{thebibliography}{}
\expandafter\ifx\csname natexlab\endcsname\relax\def\natexlab#1{#1}\fi

\bibitem[{{Adelberger} {et~al.}(2005){Adelberger}, {Shapley}, {Steidel},
  {Pettini}, {Erb}, \& {Reddy}}]{Adelberger2005}
{Adelberger}, K.~L., {Shapley}, A.~E., {Steidel}, C.~C., {et~al.} 2005, \apj,
  629, 636

\bibitem[{{Allen} {et~al.}(1998){Allen}, {Dopita}, \& {Tsvetanov}}]{Allen1998}
{Allen}, M.~G., {Dopita}, M.~A., \& {Tsvetanov}, Z.~I. 1998, \apj, 493, 571

\bibitem[{{Allen} {et~al.}(2008){Allen}, {Groves}, {Dopita}, {Sutherland}, \&
  {Kewley}}]{Allen2008}
{Allen}, M.~G., {Groves}, B.~A., {Dopita}, M.~A., {Sutherland}, R.~S., \&
  {Kewley}, L.~J. 2008, \apjs, 178, 20

\bibitem[{{Antonucci}(1993)}]{Antonucci1993}
{Antonucci}, R. 1993, \araa, 31, 473

\bibitem[{{Arrigoni Battaia} {et~al.}(2014){Arrigoni Battaia}, {Hennawi},
  {Cantalupo}, \& {Prochaska}}]{FABProceeding}
{Arrigoni Battaia}, F., {Hennawi}, J.~F., {Cantalupo}, S., \& {Prochaska},
  J.~X. 2014, Proc. IAU Symp. \#304: Multiwavelength AGN Surveys and Studies,
  Cambridge Univ. Press

\bibitem[{{Bacon} {et~al.}(2004){Bacon}, {Bauer}, {Bower}, {Cabrit},
  {Cappellari}, {Carollo}, {Combes}, {Davies}, {Delabre}, {Dekker},
  {Devriendt}, {Djidel}, {Duchateau}, {Dubois}, {Emsellem}, {Ferruit}, {Franx},
  {Gilmore}, {Guiderdoni}, {Henault}, {Hubin}, {Jungwiert}, {Kelz}, {Le
  Louarn}, {Lewis}, {Lizon}, {McDermid}, {Morris}, {Laux}, {Le F{\`e}vre},
  {Lantz}, {Lilly}, {Lynn}, {Pasquini}, {Pecontal}, {Pinet}, {Popovic},
  {Quirrenbach}, {Reiss}, {Roth}, {Steinmetz}, {Stuik}, {Wisotzki}, \& {de
  Zeeuw}}]{Bacon2004}
{Bacon}, R., {Bauer}, S.-M., {Bower}, R., {et~al.} 2004, in Society of
  Photo-Optical Instrumentation Engineers (SPIE) Conference Series, Vol. 5492,
  Ground-based Instrumentation for Astronomy, ed. A.~F.~M. {Moorwood} \&
  M.~{Iye}, 1145--1149

\bibitem[{{Barrio} {et~al.}(2008){Barrio}, {Jarvis}, {Rawlings}, {Bauer},
  {Croft}, {Hill}, {Manchado}, {McLure}, {Smith}, \& {Targett}}]{Barrio2008}
{Barrio}, F.~E., {Jarvis}, M.~J., {Rawlings}, S., {et~al.} 2008, \mnras, 389,
  792

\bibitem[{{Basu-Zych} \& {Scharf}(2004)}]{Basu2004}
{Basu-Zych}, A., \& {Scharf}, C. 2004, \apjl, 615, L85

\bibitem[{{Bertin}(2006)}]{Bertin2006}
{Bertin}, E. 2006, in Astronomical Society of the Pacific Conference Series,
  Vol. 351, Astronomical Data Analysis Software and Systems XV, ed.
  C.~{Gabriel}, C.~{Arviset}, D.~{Ponz}, \& S.~{Enrique}, 112

\bibitem[{{Bertin} {et~al.}(2002){Bertin}, {Mellier}, {Radovich}, {Missonnier},
  {Didelon}, \& {Morin}}]{Bertin2002}
{Bertin}, E., {Mellier}, Y., {Radovich}, M., {et~al.} 2002, in Astronomical
  Society of the Pacific Conference Series, Vol. 281, Astronomical Data
  Analysis Software and Systems XI, ed. D.~A. {Bohlender}, D.~{Durand}, \&
  T.~H. {Handley}, 228

\bibitem[{{Binette} {et~al.}(1985){Binette}, {Dopita}, \&
  {Tuohy}}]{Binette1985}
{Binette}, L., {Dopita}, M.~A., \& {Tuohy}, I.~R. 1985, \apj, 297, 476

\bibitem[{{Binette} {et~al.}(1996){Binette}, {Wilson}, \&
  {Storchi-Bergmann}}]{Binette1996}
{Binette}, L., {Wilson}, A.~S., \& {Storchi-Bergmann}, T. 1996, \aap, 312, 365

\bibitem[{{Bowen} {et~al.}(2006){Bowen}, {Hennawi}, {M{\'e}nard}, {Chelouche},
  {Inada}, {Oguri}, {Richards}, {Strauss}, {Vanden Berk}, \&
  {York}}]{Bowen2006}
{Bowen}, D.~V., {Hennawi}, J.~F., {M{\'e}nard}, B., {et~al.} 2006, \apjl, 645,
  L105

\bibitem[{{Cantalupo} {et~al.}(2014){Cantalupo}, {Arrigoni-Battaia},
  {Prochaska}, {Hennawi}, \& {Madau}}]{Cantalupo2014}
{Cantalupo}, S., {Arrigoni-Battaia}, F., {Prochaska}, J.~X., {Hennawi}, J.~F.,
  \& {Madau}, P. 2014, \nat, 506, 63

\bibitem[{{Cen} \& {Zheng}(2013)}]{Cen2013}
{Cen}, R., \& {Zheng}, Z. 2013, \apj, 775, 112

\bibitem[{{Christensen} {et~al.}(2006){Christensen}, {Jahnke}, {Wisotzki}, \&
  {S{\'a}nchez}}]{Christensen2006}
{Christensen}, L., {Jahnke}, K., {Wisotzki}, L., \& {S{\'a}nchez}, S.~F. 2006,
  \aap, 459, 717

\bibitem[{{Crighton} {et~al.}(2013){Crighton}, {Hennawi}, \&
  {Prochaska}}]{Crighton2013}
{Crighton}, N.~H.~M., {Hennawi}, J.~F., \& {Prochaska}, J.~X. 2013, \apjl, 776,
  L18

\bibitem[{{Crighton} {et~al.}(2014){Crighton}, {Hennawi}, {Simcoe}, {Cooksey},
  {Murphy}, {Fumagalli}, {Prochaska}, \& {Shanks}}]{Crighton2014}
{Crighton}, N.~H.~M., {Hennawi}, J.~F., {Simcoe}, R.~A., {et~al.} 2014, ArXiv
  e-prints, arXiv:1406.4239

\bibitem[{{Crighton} {et~al.}(2011){Crighton}, {Bielby}, {Shanks}, {Infante},
  {Bornancini}, {Bouch{\'e}}, {Lambas}, {Lowenthal}, {Minniti}, {Morris},
  {Padilla}, {P{\'e}roux}, {Petitjean}, {Theuns}, {Tummuangpak}, {Weilbacher},
  {Wisotzki}, \& {Worseck}}]{Crighton2011}
{Crighton}, N.~H.~M., {Bielby}, R., {Shanks}, T., {et~al.} 2011, \mnras, 414,
  28

\bibitem[{{De Breuck} {et~al.}(2000){De Breuck}, {R{\"o}ttgering}, {Miley},
  {van Breugel}, \& {Best}}]{DeBreuck2000}
{De Breuck}, C., {R{\"o}ttgering}, H., {Miley}, G., {van Breugel}, W., \&
  {Best}, P. 2000, \aap, 362, 519

\bibitem[{{Dekel} {et~al.}(2009){Dekel}, {Birnboim}, {Engel}, {Freundlich},
  {Goerdt}, {Mumcuoglu}, {Neistein}, {Pichon}, {Teyssier}, \&
  {Zinger}}]{Dekel2009}
{Dekel}, A., {Birnboim}, Y., {Engel}, G., {et~al.} 2009, \nat, 457, 451

\bibitem[{{Dey} {et~al.}(1995){Dey}, {Spinrad}, \& {Dickinson}}]{Dey1995}
{Dey}, A., {Spinrad}, H., \& {Dickinson}, M. 1995, \apj, 440, 515

\bibitem[{{Dey} {et~al.}(2005){Dey}, {Bian}, {Soifer}, {Brand}, {Brown},
  {Chaffee}, {Le Floc'h}, {Hill}, {Houck}, {Jannuzi}, {Rieke}, {Weedman},
  {Brodwin}, \& {Eisenhardt}}]{Dey2005}
{Dey}, A., {Bian}, C., {Soifer}, B.~T., {et~al.} 2005, \apj, 629, 654

\bibitem[{{Dijkstra} {et~al.}(2006){Dijkstra}, {Haiman}, \&
  {Spaans}}]{Dijkstra06}
{Dijkstra}, M., {Haiman}, Z., \& {Spaans}, M. 2006, \apj, 649, 14

\bibitem[{{Dijkstra} \& {Loeb}(2008)}]{Dijkstra2008}
{Dijkstra}, M., \& {Loeb}, A. 2008, \mnras, 386, 492

\bibitem[{{Dijkstra} \& {Loeb}(2009)}]{Dijkstra&Loeb2009}
---. 2009, \mnras, 400, 1109

\bibitem[{{Dopita} {et~al.}(2002){Dopita}, {Groves}, {Sutherland}, {Binette},
  \& {Cecil}}]{Dopita2002}
{Dopita}, M.~A., {Groves}, B.~A., {Sutherland}, R.~S., {Binette}, L., \&
  {Cecil}, G. 2002, \apj, 572, 753

\bibitem[{{Elvis}(2000)}]{Elvis2000}
{Elvis}, M. 2000, \apj, 545, 63

\bibitem[{{Erb} {et~al.}(2010){Erb}, {Pettini}, {Shapley}, {Steidel}, {Law}, \&
  {Reddy}}]{Erb2010}
{Erb}, D.~K., {Pettini}, M., {Shapley}, A.~E., {et~al.} 2010, \apj, 719, 1168

\bibitem[{{Fabian}(1999)}]{Fabian1999}
{Fabian}, A.~C. 1999, \mnras, 308, L39

\bibitem[{{Fardal} {et~al.}(2001){Fardal}, {Katz}, {Gardner}, {Hernquist},
  {Weinberg}, \& {Dav{\'e}}}]{Fardal2001}
{Fardal}, M.~A., {Katz}, N., {Gardner}, J.~P., {et~al.} 2001, \apj, 562, 605

\bibitem[{{Farina} {et~al.}(2013){Farina}, {Falomo}, {Decarli}, {Treves}, \&
  {Kotilainen}}]{Farina2013}
{Farina}, E.~P., {Falomo}, R., {Decarli}, R., {Treves}, A., \& {Kotilainen},
  J.~K. 2013, \mnras, 429, 1267

\bibitem[{{Faucher-Gigu{\`e}re} {et~al.}(2010){Faucher-Gigu{\`e}re}, {Kere{\v
  s}}, {Dijkstra}, {Hernquist}, \& {Zaldarriaga}}]{Faucher2010}
{Faucher-Gigu{\`e}re}, C.-A., {Kere{\v s}}, D., {Dijkstra}, M., {Hernquist},
  L., \& {Zaldarriaga}, M. 2010, \apj, 725, 633

\bibitem[{{Ferland}(2003)}]{FerlandARAAarticle2003}
{Ferland}, G.~J. 2003, \araa, 41, 517

\bibitem[{{Ferland} {et~al.}(1984){Ferland}, {Williams}, {Lambert}, {Slovak},
  {Gondhalekar}, {Truran}, \& {Shields}}]{Ferland1984}
{Ferland}, G.~J., {Williams}, R.~E., {Lambert}, D.~L., {et~al.} 1984, \apj,
  281, 194

\bibitem[{{Ferland} {et~al.}(2013){Ferland}, {Porter}, {van Hoof}, {Williams},
  {Abel}, {Lykins}, {Shaw}, {Henney}, \& {Stancil}}]{Ferland2013}
{Ferland}, G.~J., {Porter}, R.~L., {van Hoof}, P.~A.~M., {et~al.} 2013, \rmxaa,
  49, 137

\bibitem[{{Francis} {et~al.}(2001){Francis}, {Williger}, {Collins}, {Palunas},
  {Malumuth}, {Woodgate}, {Teplitz}, {Smette}, {Sutherland}, {Danks}, {Hill},
  {Lindler}, {Kimble}, {Heap}, \& {Hutchings}}]{Francis2001}
{Francis}, P.~J., {Williger}, G.~M., {Collins}, N.~R., {et~al.} 2001, \apj,
  554, 1001

\bibitem[{{Fumagalli} {et~al.}(2014){Fumagalli}, {Hennawi}, {Prochaska},
  {Kasen}, {Dekel}, {Ceverino}, \& {Primack}}]{Fumagalli2014}
{Fumagalli}, M., {Hennawi}, J.~F., {Prochaska}, J.~X., {et~al.} 2014, \apj,
  780, 74

\bibitem[{{Furlanetto} {et~al.}(2005){Furlanetto}, {Schaye}, {Springel}, \&
  {Hernquist}}]{Furlanetto05}
{Furlanetto}, S.~R., {Schaye}, J., {Springel}, V., \& {Hernquist}, L. 2005,
  \apj, 622, 7

\bibitem[{{Geach} {et~al.}(2007){Geach}, {Smail}, {Chapman}, {Alexander},
  {Blain}, {Stott}, \& {Ivison}}]{Geach2007}
{Geach}, J.~E., {Smail}, I., {Chapman}, S.~C., {et~al.} 2007, \apjl, 655, L9

\bibitem[{{Geach} {et~al.}(2009){Geach}, {Alexander}, {Lehmer}, {Smail},
  {Matsuda}, {Chapman}, {Scharf}, {Ivison}, {Volonteri}, {Yamada}, {Blain},
  {Bower}, {Bauer}, \& {Basu-Zych}}]{Geach2009}
{Geach}, J.~E., {Alexander}, D.~M., {Lehmer}, B.~D., {et~al.} 2009, \apj, 700,
  1

\bibitem[{{Goerdt} {et~al.}(2010){Goerdt}, {Dekel}, {Sternberg}, {Ceverino},
  {Teyssier}, \& {Primack}}]{Goerdt2010}
{Goerdt}, T., {Dekel}, A., {Sternberg}, A., {et~al.} 2010, \mnras, 407, 613

\bibitem[{{Groves} {et~al.}(2004){Groves}, {Dopita}, \&
  {Sutherland}}]{Groves2004}
{Groves}, B.~A., {Dopita}, M.~A., \& {Sutherland}, R.~S. 2004, \apjs, 153, 75

\bibitem[{{Haiman} \& {Rees}(2001)}]{HaimanRees2001}
{Haiman}, Z., \& {Rees}, M.~J. 2001, \apj, 556, 87

\bibitem[{{Haiman} {et~al.}(2000){Haiman}, {Spaans}, \&
  {Quataert}}]{Haiman2000}
{Haiman}, Z., {Spaans}, M., \& {Quataert}, E. 2000, \apjl, 537, L5

\bibitem[{{Hayashino} {et~al.}(2004){Hayashino}, {Matsuda}, {Tamura},
  {Yamauchi}, {Yamada}, {Ajiki}, {Fujita}, {Murayama}, {Nagao}, {Ohta},
  {Okamura}, {Ouchi}, {Shimasaku}, {Shioya}, \& {Taniguchi}}]{Hayashino2004}
{Hayashino}, T., {Matsuda}, Y., {Tamura}, H., {et~al.} 2004, \aj, 128, 2073

\bibitem[{{Hayes} {et~al.}(2011){Hayes}, {Scarlata}, \& {Siana}}]{Hayes2011}
{Hayes}, M., {Scarlata}, C., \& {Siana}, B. 2011, \nat, 476, 304

\bibitem[{{Heckman} {et~al.}(1991{\natexlab{a}}){Heckman}, {Lehnert}, {Miley},
  \& {van Breugel}}]{Heckman1991spec}
{Heckman}, T.~M., {Lehnert}, M.~D., {Miley}, G.~K., \& {van Breugel}, W.
  1991{\natexlab{a}}, \apj, 381, 373

\bibitem[{{Heckman} {et~al.}(1991{\natexlab{b}}){Heckman}, {Miley}, {Lehnert},
  \& {van Breugel}}]{Heckman1991}
{Heckman}, T.~M., {Miley}, G.~K., {Lehnert}, M.~D., \& {van Breugel}, W.
  1991{\natexlab{b}}, \apj, 370, 78

\bibitem[{{Hennawi} \& {Prochaska}(2007)}]{Hennawi2007}
{Hennawi}, J.~F., \& {Prochaska}, J.~X. 2007, \apj, 655, 735

\bibitem[{{Hennawi} \& {Prochaska}(2013)}]{Hennawi2013}
---. 2013, \apj, 766, 58

\bibitem[{{Hennawi} {et~al.}(2006){Hennawi}, {Prochaska}, {Burles}, {Strauss},
  {Richards}, {Schlegel}, {Fan}, {Schneider}, {Zakamska}, {Oguri}, {Gunn},
  {Lupton}, \& {Brinkmann}}]{Hennawi2006}
{Hennawi}, J.~F., {Prochaska}, J.~X., {Burles}, S., {et~al.} 2006, \apj, 651,
  61

\bibitem[{{Hopkins} {et~al.}(2007){Hopkins}, {Richards}, \&
  {Hernquist}}]{HRH2007}
{Hopkins}, P.~F., {Richards}, G.~T., \& {Hernquist}, L. 2007, \apj, 654, 731

\bibitem[{{Hu} \& {Cowie}(1987)}]{HuCowie1987}
{Hu}, E.~M., \& {Cowie}, L.~L. 1987, \apjl, 317, L7

\bibitem[{{Humphrey} {et~al.}(2013){Humphrey}, {Binette},
  {Villar-Mart{\'{\i}}n}, {Aretxaga}, \& {Papaderos}}]{Humphrey2013}
{Humphrey}, A., {Binette}, L., {Villar-Mart{\'{\i}}n}, M., {Aretxaga}, I., \&
  {Papaderos}, P. 2013, \mnras, 428, 563

\bibitem[{{Humphrey} {et~al.}(2006){Humphrey}, {Villar-Mart{\'{\i}}n},
  {Fosbury}, {Vernet}, \& {di Serego Alighieri}}]{Humphrey2006}
{Humphrey}, A., {Villar-Mart{\'{\i}}n}, M., {Fosbury}, R., {Vernet}, J., \& {di
  Serego Alighieri}, S. 2006, \mnras, 369, 1103

\bibitem[{{Humphrey} {et~al.}(2008){Humphrey}, {Villar-Mart{\'{\i}}n},
  {Vernet}, {Fosbury}, {di Serego Alighieri}, \& {Binette}}]{Humphrey2008}
{Humphrey}, A., {Villar-Mart{\'{\i}}n}, M., {Vernet}, J., {et~al.} 2008,
  \mnras, 383, 11

\bibitem[{{Keel} {et~al.}(1999){Keel}, {Cohen}, {Windhorst}, \&
  {Waddington}}]{Keel1999}
{Keel}, W.~C., {Cohen}, S.~H., {Windhorst}, R.~A., \& {Waddington}, I. 1999,
  \aj, 118, 2547

\bibitem[{{King}(2003)}]{King2003}
{King}, A. 2003, \apjl, 596, L27

\bibitem[{{Matsuda} {et~al.}(2006){Matsuda}, {Yamada}, {Hayashino}, {Yamauchi},
  \& {Nakamura}}]{Matsuda2006}
{Matsuda}, Y., {Yamada}, T., {Hayashino}, T., {Yamauchi}, R., \& {Nakamura}, Y.
  2006, \apjl, 640, L123

\bibitem[{{Matsuda} {et~al.}(2004){Matsuda}, {Yamada}, {Hayashino}, {Tamura},
  {Yamauchi}, {Ajiki}, {Fujita}, {Murayama}, {Nagao}, {Ohta}, {Okamura},
  {Ouchi}, {Shimasaku}, {Shioya}, \& {Taniguchi}}]{Matsuda2004}
{Matsuda}, Y., {Yamada}, T., {Hayashino}, T., {et~al.} 2004, \aj, 128, 569

\bibitem[{{Matsuda} {et~al.}(2011){Matsuda}, {Yamada}, {Hayashino}, {Yamauchi},
  {Nakamura}, {Morimoto}, {Ouchi}, {Ono}, {Kousai}, {Nakamura}, {Horie},
  {Fujii}, {Umemura}, \& {Mori}}]{Matsuda2011}
---. 2011, \mnras, 410, L13

\bibitem[{{Matsuoka} {et~al.}(2009){Matsuoka}, {Nagao}, {Maiolino}, {Marconi},
  \& {Taniguchi}}]{Matsuoka2009}
{Matsuoka}, K., {Nagao}, T., {Maiolino}, R., {Marconi}, A., \& {Taniguchi}, Y.
  2009, \aap, 503, 721

\bibitem[{{McCarthy}(1993)}]{McCarthy1993}
{McCarthy}, P.~J. 1993, \araa, 31, 639

\bibitem[{{McLinden} {et~al.}(2013){McLinden}, {Malhotra}, {Rhoads}, {Hibon},
  {Weijmans}, \& {Tilvi}}]{McLinden2013}
{McLinden}, E.~M., {Malhotra}, S., {Rhoads}, J.~E., {et~al.} 2013, \apj, 767,
  48

\bibitem[{{Miley} \& {De Breuck}(2008)}]{Miley2008}
{Miley}, G., \& {De Breuck}, C. 2008, \aapr, 15, 67

\bibitem[{{Miyazaki} {et~al.}(2002){Miyazaki}, {Komiyama}, {Sekiguchi},
  {Okamura}, {Doi}, {Furusawa}, {Hamabe}, {Imi}, {Kimura}, {Nakata}, {Okada},
  {Ouchi}, {Shimasaku}, {Yagi}, \& {Yasuda}}]{Miyazaki2002}
{Miyazaki}, S., {Komiyama}, Y., {Sekiguchi}, M., {et~al.} 2002, \pasj, 54, 833

\bibitem[{{Mori} \& {Umemura}(2006)}]{Mori2006}
{Mori}, M., \& {Umemura}, M. 2006, \nar, 50, 199

\bibitem[{{Nagao} {et~al.}(2006){Nagao}, {Maiolino}, \& {Marconi}}]{Nagao2006}
{Nagao}, T., {Maiolino}, R., \& {Marconi}, A. 2006, \aap, 459, 85

\bibitem[{{Nesvadba} {et~al.}(2006){Nesvadba}, {Lehnert}, {Eisenhauer},
  {Gilbert}, {Tecza}, \& {Abuter}}]{Nesvadba2006}
{Nesvadba}, N.~P.~H., {Lehnert}, M.~D., {Eisenhauer}, F., {et~al.} 2006, \apj,
  650, 693

\bibitem[{{Nilsson} {et~al.}(2006){Nilsson}, {Fynbo}, {M{\o}ller},
  {Sommer-Larsen}, \& {Ledoux}}]{Nilsson2006}
{Nilsson}, K.~K., {Fynbo}, J.~P.~U., {M{\o}ller}, P., {Sommer-Larsen}, J., \&
  {Ledoux}, C. 2006, \aap, 452, L23

\bibitem[{{North} {et~al.}(2012){North}, {Courbin}, {Eigenbrod}, \&
  {Chelouche}}]{North2012}
{North}, P.~L., {Courbin}, F., {Eigenbrod}, A., \& {Chelouche}, D. 2012, \aap,
  542, A91

\bibitem[{{Ohyama} {et~al.}(2003){Ohyama}, {Taniguchi}, {Kawabata}, {Shioya},
  {Murayama}, {Nagao}, {Takata}, {Iye}, \& {Yoshida}}]{Ohyama2003}
{Ohyama}, Y., {Taniguchi}, Y., {Kawabata}, K.~S., {et~al.} 2003, \apjl, 591, L9

\bibitem[{{Oke}(1974)}]{Oke1974}
{Oke}, J.~B. 1974, \apjs, 27, 21

\bibitem[{{Ouchi} {et~al.}(2009){Ouchi}, {Ono}, {Egami}, {Saito}, {Oguri},
  {McCarthy}, {Farrah}, {Kashikawa}, {Momcheva}, {Shimasaku}, {Nakanishi},
  {Furusawa}, {Akiyama}, {Dunlop}, {Mortier}, {Okamura}, {Hayashi},
  {Cirasuolo}, {Dressler}, {Iye}, {Jarvis}, {Kodama}, {Martin}, {McLure},
  {Ohta}, {Yamada}, \& {Yoshida}}]{Ouchi2009}
{Ouchi}, M., {Ono}, Y., {Egami}, E., {et~al.} 2009, \apj, 696, 1164

\bibitem[{{Overzier} {et~al.}(2013){Overzier}, {Nesvadba}, {Dijkstra}, {Hatch},
  {Lehnert}, {Villar-Mart{\'{\i}}n}, {Wilman}, \& {Zirm}}]{Overzier2013}
{Overzier}, R.~A., {Nesvadba}, N.~P.~H., {Dijkstra}, M., {et~al.} 2013, \apj,
  771, 89

\bibitem[{{Prescott} {et~al.}(2009){Prescott}, {Dey}, \&
  {Jannuzi}}]{Prescott2009}
{Prescott}, M.~K.~M., {Dey}, A., \& {Jannuzi}, B.~T. 2009, \apj, 702, 554

\bibitem[{{Prescott} {et~al.}(2012){Prescott}, {Dey}, \&
  {Jannuzi}}]{Prescott2012a}
---. 2012, \apj, 748, 125

\bibitem[{{Prescott} {et~al.}(2013){Prescott}, {Dey}, \&
  {Jannuzi}}]{Prescott2013}
---. 2013, \apj, 762, 38

\bibitem[{{Prochaska} {et~al.}(2014){Prochaska}, {Lau}, \&
  {Hennawi}}]{Prochaska2014sub}
{Prochaska}, J., {Lau}, M., \& {Hennawi}, J. 2014, \mnras

\bibitem[{{Prochaska} \& {Hennawi}(2009)}]{Prochaska2009}
{Prochaska}, J.~X., \& {Hennawi}, J.~F. 2009, \apj, 690, 1558

\bibitem[{{Prochaska} {et~al.}(2013{\natexlab{a}}){Prochaska}, {Hennawi}, \&
  {Simcoe}}]{Prochaska2013}
{Prochaska}, J.~X., {Hennawi}, J.~F., \& {Simcoe}, R.~A. 2013{\natexlab{a}},
  \apjl, 762, L19

\bibitem[{{Prochaska} {et~al.}(2013{\natexlab{b}}){Prochaska}, {Hennawi},
  {Lee}, {Cantalupo}, {Bovy}, {Djorgovski}, {Ellison}, {Lau}, {Martin},
  {Myers}, {Rubin}, \& {Simcoe}}]{Prochaska2013b}
{Prochaska}, J.~X., {Hennawi}, J.~F., {Lee}, K.-G., {et~al.}
  2013{\natexlab{b}}, \apj, 776, 136

\bibitem[{{Rakic} {et~al.}(2012){Rakic}, {Schaye}, {Steidel}, \&
  {Rudie}}]{Rakic2012}
{Rakic}, O., {Schaye}, J., {Steidel}, C.~C., \& {Rudie}, G.~C. 2012, \apj, 751,
  94

\bibitem[{{Rees}(1988)}]{Rees1988}
{Rees}, M.~J. 1988, \mnras, 231, 91P

\bibitem[{{Reuland} {et~al.}(2003){Reuland}, {van Breugel}, {R{\"o}ttgering},
  {de Vries}, {Stanford}, {Dey}, {Lacy}, {Bland-Hawthorn}, {Dopita}, \&
  {Miley}}]{Reuland2003}
{Reuland}, M., {van Breugel}, W., {R{\"o}ttgering}, H., {et~al.} 2003, \apj,
  592, 755

\bibitem[{{Reuland} {et~al.}(2007){Reuland}, {van Breugel}, {de Vries},
  {Dopita}, {Dey}, {Miley}, {R{\"o}ttgering}, {Venemans}, {Stanford}, {Lacy},
  {Spinrad}, {Dawson}, {Stern}, \& {Bunker}}]{Reuland2007}
{Reuland}, M., {van Breugel}, W., {de Vries}, W., {et~al.} 2007, \aj, 133, 2607

\bibitem[{{Rosdahl} \& {Blaizot}(2012)}]{Rosdahl12}
{Rosdahl}, J., \& {Blaizot}, J. 2012, \mnras, 423, 344

\bibitem[{{Rudie} {et~al.}(2012){Rudie}, {Steidel}, {Trainor}, {Rakic},
  {Bogosavljevi{\'c}}, {Pettini}, {Reddy}, {Shapley}, {Erb}, \&
  {Law}}]{Rudie2012}
{Rudie}, G.~C., {Steidel}, C.~C., {Trainor}, R.~F., {et~al.} 2012, \apj, 750,
  67

\bibitem[{{Saito} {et~al.}(2006){Saito}, {Shimasaku}, {Okamura}, {Ouchi},
  {Akiyama}, \& {Yoshida}}]{Saito2006}
{Saito}, T., {Shimasaku}, K., {Okamura}, S., {et~al.} 2006, \apj, 648, 54

\bibitem[{{Shapley} {et~al.}(2003){Shapley}, {Steidel}, {Pettini}, \&
  {Adelberger}}]{Shapley2003}
{Shapley}, A.~E., {Steidel}, C.~C., {Pettini}, M., \& {Adelberger}, K.~L. 2003,
  \apj, 588, 65

\bibitem[{{Silk} \& {Rees}(1998)}]{SilkRees1998}
{Silk}, J., \& {Rees}, M.~J. 1998, \aap, 331, L1

\bibitem[{{Smith} \& {Jarvis}(2007)}]{Smith2007}
{Smith}, D.~J.~B., \& {Jarvis}, M.~J. 2007, \mnras, 378, L49

\bibitem[{{Smith} {et~al.}(2009){Smith}, {Jarvis}, {Simpson}, \&
  {Mart{\'{\i}}nez-Sansigre}}]{Smith2009}
{Smith}, D.~J.~B., {Jarvis}, M.~J., {Simpson}, C., \&
  {Mart{\'{\i}}nez-Sansigre}, A. 2009, \mnras, 393, 309

\bibitem[{{Steidel} {et~al.}(2000){Steidel}, {Adelberger}, {Shapley},
  {Pettini}, {Dickinson}, \& {Giavalisco}}]{Steidel2000}
{Steidel}, C.~C., {Adelberger}, K.~L., {Shapley}, A.~E., {et~al.} 2000, \apj,
  532, 170

\bibitem[{{Steidel} {et~al.}(2011){Steidel}, {Bogosavljevi{\'c}}, {Shapley},
  {Kollmeier}, {Reddy}, {Erb}, \& {Pettini}}]{Steidel2011}
{Steidel}, C.~C., {Bogosavljevi{\'c}}, M., {Shapley}, A.~E., {et~al.} 2011,
  \apj, 736, 160

\bibitem[{{Steidel} {et~al.}(2010){Steidel}, {Erb}, {Shapley}, {Pettini},
  {Reddy}, {Bogosavljevi{\'c}}, {Rudie}, \& {Rakic}}]{Steidel2010}
{Steidel}, C.~C., {Erb}, D.~K., {Shapley}, A.~E., {et~al.} 2010, \apj, 717, 289

\bibitem[{{Stern} {et~al.}(2014){Stern}, {Laor}, \& {Baskin}}]{Stern2014}
{Stern}, J., {Laor}, A., \& {Baskin}, A. 2014, \mnras, 438, 901

\bibitem[{{Taniguchi} \& {Shioya}(2000)}]{Taniguchi&Shioya2000}
{Taniguchi}, Y., \& {Shioya}, Y. 2000, \apjl, 532, L13

\bibitem[{{Taniguchi} {et~al.}(2001){Taniguchi}, {Shioya}, \&
  {Kakazu}}]{Taniguchi2001}
{Taniguchi}, Y., {Shioya}, Y., \& {Kakazu}, Y. 2001, \apjl, 562, L15

\bibitem[{{Telfer} {et~al.}(2002){Telfer}, {Zheng}, {Kriss}, \&
  {Davidsen}}]{Telfer2002}
{Telfer}, R.~C., {Zheng}, W., {Kriss}, G.~A., \& {Davidsen}, A.~F. 2002, \apj,
  565, 773

\bibitem[{{Urry} \& {Padovani}(1995)}]{UP1995}
{Urry}, C.~M., \& {Padovani}, P. 1995, \pasp, 107, 803

\bibitem[{{van Dokkum}(2001)}]{vanDokkum2001}
{van Dokkum}, P.~G. 2001, \pasp, 113, 1420

\bibitem[{{van Ojik} {et~al.}(1997){van Ojik}, {Roettgering}, {Miley}, \&
  {Hunstead}}]{vanOjik1997}
{van Ojik}, R., {Roettgering}, H.~J.~A., {Miley}, G.~K., \& {Hunstead}, R.~W.
  1997, \aap, 317, 358

\bibitem[{{van Ojik} {et~al.}(1994){van Ojik}, {Rottgering}, {Miley}, {Bremer},
  {Macchetto}, \& {Chambers}}]{vanOjik1994}
{van Ojik}, R., {Rottgering}, H.~J.~A., {Miley}, G.~K., {et~al.} 1994, \aap,
  289, 54

\bibitem[{{Vanden Berk} {et~al.}(2001){Vanden Berk}, {Richards}, {Bauer},
  {Strauss}, {Schneider}, {Heckman}, {York}, {Hall}, {Fan}, {Knapp},
  {Anderson}, {Annis}, {Bahcall}, {Bernardi}, {Briggs}, {Brinkmann}, {Brunner},
  {Burles}, {Carey}, {Castander}, {Connolly}, {Crocker}, {Csabai}, {Doi},
  {Finkbeiner}, {Friedman}, {Frieman}, {Fukugita}, {Gunn}, {Hennessy},
  {Ivezi{\'c}}, {Kent}, {Kunszt}, {Lamb}, {Leger}, {Long}, {Loveday}, {Lupton},
  {Meiksin}, {Merelli}, {Munn}, {Newberg}, {Newcomb}, {Nichol}, {Owen}, {Pier},
  {Pope}, {Rockosi}, {Schlegel}, {Siegmund}, {Smee}, {Snir}, {Stoughton},
  {Stubbs}, {SubbaRao}, {Szalay}, {Szokoly}, {Tremonti}, {Uomoto}, {Waddell},
  {Yanny}, \& {Zheng}}]{VandenBerk2001}
{Vanden Berk}, D.~E., {Richards}, G.~T., {Bauer}, A., {et~al.} 2001, \aj, 122,
  549

\bibitem[{{Venemans} {et~al.}(2007){Venemans}, {R{\"o}ttgering}, {Miley}, {van
  Breugel}, {de Breuck}, {Kurk}, {Pentericci}, {Stanford}, {Overzier}, {Croft},
  \& {Ford}}]{Venemans2007}
{Venemans}, B.~P., {R{\"o}ttgering}, H.~J.~A., {Miley}, G.~K., {et~al.} 2007,
  \aap, 461, 823

\bibitem[{{Villar-Mart{\'{\i}}n} {et~al.}(2007){Villar-Mart{\'{\i}}n},
  {Humphrey}, {De Breuck}, {Fosbury}, {Binette}, \& {Vernet}}]{VillarM2007}
{Villar-Mart{\'{\i}}n}, M., {Humphrey}, A., {De Breuck}, C., {et~al.} 2007,
  \mnras, 375, 1299

\bibitem[{{Villar-Mart{\'{\i}}n}
  {et~al.}(2003{\natexlab{a}}){Villar-Mart{\'{\i}}n}, {Vernet}, {di Serego
  Alighieri}, {Fosbury}, {Humphrey}, \& {Pentericci}}]{VillarM2003b}
{Villar-Mart{\'{\i}}n}, M., {Vernet}, J., {di Serego Alighieri}, S., {et~al.}
  2003{\natexlab{a}}, \mnras, 346, 273

\bibitem[{{Villar-Mart{\'{\i}}n}
  {et~al.}(2003{\natexlab{b}}){Villar-Mart{\'{\i}}n}, {Vernet}, {di Serego
  Alighieri}, {Fosbury}, {Humphrey}, {Pentericci}, \& {Cohen}}]{VillarM2003}
---. 2003{\natexlab{b}}, \nar, 47, 291

\bibitem[{{White} {et~al.}(2012){White}, {Myers}, {Ross}, {Schlegel},
  {Hennawi}, {Shen}, {McGreer}, {Strauss}, {Bolton}, {Bovy}, {Fan},
  {Miralda-Escude}, {Palanque-Delabrouille}, {Paris}, {Petitjean}, {Schneider},
  {Viel}, {Weinberg}, {Yeche}, {Zehavi}, {Pan}, {Snedden}, {Bizyaev},
  {Brewington}, {Brinkmann}, {Malanushenko}, {Malanushenko}, {Oravetz},
  {Simmons}, {Sheldon}, \& {Weaver}}]{White2012}
{White}, M., {Myers}, A.~D., {Ross}, N.~P., {et~al.} 2012, \mnras, 424, 933

\bibitem[{{Wilman} {et~al.}(2005){Wilman}, {Gerssen}, {Bower}, {Morris},
  {Bacon}, {de Zeeuw}, \& {Davies}}]{Wilman2005}
{Wilman}, R.~J., {Gerssen}, J., {Bower}, R.~G., {et~al.} 2005, \nat, 436, 227

\bibitem[{{Yang} {et~al.}(2014{\natexlab{a}}){Yang}, {Walter}, {Decarli},
  {Bertoldi}, {Weiss}, {Dey}, {Prescott}, \& {B{\u a}descu}}]{Yang2014a}
{Yang}, Y., {Walter}, F., {Decarli}, R., {et~al.} 2014{\natexlab{a}}, \apj,
  784, 171

\bibitem[{{Yang} {et~al.}(2010){Yang}, {Zabludoff}, {Eisenstein}, \&
  {Dav{\'e}}}]{Yang2010}
{Yang}, Y., {Zabludoff}, A., {Eisenstein}, D., \& {Dav{\'e}}, R. 2010, \apj,
  719, 1654

\bibitem[{{Yang} {et~al.}(2011){Yang}, {Zabludoff}, {Jahnke}, {Eisenstein},
  {Dav{\'e}}, {Shectman}, \& {Kelson}}]{Yang2011}
{Yang}, Y., {Zabludoff}, A., {Jahnke}, K., {et~al.} 2011, \apj, 735, 87

\bibitem[{{Yang} {et~al.}(2009){Yang}, {Zabludoff}, {Tremonti}, {Eisenstein},
  \& {Dav{\'e}}}]{Yang2009}
{Yang}, Y., {Zabludoff}, A., {Tremonti}, C., {Eisenstein}, D., \& {Dav{\'e}},
  R. 2009, \apj, 693, 1579

\bibitem[{{Yang} {et~al.}(2006){Yang}, {Zabludoff}, {Dav{\'e}}, {Eisenstein},
  {Pinto}, {Katz}, {Weinberg}, \& {Barton}}]{Yang2006}
{Yang}, Y., {Zabludoff}, A.~I., {Dav{\'e}}, R., {et~al.} 2006, \apj, 640, 539

\bibitem[{{Yang} {et~al.}(2014{\natexlab{b}}){Yang}, {Zabludoff}, {Jahnke}, \&
  {Dav\'e}}]{Yang2014b}
{Yang}, Y., {Zabludoff}, A.~I., {Jahnke}, K., \& {Dav\'e}, R.
  2014{\natexlab{b}}, submitted to \apj

\end{thebibliography}

\clearpage

\appendix

\section{Previous observations of \heii and \civ in extended \lya\ nebulae.}

In Table \ref{Table3}, we compile the previous observations of \heii and \civ in extended \lya\ nebulae.

\begin{deluxetable*}{lccccccc}
\tablewidth{0pt}
\tabletypesize{\small}
\tabletypesize{\scriptsize}
\tablecaption{Properties of \heii and \civ emission from LABs in the literature.}
\tablehead{
\colhead{Object                    }&
\colhead{F (Ly$\alpha$)            }&
\colhead{SB (Ly$\alpha$)           }&
\colhead{Max. extent               }&
\colhead{F (CIV)                  }&
\colhead{F (HeII)                 }&
\colhead{Aperture                         }&
\colhead{Reference           }\\[0.5ex]
\colhead{}&
\colhead{(1)}&
\colhead{(2)}&
\colhead{(3)}&
\colhead{(4)}&
\colhead{(5)}&
\colhead{(6)}&
\colhead{}
}
\startdata
LABd05$^a$    &  28.9(NB)/3.10 (spectrum)  &  9.20/45.9      & 20                   &  0.42         &   0.41      &  4.5$\arcsec \times$ 1.5$\arcsec$	      & \citealt{Dey2005}		       \\
PRG1	      &  4.36			   &  58.1	     & 5.0		    &  0.21	    &	0.57	  &  5.0$\arcsec \times$ 1.5$\arcsec$	      & \citealt{Prescott2009}  	       \\
PRG2	      &  4.92			   &  41.8	     & 7.84		    &  0.18	    &	0.18	  &  7.84$\arcsec \times$ 1.5$\arcsec$        & \citealt{Prescott2013}  	       \\ 
PRG3	      &  1.02			   &  12.1	     & 5.60		    &  $<$0.08      &	$<0.09$   &  5.60$\arcsec \times$ 1.5$\arcsec$        & \citealt{Prescott2013}  	       \\ 
PRG4	      &  1.03			   &  40.9	     & 1.68		    &  $<$0.08      &	0.07	  &  1.68$\arcsec \times$ 1.5$\arcsec$        & \citealt{Prescott2013}  		
\enddata
\tablecomments{
(1) \lya line flux in $10^{-16}$ \unitcgsflux,
(2) \lya surface brightness in $10^{-18}$ \unitcgssb,
(3) maximum extent in arcsec.
(4) \civ line flux in $10^{-16}$ \unitcgssb,
(5) \heii line flux in $10^{-16}$ \unitcgssb.
(6) Apertures used to extract the values by the authors in the references.
}
\tablenotetext{a}{The author of the reference quoted a conservative aperture of 10 arcsec radius in which they calculated all their quantities in the narrow-band (NB) image.}
\label{Table3}
\end{deluxetable*}

\section{Photoionization Modeling}
In this work we have presented results of photoionization
models of LABs.  A complete description and more detailed analysis of 
dependence of our models on the input parameters will be presented in a 
future paper (Arrigoni-Battaia in prep.). In this Appendix, we provide additional 
information on how the parameters of the photoionization models were 
chosen.

Our photoionization modeling was restricted to cloud column densities
of $\log N_{\rm H}$ $\leq$ 22 because for larger columns the implied
total gas mass of the nebula alone becomes too large. Quasars at
$z\sim 2-3$ are hosted by dark matter halos of
$M_{DM}=10^{12.5}M_{\odot}$ (\citealt{White2012}), and there is
circumstantial evidence based on the strong clustering of LABs that
they inhabit a similar mass scale (\citealt{Yang2010}).
The total mass of cool ($\sim 10^4$ K) gas in our simple model can be
shown to be (\citealt{Hennawi2013}):
\begin{equation}
  M_c = 3.3\times10^{10}\, \left(\frac{R}{100 {\rm kpc}}\right)^2 \left(\frac{f_C}{1.0}\right) \left(\frac{N_{\rm H}}{10^{20}\, {\rm cm}^{-2}}\right) {\rm M}_{\odot}.
\end{equation}
Note that this value is reasonable, given the recent estimate by 
\citet{Prochaska2013} that show that the cool gas mass of the CGM of such massive
halos is  $M_c>10^{10}$ M$_{\odot}$, based on absorption line spectroscopy. 
As the smooth morphology of LAB emission 
constraints the covering factor to be $f_{\rm C} > 0.5$, 
we consider models up to $\log N_{\rm H}$ = 22, which would result in 
very high cool gas masses $M_c = 10^{12.2}M_{\odot}$, for the lowest covering factor, $f_C=0.5$. 

Additionally, we limit $n_{\rm H}$ to be $\leq$\,100\,cm$^{-3}$. Although
such high densities are typically adopted in the previous modeling of EELR
around HzRGs (e.g., \citealt{Humphrey2008}, \citealt{Matsuoka2009}),
for halo gas on a scales of $\sim 100\,{\rm kpc}$, i.e. in the
so-called circumgalactic medium (CGM), this would represent an extreme
gas densities. Indeed, for gas in the CGM of QSO halos, gas densities
this high can be ruled out by absorption line observations using
background QSOs (e.g. \citealt{Hennawi2006, Hennawi2007}). For example
\citet{Prochaska2009} used absorption in the collisionally excited
\ion{C}{2}$^\ast$ fine-structure line to obtain an estimate of $n_{\rm
  H}\simeq 1\,{\rm cm^{-3}}$ at an impact parameter of $R_{\perp}=
108\,{\rm kpc}$, however weak or absent \ion{C}{2}$^\ast$ in the
majority of sightlines probing the QSO CGM suggests that even $n_{\rm
  H}= 1\,{\rm cm^{-3}}$ is an extreme value. Note further that the
ratio $N_{\rm H}/N_{\rm H}$ is roughly the size of the emitting clouds, and even
for the largest values of $N_{\rm H}\sim10^{21}$ cm$^{-2}$, densities as
large as $N_{\rm H}=100$ cm$^{-3}$ would imply extremely small cloud sizes
of the order of parsecs, and even more implausibly small values for
lower $N_{\rm H}$.  These limits on $n_{\rm H}$ and $N_{\rm H}$ are
particularly important in the optically thin regime where $SB_{\rm
  Ly\alpha}^{\rm thin} \propto n_{\rm H} N_{\rm H}$.

For the luminosity of the central QSO, we limit the models to $i > 16$
mag because the number density of sources with brighter ionizing
fluxes is much less than the observed number density of the LABs that
we study.  At $z \sim 3$, QSOs with $i < 17$ have a number density of
$1.16 \times 10^{-9}$ Mpc$^{-3}$ in comoving units
(\citealt{HRH2007}), whereas, although current estimates are fairly
rough, bright \lya blobs with sizes of $\sim$ 100 kpc are much more
abundant ($n$ $\sim$ $10^{-5}$--$10^{-6}$ Mpc$^{-3}$;
\citealt{Yang2009}, \citealt{Yang2010}).  For reference, the quasar
luminosity function of \citet{HRH2007}, implies that QSOs with $23 < i
< 21$ have a number density of $\sim3\times10^{-6} $ Mpc$^{-3}$ at
$z=3.1$, comparable to that of LABs.

Our photoionization models assume a single population of clouds with
the same properties, and we vary the ionization parameter (by changing
$N_{\rm H}$ and the source luminosity).  However, it has been 
argued  that a single population of constant-density clouds is not able
to simultaneously explain both the high and low ionization lines
around HzRGs, and instead a mixed population of completely
ionized clouds and partially ionized clouds is invoked (e.g.,
\citealt{Binette1996}), or the clouds are assumed to be in pressure
equilibrium with the ionizing radiation (\citealt{Dopita2002},
\citealt{Stern2014}).  It is unclear whether multiple cloud populations
need to be invoked to explain the LABs, given the sparseness of the current
data on emission line ratios, and this issue clearly goes beyond the scope of
the current work, but should be revisited when more data are available.

\clearpage

\end{document}